\documentclass[%
 reprint,
superscriptaddress,
floatfix,
 amsmath,amssymb,
 aps
]{revtex4-1}

\usepackage[utf8]{inputenc}
\usepackage{amsmath,esint}
\usepackage{mathtools}
\usepackage[colorlinks=true,linkcolor=blue,citecolor=blue,urlcolor=blue]{hyperref}
\usepackage{amsfonts}
\usepackage{amssymb}
\usepackage{bm}
\usepackage{textcomp}
\usepackage{empheq}
\usepackage{siunitx}
\usepackage[titletoc,title]{appendix}
\usepackage{graphicx}
\usepackage{dcolumn}
\usepackage{bm}
\usepackage{subcaption}
\usepackage{xcolor,soul}

\renewcommand{\Re}{\operatorname{Re}}
\renewcommand{\Im}{\operatorname{Im}}

\DeclareMathOperator{\Tr}{Tr}
\newcommand{\citeasnoun}[1]{Ref.~\cite{#1}}

\newcommand{\Figref}[1]{Figure~\ref{fig:#1}}
\newcommand{\figref}[1]{Fig.~\ref{fig:#1}}
\renewcommand{\eqref}[1]{Eq.~(\ref{eq:#1})}
\newcommand{\Eqref}[1]{Equation~(\ref{eq:#1})}
\newcommand{\eqreftwo}[2]{Eqs.~(\ref{eq:#1},\ref{eq:#2})}
\newcommand{\eqrefrange}[2]{Eqs.~(\ref{eq:#1})--(\ref{eq:#2})}

\newcommand{\tens}[1]{\bm{#1}}
\newcommand{\cc}[1]{\overline{#1}}
\newcommand{\secref}[1]{Sec.~\ref{sec:#1}}

\newcommand*{\Ev}{\mathbf{E}}
\newcommand*{\Hv}{\mathbf{H}}
\newcommand*{\Dv}{\mathbf{D}}
\newcommand*{\Bv}{\mathbf{B}}
\newcommand*{\xv}{\mathbf{x}}
\newcommand*{\pv}{\mathbf{p}}
\newcommand*{\mv}{\mathbf{m}}
\newcommand*{\Pv}{\mathbf{P}}
\newcommand*{\Mv}{\mathbf{M}}

\newcommand*{\epso}{\varepsilon_0}

\newcommand*{\SM}{SM}
\newcommand*{\alphaL}{\alpha_{\rm LDOS}}
\newcommand*{\avg}[1]{\left\langle #1 \right\rangle}
 
\newcommand{\planck}{\Theta}
\newcommand*{\alphaC}{\alpha_{\rm CDOS}}
  
\begin{document}

\preprint{APS/123-QED}

\title{Fundamental limits to near-field optical response, over any bandwidth}

\author{Hyungki Shim}
\email{hyungki.shim@yale.edu}
\affiliation{Department of Applied Physics and Energy Sciences Institute, Yale University, New Haven, Connecticut 06511, USA}
\affiliation{Department of Physics, Yale University, New Haven, Connecticut 06511, USA}
\author{Lingling Fan}
\affiliation{Department of Applied Physics and Energy Sciences Institute, Yale University, New Haven, Connecticut 06511, USA}
\affiliation{School of Physics and National Laboratory of Solid State Microstructures, Nanjing University, Nanjing 210093, China}
\author{Steven G. Johnson}
\affiliation{Department of Mathematics, Massachusetts Institute of Technology, Cambridge, Massachusetts 02139, USA}
\author{Owen D. Miller}
\email{owen.miller@yale.edu}
\affiliation{Department of Applied Physics and Energy Sciences Institute, Yale University, New Haven, Connecticut 06511, USA}

\date{\today}

\begin{abstract}
    We develop an analytical framework to derive upper bounds to light--matter interactions in the optical near field, where applications ranging from spontaneous-emission amplification to greater-than-blackbody heat transfer show transformative potential. Our framework connects the classic complex-analytic properties of causal fields with newly developed energy-conservation principles, resulting in a new class of \emph{power--bandwidth limits}. These limits demonstrate the possibility of orders-of-magnitude enhancement in near-field optical response with the right combination of material and geometry. At specific frequency and bandwidth combinations, the bounds can be closely approached by canonical plasmonic geometries, with the opportunity for new designs to emerge away from those frequency ranges. Embedded in the bounds is a material ``figure of merit,'' which determines the maximum response of any material (metal/dielectric, bulk/2D, etc.), for any frequency and bandwidth. Our bounds on local density of states (LDOS) represent maximal spontaneous-emission enhancements, our bounds on cross density of states (CDOS) limit electromagnetic-field correlations, and our bounds on radiative heat transfer (RHT) represent the first such analytical rule, revealing fundamental limits relative to the classical Stefan--Boltzmann law.
\end{abstract}

\pacs{Valid PACS appear here}
\maketitle

The electromagnetic \emph{near field} comprises large-amplitude evanescent fields that can be harnessed to amplify spontaneous-emission rates~\cite{okamoto_niki_2004, englund_fattal_2005, noda_fujita_asano_2007,russell_liu_cui_hu_2012, novotny_hecht_2012,akselrod2014}, Casimir forces~\cite{casimir_1948, lifshitz_1956, lamoreaux_2004, munday2009, rodriguez2011}, Raman scattering~\cite{moskovits_1985, nie_emory_1997, kneipp_wang_kneipp_perelman_itzkan_dasari_feld_1997,zhang2013,vendrell2013,thacker2014}, and greater-than-blackbody transfer of thermal energy~\cite{polder_hove_1971, rytov_kravtsov_tatarskii_1988, pendry_1999, kim2015, song2015, stgelais2016,miller_johnson_rodriguez_2015}. Yet little is known about \emph{maximal} response---how large can such enhancements be? In this Article, we derive fundamental limits for local density of states (LDOS), the prototypical near-field optical response, for any bandwidth of interest and for any material platform. We use these bounds to then derive fundamental limits for emerging quantities of interest---cross density of states (CDOS, a useful field-correlation measure), and radiative heat transfer (RHT), where our bounds depend only on the temperatures, materials, and separation distance of the bodies involved. Conceptually, the bounds arise because LDOS and related near-field quantities are given by the real (or imaginary) parts of causal linear-response functions. We use the complex-analytic properties of such functions to transform bandwidth-averaged response to that at a single, \emph{complex}-valued frequency, where we develop generalized energy-conservation constraints, ultimately leading to bounds over arbitrary bandwidths. A distinctive feature of our arbitrary-bandwidth approach is that it predicts a simple material figure of merit (FOM) that determines the maximum possible response of \emph{any} material (metal, dielectric, 2D, 3D, etc.), for \emph{any} frequency and bandwidth. In the case of RHT, this FOM provides insight into which materials can facilitate optimal heat transfer for any temperature. There is significant ongoing debate about whether a plasmonic or an all-dielectric approach is better, and in which scenarios 2D materials might be better than conventional bulk materials. Unlike all previous bounds and sum rules~\cite{gordon_1963, drell_hearn_1966, purcell_1969, detar_freedman_veneziano_1971, aitchison_1972, mckellar_box_bohren_1982, sohl_gustafsson_kristensson_2007, carminati_saenz_2009, jackson_2013, barnett_loudon_1996, sanders_manjavacas_2018,miller_johnson_rodriguez_2015,miller_polimeridis_reid_hsu_delacy_joannopoulos_soljacic_johnson_2016, miller_ilic_christensen_reid_atwater_joannopoulos_soljacic_johnson_2017,kwon_pozar_2009,ruan_fan_2011, liberal_ederra_gonzalo_ziolkowski_2014,hugonin_besbes_ben-abdallah_2015,yang_miller_christensen_joannopoulos_soljacic_2017,Yang2018}, the material figure of merit we derive here enables general quantitative answers to these questions. In a frequency--bandwidth phase space, we map out which materials are optimal and where the critical thresholds, from dielectric to plasmonic and bulk to 2D, occur. The techniques developed herein for LDOS, CDOS, and radiative heat transfer should be extensible to other near-field effects ranging from engineered Lamb shifts~\cite{fussell_mcphedran_sterke_2005, chang_rivera_joannopoulos_soljacic_kaminer_2017} and F{\"o}rster resonance energy transfer~\cite{dung_knoll_welsch_2002, gonzaga-galeana_zurita-sanchez_2013} to nonlinear (Raman) or fluctuation-induced (Casimir) phenomena.

Near-field electromagnetism, in which localized sources interact with scatterers separated by less than an optical wavelength, offers transformative potential for wide-ranging applications. Quantum emitters that only weakly couple to the radiation continuum can be dramatically amplified by near-field engineering: optical antennas offer prospects for imaging single molecules~\cite{nie_emory_1997, kneipp_wang_kneipp_perelman_itzkan_dasari_feld_1997, zanten_cambi_koopman_joosten_figdor_garcia-parajo_2009, schermelleh_heintzmann_leonhardt_2010} or for designing nanoscale light-emitting diodes that are faster than lasers~\cite{lau_lakhani_tucker_wu_2009}. Nonlinear emitters such as Raman-active molecules~\cite{barron_buckingham_1971, lim_jeon_kim_nam_suh_2009} experience even more dramatic enhancements: surface-enhanced Raman scattering (SERS)~\cite{moskovits_1985, nie_emory_1997, kneipp_wang_kneipp_perelman_itzkan_dasari_feld_1997}, for example, scales with the square of the spontaneous-emission enhancement rate. Thermal emission can be accessed and controlled for the productive transfer of heat energy, at rates many orders of magnitude beyond classical blackbodies~\cite{polder_hove_1971, rytov_kravtsov_tatarskii_1988, pendry_1999, kim2015, song2015, stgelais2016,miller_johnson_rodriguez_2015}. The emission can be stimulated by the vacuum itself: the field of Casimir physics is probing a vast expanse of materials and structures to explore how vacuum forces can be controlled and manipulated~\cite{casimir_1948, lifshitz_1956, spruch_1996, bordag_mohideen_mostepanenko_2001, lamoreaux_2004, ball_2007, rodriguez_capasso_johnson_2011, johnson_2011, reid_white_johnson_2013}.

\begin{figure*} [t!]
    \includegraphics[width=1\textwidth]{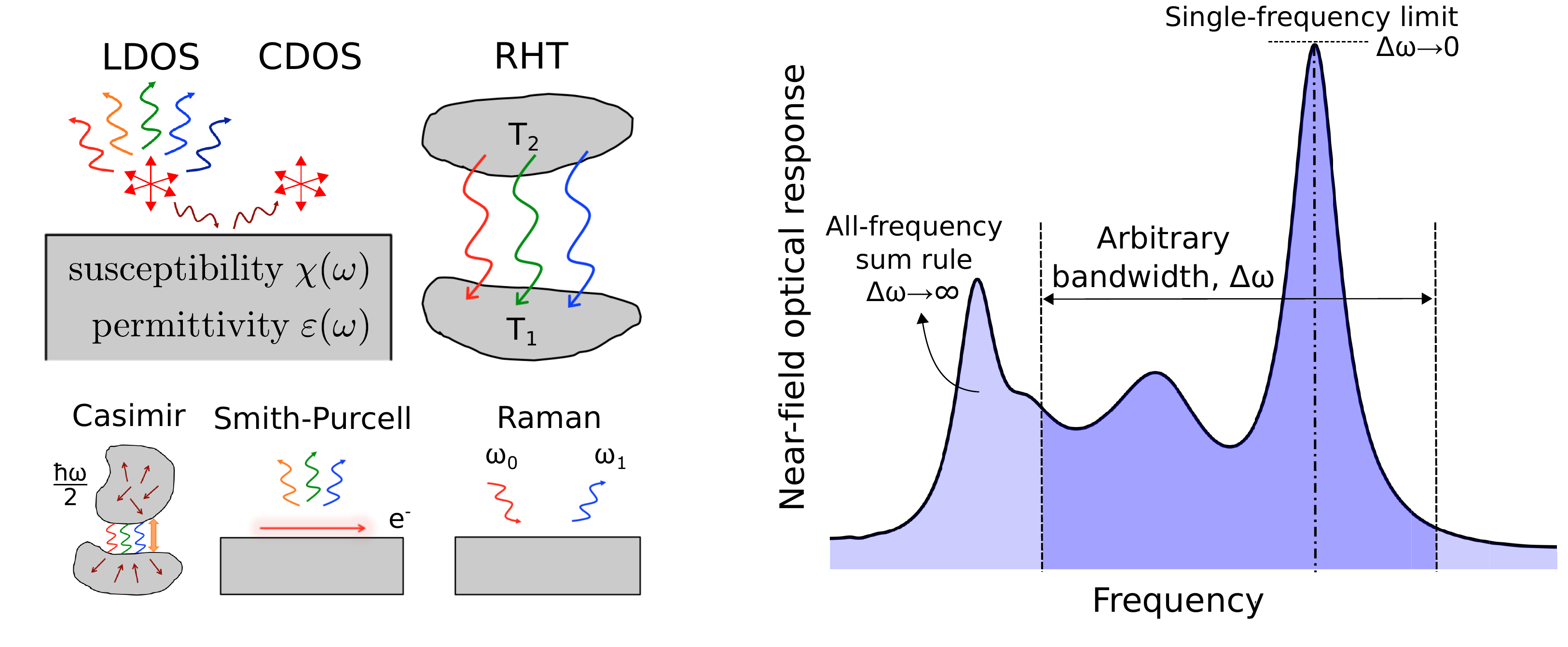} 
    \caption{Near-field optics includes phenomena ranging from spontaneous-emission enhancements through local density of states (LDOS) engineering, field-correlation phenomena as measured by cross density of states (CDOS), radiative heat transfer (RHT), Casimir effects, Smith--Purcell radiation, and Raman scattering. Our theoretical framework, connecting causality principles with energy-conservation constraints, yields bounds over any arbitrary bandwidth. In the limit of zero bandwidth, we obtain recently discovered single-frequency bounds \cite{miller_polimeridis_reid_hsu_delacy_joannopoulos_soljacic_johnson_2016, miller_ilic_christensen_reid_atwater_joannopoulos_soljacic_johnson_2017}; in the infinite-bandwidth limit, we arrive at a sum rule for integrated all-frequency response.}  
    \label{fig:cartoon} 
\end{figure*} 

For such a broad scope of applications, there is a fundamental theoretical question that remains unanswered: for a given bandwidth of interest, what is the maximum near-field response that is possible? Sum rules enable at least a partial answer. Relying on the analytic properties embedded within particular response functions---such as susceptibility and cross-sections---sum rules relate integrated response over \emph{all frequencies} to behavior at a single frequency, and have been derived in a variety of classical and quantum frameworks~\cite{gordon_1963, drell_hearn_1966, purcell_1969, detar_freedman_veneziano_1971, aitchison_1972, mckellar_box_bohren_1982, sohl_gustafsson_kristensson_2007, carminati_saenz_2009, jackson_2013}. In the near field, there is a well-known sum rule for spontaneous emission~\cite{barnett_loudon_1996} that suggests spontaneous-emission enhancements average out to zero over all frequencies. Yet this sum rule neglects the longitudinal component of the Green's function, thus neglecting the near-to-far-field coupling that is crucial for spontaneous-emission engineering (and hence recovering a far-field refractive-index sum rule). Very recently, a near-field sum rule for electric LDOS has been derived~\cite{sanders_manjavacas_2018}, which represents a specialized version of a sum rule that we derive in \eqref{Vint}. However, sum rules make no predictions as to how a \emph{finite} available bandwidth affects maximal response. The difference between infinite and finite bandwidth is stark for ``dielectric'' ($\Re \epsilon > 0$) materials, because infinite-bandwidth sum rules include technologically irrelevant high-frequency contributions that dominate the integrated response. For example, when applied to Silicon, the sum rule for plane-wave scattering is dominated by contributions at energies on the order of {\SI{100}{eV}}~\cite{yang_antosiewicz_verre_abajo_apell_kall_2015}, thus providing little insight into maximal response over typical bandwidths of interest (infrared, visible, etc.).

At the other extreme, \emph{single-frequency} limits to power extinction and other physical observables have been discovered in both the near field~\cite{miller_johnson_rodriguez_2015, miller_polimeridis_reid_hsu_delacy_joannopoulos_soljacic_johnson_2016, miller_ilic_christensen_reid_atwater_joannopoulos_soljacic_johnson_2017,Yang2018} and far field \cite{hamam_karalis_joannopoulos_soljacic_2007, kwon_pozar_2009, ruan_fan_2011, liberal_ederra_gonzalo_ziolkowski_2014, hugonin_besbes_ben-abdallah_2015, yang_miller_christensen_joannopoulos_soljacic_2017} based on energy-conservation principles, but they necessarily fail to account for the effects of nonzero bandwidth. (As an example, they predict infinite maximal response for any lossless material. Such a prediction is in fact correct---it is possible to make LDOS arbitrarily large~\cite{liang_johnson_2013}. But the bounds developed herein show that the average response over any \emph{non}zero bandwidth has a finite upper bound). Thus, previous approaches do not provide any meaningful metric for lossless dielectrics at optical frequencies, either for quantitative comparisons amongst each other, or to plasmonic and other metallic systems.

The key idea of our work is that two seemingly independent ideas---causality for sum rules and energy-conservation principles for single-frequency bounds---can be unified into a single framework that yield bounds for \emph{any} bandwidth, as illustrated in \figref{cartoon}. In this framework, single-frequency bounds and all-frequency sum rules emerge as asymptotic limits of a more general arbitrary-bandwidth approach. Our bounds over arbitrary bandwidths, which we term ``power--bandwidth limits,'' arise by connecting the properties that enable sum rules to those that enable energy-conservation principles. Sum rules for power quantities (such as optical cross-section) require one to be able to compute the quantity by taking the imaginary (or real) part of some \emph{amplitude}---for extinction, the optical theorem~\cite{newton_1976, bohren_clothiaux_huffman_1983, lytle_carney_schotland_wolf_2005, jackson_2013} guarantees such a form in terms of the \emph{scattering amplitude}. The amplitude is a causal linear-response function and thus analytic in the upper-half of the complex-frequency plane~\cite{nussenzveig_1972}. With suitable boundary conditions, a Hilbert transform (i.e., a Kramers--Kronig-like transform) then enables a sum rule, relating integrated response over all frequencies to that of a single frequency. Conversely, energy-conservation bounds---recognized primarily within the past decade~\cite{miller_johnson_rodriguez_2015, miller_polimeridis_reid_hsu_delacy_joannopoulos_soljacic_johnson_2016, miller_ilic_christensen_reid_atwater_joannopoulos_soljacic_johnson_2017, hamam_karalis_joannopoulos_soljacic_2007, kwon_pozar_2009, ruan_fan_2011, liberal_ederra_gonzalo_ziolkowski_2014, hugonin_besbes_ben-abdallah_2015, yang_miller_christensen_joannopoulos_soljacic_2017,Yang2018}---exploit the power-quantity-by-amplitude form in a different way. In writing a power quantity as the imaginary part of an amplitude, the amplitude itself is \emph{linear} in the electromagnetic fields and/or currents (holding the incident field fixed). By contrast, many power quantities---absorption, scattering, radiation, etc.---are explicitly \emph{quadratic} functionals of the fields and/or currents. If it can be shown that the linear quantity must be larger than the quadratic one, then an upper bound can be derived. Remarkably, it is the same optical theorem that provides such a constraint. Thus we see that sum rules and single-frequency bounds both start with particular response functions that can be written as the imaginary part of an amplitude, but diverge in their approaches thereafter. (For response functions that do not admit such expressions, the power--bandwidth limits established above can be applied by transformation to the real/imaginary part of a causal linear-response function.)

We connect the sum-rule and single-frequency approaches through use of a ``window function,'' an averaging function over a prescribed bandwidth. In general, such a function will have one (or multiple) poles in the upper-half plane (UHP), and thus the typical contour-integral analysis of a given power quantity requires the computation of residues not at a single real frequency (as in sum rules), but at a discrete set of \emph{complex} frequencies. At this juncture we identify the energy-conservation constraints at those complex frequencies, deriving bounds to how large they can be. This multistep procedure (fleshed out in detail below) thus provides perhaps the first approach to arbitrary-bandwidth bounds. For maximal clarity, we start with local density of states---the prototypical optical response---in \secref{ldos}. We first derive near-field sum rules for LDOS (\secref{sumrule}), showing that near-field response integrated over all frequencies must equal a new electrostatic constant, $\alphaL$. Then we use geometric perturbation theory to prove a monotonicity theorem that enables us to bound the electrostatic constant itself in terms of only the material permittivity and near-field separation distance (\secref{mono}). We introduce the window function in \secref{power}, combining it with the complex-frequency energy-conservation idea to develop arbitrary-bandwidth bounds and show how closely they can be approached for specific choices of frequency and bandwidth by canonical structures. Having established the theoretical bound framework, we derive general bounds for cross density of states (\secref{cdos}) and near-field radiative heat transfer (\secref{RHT}). Emerging within all these bounds is a common material figure of merit, and in \secref{materials} we discuss the physical intuition of the FOM and compare a wide variety of materials across frequency and bandwidth. Finally, in \secref{extensions}, we discuss extensions of our formulation to near-field phenomena such as Lamb shifts, Raman scattering, Casimir forces, and more.

\section{Local density of states (LDOS)} \label{sec:ldos}
In this section, we develop a theoretical framework for upper bounds to  near-field optical-response functions. The prototypical near-field interaction is the alteration---and potentially dramatic enhancement---of spontaneous emission from a two-level dipolar transition in a quantum emitter by an inhomogeneous environment. The power radiated by such an emitter, and thus the spontaneous-emission rate enhancement, is proportional to the \emph{local density of states} (LDOS)~\cite{martin_piller_1998, daguanno_2004, joulain_carminati_mulet_greffet_2003, scheel_2008, novotny_hecht_2012, taflove_oskooi_johnson_2013, liang_johnson_2013}. It has long been recognized that changing the electromagnetic environment of an emitter alters its spontaneous-emission rate~\cite{purcell_torrey_pound_1946, drexhage_1974, yablonovitch_1987}; applications where such enhancements play an important role include single-molecule imaging~\cite{nie_emory_1997, kneipp_wang_kneipp_perelman_itzkan_dasari_feld_1997}, micro-LED design~\cite{Eggleston2015}, and photovoltaics~\cite{Callahan2012}. Mathematically, the spontaneous-emission rate is determined by the imaginary part of the total Green's function~\cite{Wijnands1997, xu_lee_yariv_2000, taflove_oskooi_johnson_2013}. To avoid unwieldy expressions and derivations, we will use six-vector notation for fields and currents, treating the electric and magnetic fields, and electric/magnetic dipolar transitions, on equal footing. (We take the background to be vacuum throughout this paper and work in dimensionless units such that $\epso=\mu_0=1$, with generalization to non-vacuum background in the \SM.) We denote the fields $\psi$, the polarization currents $\phi$, and dipolar sources $\xi$:
\begin{align}
    \psi = \begin{pmatrix}
        \Ev \\
        \Hv
    \end{pmatrix}, \quad
    \phi = \begin{pmatrix}
        \Pv \\
        \Mv
    \end{pmatrix}, \textrm{ and } \quad
    \xi = \begin{pmatrix}
        \pv \\
        \mv
    \end{pmatrix}.
    \label{eq:defs}
\end{align}
Then the spontaneous-emission rate of randomly oriented electric ($\pv$) and magnetic ($\mv$) dipoles at a point $\xv_0$ is modified relative to its free-space rate by the scattered-field contribution to the LDOS:
\begin{align}
    \rho(\xv_0,\omega) &= \Im \sum_j \left[ \frac{1}{\pi \omega} \left(\cc{\pv}_j \cdot \Ev_{\textrm{s},j}(\xv_0)  + \cc{\mv}_j \cdot \Hv_{\textrm{s},j}(\xv_0) \right)\right] \nonumber \\
    &= \Im \underbrace{\sum_j \left[ \frac{1}{\pi \omega} \xi_j^\dagger(\xv_0) \psi_{\textrm{s},j}(\xv_0) \right]}_{s(\omega)} \nonumber \\
    &= \Im \Tr \left[ \frac{1}{\pi\omega} \tens{\Gamma}_{\rm s}(\xv_0,\xv_0,\omega) \right], \label{eq:rho}
\end{align}
where $\tens{\Gamma}$ denotes the $6\times 6$ dyadic Green's function, the ``s'' subscripts denote scattered-field contributions (thus $\tens{\Gamma}_s$ is the total Green's function minus the free-space Green's function), and we use the convention~\cite{joulain_carminati_mulet_greffet_2003} that LDOS is the sum of electric-dipole and magnetic-dipole contributions. It is important to subtract off the free-space rate, and consider only the scattered-field contribution, to ensure sufficiently fast decay at high frequency and thus convergence of integrals over frequency. The random dipole orientation (for dipoles of unit amplitude, i.e. $\xi^\dagger \xi = 1$) is encoded in the summation over directions $j=\{x,y,z\}$ and ultimately the trace of the Green's function. In \eqref{rho} we denote a term $s(\omega)$ (suppressing the implicit position dependence) that we identify as a \emph{near-field scattering amplitude}, as measured at the location of the emitter. It is this term that enables the sum rule and the power--bandwidth limits.

Maxwell's equations do not prevent us from taking the frequency to be complex. In the complex-$\omega$ plane, we can define the complex extension of the near-field scattering amplitude as
\begin{align} 
    s(\omega) =  \sum_j \frac{1}{\pi \omega} \xi_j^T(\xv_0) \psi_{\textrm{s},j}(\xv_0), \quad \omega \in \mathbb{C},  \label{eq:s(w)}
\end{align}
where we have made the typical assumption that the dipole amplitudes are real-valued, such that $\xi^\dagger = \xi^T$. (For complex dipole amplitudes, a few additional steps in the derivations below are needed, but the results remain unchanged.) Since $\xi$ is constant (analytic everywhere), and the scattered field $\psi_{\rm s}$ is a causal linear-response function~\cite{nussenzveig_1972}, the amplitude $s(\omega)$ is analytic in the upper half of the complex-$\omega$ plane. This is analogous to the classical result that quantities such as refractive index and far-field scattering amplitudes are analytic in the upper half plane (UHP)~\cite{nussenzveig_1972}. On the real line, $s(\omega)$ has a pole at the origin, due to the singularity in the $1/\pi\omega$ prefactor and the fact that $\xi^T \psi_s$ (usually) has a nonzero value in the zero-frequency limit.

\subsection{Sum Rules} \label{sec:sumrule}
\begin{figure*} [t!]
    \includegraphics[width=1\linewidth]{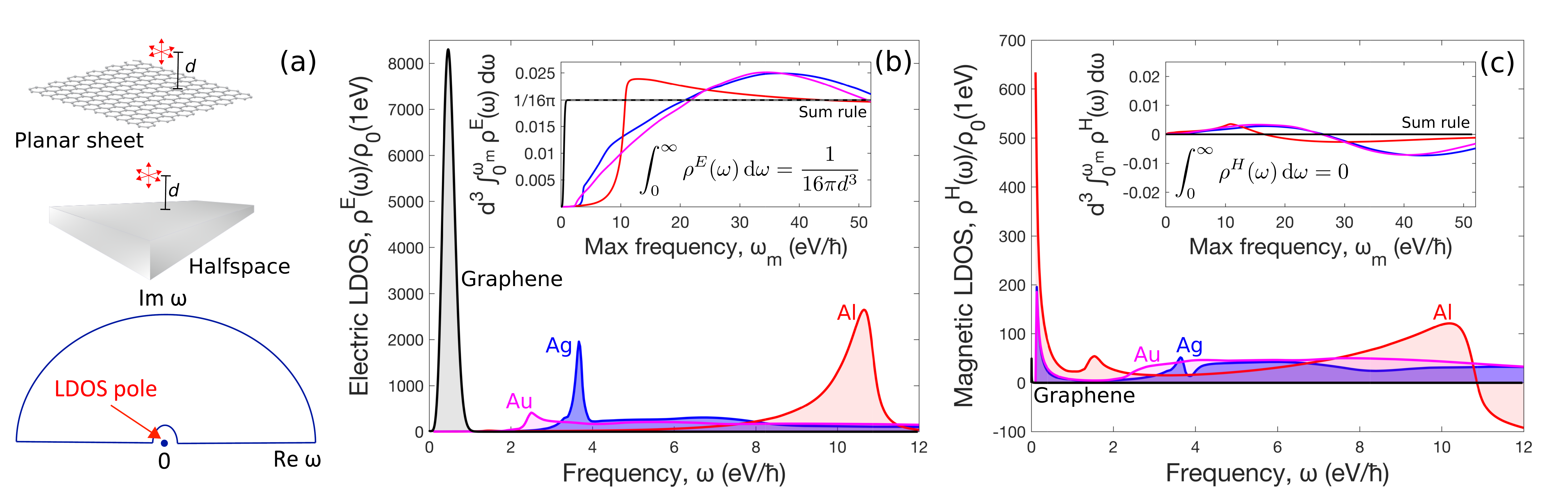} 
    \caption{(a) Contour of integration in the complex-$\omega$ plane used to obtain the sum rule for LDOS, which has a simple pole at the origin and is analytic everywhere on the upper half plane. For high-symmetry geometries such as a halfspace or planar sheet, the sum-rule constant is known analytically. (b) Electric and (c) magnetic LDOS for Ag, Al, Au (halfspaces), and graphene (planar sheet). Although they all have resonant peaks at different frequencies with varying amplitudes and widths, their integrated response converges to the same constant, $1/16\pi d^3$, as shown in the inset. The emitter-scatterer distance \textit{d} is set to 10nm, and the Fermi level for graphene set to 0.6eV.}  
    \label{fig:sumrule} 
\end{figure*}

Thus a sum rule can be derived for $\rho(\xv_0,\omega)$ through contour integration of $s(\omega)$ in the UHP. If we enclose the UHP in a typical contour that is semicircular going to infinity, and follows the real line with a ``bump'' at the origin (see \figref{sumrule}(a)), then analyticity ensures that the total integral is zero. For local, linear susceptibilities, $s(\omega)$ falls off sufficiently rapidly as $\omega \rightarrow \infty$ (because the free-space contribution was subtracted out), such that the real-line integral is well-defined and the contour at infinity does not contribute (shown explicitly in the \SM). Thus there are two contributions to the integral: the (principal value of the) integral over the real line, and the residue of the simple pole at zero. Because negative-real-frequency fields are conjugates of their positive counterparts~\cite{landau_lifshitz_1984}, $\rho(-\omega) = \rho(\omega)$, such that the real-line integral can be reduced to only positive frequencies, after which a few algebraic steps (\SM) yields a general expression for the value of $\rho(\xv_0,\omega)$ integrated over all frequencies:
\begin{align}
    \int_0^\infty \rho(\omega) \,{\rm d}\omega = \alphaL, \label{eq:Vint}
\end{align}
where  $\alphaL = \frac{1}{2} \Re \left[\Tr \tens{\Gamma}_s(\xv_0,\xv_0)\big\rvert_{\omega=0}\right]$. The electrostatic constant $\alphaL$ measures the scattered field at the position of the dipole source, and is shown in the {\SM} to always be positive. \Eqref{Vint} marks our first result: over all frequencies, integrated LDOS must equal an electrostatic constant. (An electric-only specialization of \eqref{Vint} was very recently discovered~\cite{sanders_manjavacas_2018}, albeit without the analytical bounds to follow.) For materials with an electrical conductivity (e.g.  metals), $\alphaL$ is \emph{independent} of the value of the conductivity, for \textit{any} geometry. More generally, \eqref{Vint} applies to any material (whose susceptibility decays in the limit $\omega \rightarrow \infty$), including the wide array of newly emerging 2D materials. 

Some care is required with \eqref{Vint} in singular situations. At high-symmetry points near high-symmetry scatterers, e.g. at the center of a hemispherical bowl, LDOS may appear to not decay at frequencies going to infinity because the optical rays reflect off the perfect spherical interface and constructively interfere at the origin. However, \emph{any} random deviation from a hemisphere, no matter how small, would destroy such interference at high enough frequencies, restoring the natural, sufficiently rapid decay. Hence the correct approach to regularizing such singularities is to compute the sum rule for a hemisphere with imperfections, and then take the limit as the imperfections go to zero, such that \eqref{Vint} still applies. (Such a procedure is a geometrical analog of the limiting absorption principle~\cite{schulenberger1971}, defining ``lossless'' materials in the limit as loss goes to zero from above.)

For any scatterer, the constant $\alphaL$ can be found from an electrostatic calculation. In high-symmetry geometries the calculation may be analytically tractable. Consider a halfspace of permittivity $\varepsilon$ and permeability $\mu$. (In this paper, we will consider materials that have scalar material susceptibilities, with tensor-valued generalizations in the \SM.) The value of $\alphaL$ at a separation $d$ can be computed via the image charge within the halfspace, leading to an expression (derived in the \SM) for $\alphaL$, and thus of the frequency-integrated LDOS, of
\begin{align}
    \int_0^\infty \rho(\omega) \,{\rm d}\omega = \frac{1}{16\pi d^3}\left[ \frac{\varepsilon(0)-1}{\varepsilon(0)+1} + \frac{\mu(0)-1}{\mu(0)+1} \right] \label{eq:canonical} 
\end{align}
where $\varepsilon(0)$ and $\mu(0)$ are the zero-frequency (electro-/magnetostatic) permittivities and permeabilities, respectively. For metals and any material with a conductivity, the permittivity and/or permeability diverges in the zero-frequency limit, such that the corresponding term in square brackets in \eqref{canonical} simplifies to 1. This is also the case for any 2D conductive sheet, which at zero frequency represents the same perfect-conductor boundary condition as a conductive 3D halfspace. In the \SM, we also derive the simple $\alphaL$ expression for conductive spheres. 

An interesting feature of the LDOS sum rule is that it depends only on zero-frequency behavior, where electric and magnetic fields decouple. For the nonmagnetic materials that are ubiquitous at optical frequencies, this implies very different behavior for electric-dipole sources (i.e. electric-dipole transitions) as compared to magnetic-dipole sources (magnetic-dipole transitions). To illustrate the difference, one can separately define electric LDOS, $\rho^E$, as the component arising from the electric sources only, and magnetic LDOS, $\rho^H$, as arising from the magnetic sources only:
\begin{align}
    \rho(\omega) &= \rho^E(\omega) + \rho^M(\omega) \nonumber \\
    &= \Im \left[ \frac{1}{\pi \omega} \left( \Tr \tens{G}^{EE} + \Tr \tens{G}^{HH}\right) \right] \nonumber
\end{align}
where $\tens{G}^{EE}$ and $\tens{G}^{HH}$ are the electric and magnetic dyadic Green's functions, respectively. The two terms in the square brackets in \eqref{canonical} correspond to these individual LDOS constituents. For a nonmagnetic halfspace (which we take as conductive just for simplicity), the electric and magnetic LDOS sum rules are:
\begin{equation}
\begin{aligned}
    \int_0^\infty \rho^E(\omega) \,{\rm d}\omega &= \frac{1}{16\pi d^3} \\
    \int_0^\infty \rho^H(\omega) \,{\rm d}\omega &= 0. \label{eq:EM_LDOS}
\end{aligned}
\end{equation}
The magnetic LDOS must average out to zero, because in the zero-frequency limit a magnetic dipole does not interact with a nonmagnetic medium. 

\Figref{sumrule}(b,c) illustrates the generality of the electric/magnetic LDOS sum rules for bulk metals (Ag, Al, Au)~\cite{palik_2003} as well as 2D materials such as graphene (material model from~\cite{jablan_buljan_soljacic_2009, christensen_jauho_wubs_mortensen_2015}). In our figures, we normalize electric/magnetic LDOS by the free-space electric (\emph{or} magnetic) LDOS, $\rho_0 = \frac{\omega^2}{2\pi^2 c^3}$ (which is half that from Ref.~\cite{joulain_carminati_mulet_greffet_2003} as they consider total free-space LDOS). Each of these materials supports surface plasmon-polaritons~\cite{maier_2007}, which are excited by near-field sources. These materials exhibit very different resonant frequencies and linewidths, as seen in \figref{sumrule}(b,c). In terms of electric LDOS in \figref{sumrule}(b), graphene exhibits a very large and narrow-band response at infrared frequencies, whereas metals exhibit varying levels of maximum response, with inversely proportional bandwidths, at visible or ultraviolet energies. In contrast to the large order-of-magnitude enhancements for electric LDOS, the magnetic LDOS in \figref{sumrule}(c) shows only limited response---note the scale of the y-axis in \figref{sumrule}(c) relative to \figref{sumrule}(b). The modest, fluctuating magnetic LDOS arises because of the small electric field generated by the magnetic source, or, equivalently, because the magnetic source cannot induce any magnetization in nonmagnetic media. Across the wide variations of response seen for both electric and magnetic LDOS, for systems with different materials and dimensionality, the all-frequency response must converge to the sum rules of \eqrefrange{Vint}{EM_LDOS}, as shown in the insets of \figref{sumrule}(b,c).

\subsection{All-Frequency Bounds} \label{sec:mono}
\Eqref{Vint} is an \emph{equality} for any geometry. For structures without a high degree of symmetry, the electrostatic constant $\alphaL$ would typically require an electrostatic computation. In this section, we use perturbation theory to derive a ``monotonicity theorem,'' showing that if one material body (with static permittivity and permeability greater than 1) encloses a second body of the same material, the electrostatic constant $\alphaL$ must be larger for the first than for the second. With this result, we can \emph{bound} the all-frequency integrated response for \emph{any} geometry in terms of the analytically known $\alphaL$ for high-symmetry enclosures, yielding general analytical bounds.

Any outward deformation can be broken down into a sequence of outward perturbations. Thus if one can show that any outward perturbation of a geometry must increase some response function, a ``mononicity theorem'' has been proved, guaranteeing that such a function increases for any outward deformation. Such theorems are known for electrostatic polarizability under plane-wave excitations~\cite{jones_1985, sohl_gustafsson_kristensson_2007, sjoberg_2009}. To understand how $\alphaL$ changes under geometrical perturbations, we use a variational-calculus approach applicable to any frequency and then isolate the electrostatic behavior. Within the variational-calculus approach, quantities incorporating the displacement fields $\Dv$ and $\Bv$ as well as the scalar permittivity and permeability will be necessary, for which we define the six-vector field $\Psi$ and tensor $\nu$:
\begin{align}
    \Psi = \begin{pmatrix}
        \Dv \\
        \Bv
    \end{pmatrix}, \quad
    \nu = \begin{pmatrix}
        \varepsilon \mathcal{I} & \\
        & \mu \mathcal{I}
    \end{pmatrix}, 
    \label{eq:defs2}
\end{align}
where $\mathcal{I}$ is the 3$\times$3 identity matrix (and as discussed above, generalizations to anisotropic materials are included in the \SM). 

Consider a scattering figure of merit (FOM) such as LDOS. Under geometrical perturbations, the variation in the FOM can be written as an overlap integral between two fields: (1) a ``direct'' field, which is the response of the unperturbed geometry to the original source (e.g. a nearby dipolar emitter), and (2) an ``adjoint'' field, which is the response of the same unperturbed geometry to a source whose phase, amplitude, and position depends on the precise FOM~\cite{johnson_ibanescu_skorobogatiy_weisberg_joannopoulos_fink_2002, miller_2012}. For any FOM, if we write the displacement of the boundary in the normal direction as $\Delta h_n(\xv)$, the variation in the FOM can generally be written as an overlap integral of the direct and adjoint fields over the geometrical boundary~\cite{miller_2012}:
\begin{align}
\Delta \alphaL &= 2 \operatorname{Re} \int \Delta h_n
\bigg[ \Delta\varepsilon \mathbf{E_{\parallel}} \cdot \mathbf{E}^{\rm (adj)}_{\parallel} -  \Delta (\varepsilon^{-1}) \mathbf{D_{\perp}} \cdot \mathbf{D}^{\rm (adj)}_{\perp} \nonumber \\ 
& \qquad \qquad \quad + \Delta \mu \mathbf{H_{\parallel}} \cdot \mathbf{H}^{\rm (adj)}_{\parallel} - \Delta (\mu^{-1}) \mathbf{B_{\perp}} \cdot \mathbf{B}^{\rm (adj)}_{\perp} \bigg] \nonumber \\
&= 2 \Re \int \Delta h_n \left[ \psi^T_{\parallel} \Delta \nu \psi^{\rm (adj)}_{\parallel} + \Psi^T_{\perp} \nu_1^{-1} \Delta \nu \nu_0^{-1} \Psi^{\rm (adj)}_{\perp} \right] 
\label{eq:ShapeDeriv}
\end{align}
where $\Delta \varepsilon = \varepsilon_1 - \varepsilon_0$, $\Delta(\varepsilon^{-1}) = \varepsilon_1^{-1} - \varepsilon_0^{-1}$ (similarly for $\mu$), $\Delta \nu = \nu_1 - \nu_0$, $\nu_1$ and $\nu_0$ represent the material properties of the scatterer and its surroundings, respectively, while the ``$\rm{adj}$" superscript denotes adjoint field solutions. Implied in the above integral over the boundary is multiplication with an area element $\rm{d}A$ along the boundary, where $\psi_{\parallel}$ and $\psi_{\perp}$ denotes the field components parallel and perpendicular to the locally flat boundary, respectively. We have explicitly used the electrostatic constant $\alphaL$ for the figure of merit since it is that constant for which monotonicity will apply. For any figure of merit $f$, the adjoint fields are a solution with source currents given by the functional derivatives $\partial f / \partial \psi^T$. From \eqref{s(w)} and the discussion following \eqref{Vint}, we know that $\alphaL$ is given by
\begin{align}
    \alphaL = \frac{1}{2} \Re \sum_j \xi_j^T(\xv_0) \psi_{\textrm{s},j}(\xv_0),
\end{align}
which means that for any given dipole orientation $j$, the adjoint source field is given by $\partial f / \partial \psi_j^T = \frac{1}{2} \xi_j(\xv_0)$. This shows a unique circumstance: the dipolar sources for the adjoint field are exactly half of $\xi_j(\xv_0)$, which were the original LDOS dipolar excitations, such that $\psi^{\rm (adj)} = \frac{1}{2}\psi$ and $\Psi^{\rm (adj)} = \frac{1}{2} \Psi$. Moreover, at zero frequency, without any material or radiative losses, the fields can be chosen real-valued. Thus, for materials with positive static permittivities and permeabilities that are greater than those of their surroundings at zero frequency ($\nu_0$, $\nu_1$, and $\Delta \nu$ positive-definite), \eqref{ShapeDeriv} can be written as the integral of a positive quantity,
\begin{alignat}{2}
\label{eq:ShapeDeriv2}
    \Delta \alphaL &= \Re \int &&\Delta h_n \left[ \psi^T \left(\Delta \nu\right) \psi + \Psi^T_\perp \left(\nu_1^{-1} \Delta \nu \nu_0^{-1} \right) \Psi_\perp \right] \nonumber \\ 
    & > 0.&&
\end{alignat}
\Eqref{ShapeDeriv2} ensures that $\Delta \alphaL$ is positive for any outward deformation ($\Delta h_n > 0$ everywhere on the boundary). Regardless of the size and shape of a given scatterer $\Omega_1$ with constant $\alphaL^{(1)}$, we can always enclose it by another object, $\Omega_2$, whose constant $\alphaL^{(2)}$ must be larger than $\alphaL^{(1)}$, proving our monotonicity theorem:
\begin{align}
\alphaL^{(2)} >  \alphaL^{(1)} \quad \text{for} \ \Omega_2 \supset \Omega_1. \label{eq:mono}
\end{align}
By connecting this monotonicity theorem to the sum rule in \eqref{Vint}, one can see that the integrated LDOS near a scatterer must increase with the size/shape of that scatterer. Note that although our derivation does not strictly apply to the limiting case of a conductive material with $\Delta \nu \rightarrow \infty$ (because $\varepsilon\rightarrow \infty$ or $\mu \rightarrow \infty$) in the zero-frequency limit, it does apply for any arbitrarily large but finite material susceptibility, and in the \SM\ we provide an alternative proof that confirms the validity of \eqref{mono} for conductive materials. 

The use of \eqref{ShapeDeriv} assumes a smooth perturbation of the boundary. In the case of a region with ``kinks,'' or sharp corners, such a boundary represents the limit of smooth deformations. Since the fields are finite and the discontinuity is a region of zero measure, it would not contribute to first order~\cite{johnson_ibanescu_skorobogatiy_weisberg_joannopoulos_fink_2002}, and monotonicity would hold. (It is not clear whether monotonicity would hold for fractal surfaces.) Although not covered explicitly by our derivation, monotonicity would also hold for a gradient-index (at zero frequency) medium, if the increase in index is nonnegative everywhere (and positive over some region) across the medium.

A simple application of the monotonicity theorem is to enclose the scatterer within some larger halfspace, which is possible as long as there is a separating plane between the emitter and the scatterer. The electrostatic constant of any halfspace is given in \eqref{canonical}, involving only the separation distance and the zero-frequency material parameters. Since the material factors $(\varepsilon-1)/(\varepsilon+1)$ and $(\mu-1)/(\mu+1)$ are bounded above by 1 (for static material constants larger than 1), we can replace them with one in upper bounds. By the monotonicity theorem, the all-frequency integrated LDOS for any structure enclosed by a halfspace (which is separated from the emitter by a minimum distance $d$) must obey the general bounds, for \emph{any} material:
\begin{align}
    \int_0^\infty \rho(\omega) \,{\rm d}\omega = \int_0^\infty \left[ \rho^E(\omega) + \rho^H(\omega) \right] &\leq \frac{1}{8\pi d^3}, \label{eq:hsbound0}
\end{align}
and, for \emph{nonmagnetic} materials:
\begin{equation}
\begin{aligned}
    \int_0^\infty \rho^E(\omega) \,{\rm d}\omega &\leq \frac{1}{16\pi d^3} \\ 
    \int_0^\infty \rho^H(\omega) \,{\rm d}\omega &= 0 \label{eq:hsbound}
\end{aligned}
\end{equation}
If the scatterer in question can be more tightly enclosed by another shape, such as a sphere or two halfspaces, one can replace the RHS of \eqreftwo{hsbound0}{hsbound} with the respective electrostatic constants to obtain a tighter bound on $\alphaL$.

\begin{figure} [t!]
    \includegraphics[width=1\linewidth]{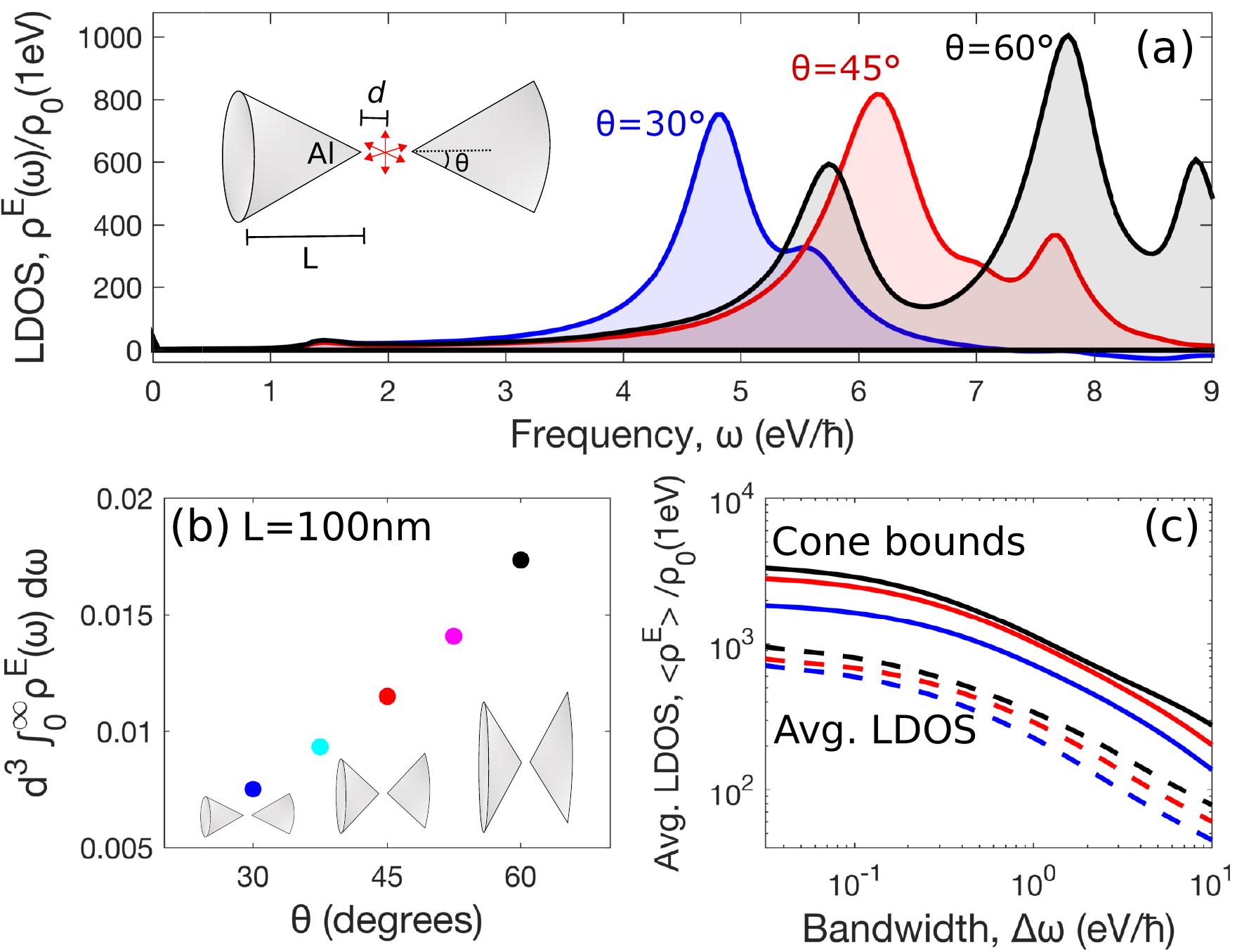} 
    \caption{(a) Electric LDOS for a double cone made of Al with $d$=10nm (L=15, 25, 35nm for $\theta$=30, 45, 60$^\circ$ respectively). (b) As the opening angle increases, the all-frequency integrated electric LDOS increases in confirmation of the monotonicity theorem. (c) Average electric LDOS centered at their peaks ($\omega_0$=4.8, 6.2, 7.8eV/$\hbar$ for $\theta$=30, 45, 60$^\circ$ respectively) for each angles are shown along with their respective bounds. The double cones approach within almost a factor of 2 their upper bounds, over a large range of bandwidths.}  
    \label{fig:mono} 
\end{figure} 

\Figref{mono} shows the electric LDOS for an emitter at the center of an aluminum double cone (similar to a bowtie antenna~\cite{grober1997}), computed with an open-source software implementation~\cite{reid_johnson_2015, reidScuffEM} of the boundary element method (BEM)~\cite{harrington_2000}. \Figref{mono}(a) shows that for an emitter--antenna separation of $d=\SI{10}{nm}$ (a cone--cone separation of \SI{20}{nm}), three orders of magnitude enhancements in electric LDOS are possible, at resonant frequencies determined by the geometry and the opening angle $\theta$. Enlarging the opening angle represents a way to increase the size of the scatterer, and \figref{mono}(b) demonstrates a monotonic increase in integrated electric LDOS, as expected from the monotonicity theorem. In conjunction, the sum rule, \eqref{Vint}, and the monotonicity theorem, \eqref{mono}, suggest a critical takeaway: isolated sharp tips do not occupy enough of the near-field to maximize electric LDOS; instead, large-area structured surfaces offer significantly greater potential.

Our identification of the causal linear-response function $s(\omega)$ defined in \eqref{rho} as the underpinning of near-field sum rules ultimately yields a relation between all-frequency response and the single pole at the origin. Such relations form the crux of all sum rules~\cite{nussenzveig_1972}, where the pole is almost always chosen at the origin or in the limit $\omega \rightarrow \infty$ (on the real line), because the response at those two particular frequencies can often be simplified: $\omega=0$ is the regime of electrostatics, while electromagnetism in the $\omega \rightarrow \infty$ limit is perturbative, as material susceptibilities converge to zero. From a theoretical viewpoint, of course, a pole can be introduced anywhere on the real line (not just $\omega\rightarrow 0, \infty$), or even anywhere in the UHP. Typically, however, one cannot make general statements about the response at arbitrary frequencies. In the next section, we show how to employ recently developed energy-conservation techniques to derive general bounds at such frequencies, moving beyond the single-frequency/all-frequency dichotomy to a framework that work for \emph{any} bandwidth.

\subsection{Power--Bandwidth Limits} \label{sec:power}
In this section, we introduce two ideas that transform the sum-rule approach of \secref{sumrule} to an approach that bounds response over any bandwidth: (1) we use the notion of a window function to connect average response over some bandwidth to discrete frequencies in the upper half plane (at the window-function's poles), and (2) we show how energy-conservation and passivity constraints can be applied at those complex frequencies, yielding analytical bounds on the bandwidth-averaged response.

Bandwidth plays a key role in any electromagnetic scattering problem, whether arising from the intrinsic linewidth of a source or as a primary technological constraint. For example, enhancements in LDOS over a broad spectrum could enable simultaneous imaging of multiple molecular species. Also, broadband LDOS enhancements provide a key criterion in designing optimal photovoltaic structures capable of maximal absorption enhancements~\cite{callahan_munday_atwater_2012}.

There are many ways in which one might want to average response over bandwidth (equal weighting, Lorentzian weighting, etc.), and one can accommodate almost any by prescribing a window function $H_{\omega_0,\Delta\omega}(\omega)$ that serves as a weighting function---it is concentrated around a center frequency $\omega_0$, is defined by a frequency width $\Delta \omega$, and is normalized ($\int_{-\infty}^\infty H_{\omega_0,\Delta\omega}(\omega) \,{\rm d}\omega = 1$). Then the average LDOS over that range of frequencies, which we denote $\langle \rho \rangle$, can be defined by:
\begin{align}
    \langle \rho \rangle \equiv \int_{-\infty}^\infty \rho(\omega) H_{\omega_0,\Delta\omega}(\omega) \,{\rm d}\omega.
\label {eq:avgdef} 
\end{align}
For the remainder of this section, we choose a Lorentzian function for $H_{\omega_0,\Delta\omega}$:
\begin{align}
    H_{\omega_0, \Delta \omega}(\omega)= \frac{\Delta \omega / \pi}{(\omega - \omega_0)^2 + (\Delta\omega)^2}. \label {eq:lor}
\end{align}
where $\Delta\omega$ is the half-width at half-maximum. We use a Lorentzian for simplicity: $H_{\omega_0, \Delta \omega}(\omega)$ extended into the UHP has only a single pole at $\omega = \omega_0 + i\Delta \omega$. Other window functions can be used with the simple addition of extra (or higher-order~\cite{liang_johnson_2013}) poles. For example, one can approximate a rect function (where $H_{\omega_0, \Delta \omega}$ is constant over the bandwidth of interest and zero elsewhere) by the generalization $H_{\omega_0, \Delta \omega}(\omega)=\frac{c \left(\Delta\omega\right)^{2n-1}}{\left(\omega - \omega_0\right)^{2n} + \left(\Delta\omega\right)^{2n}}$, which has $n$ poles in the UHP. The bounds below at a single pole would then become a linear combination of the bounds at the new poles, with slightly modified numerical factors but the same physical ramifications for material and structural design.

\begin{figure*} [t!]
    \includegraphics[width=1\linewidth]{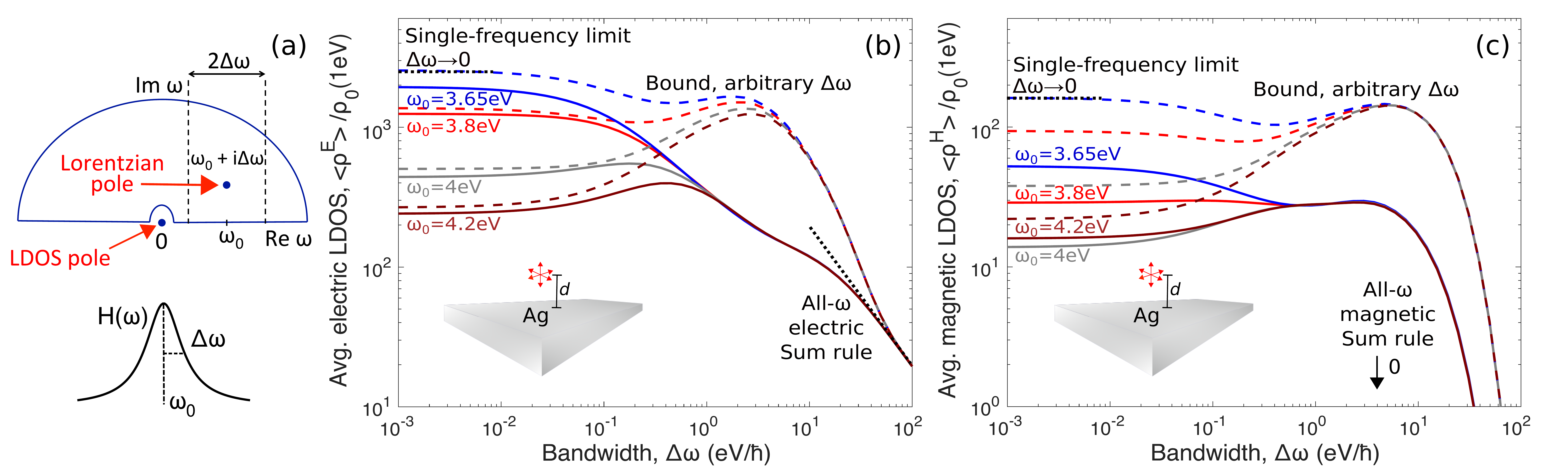} 
    \caption{(a) Contour of integration in the complex-$\omega$ plane used to obtain the average LDOS, which contains singularities at the origin (``LDOS pole'') that is intrinsic to LDOS and at a complex frequency (``Lorentzian pole'') determined by the parameters chosen for the Lorentzian window function $H_{\omega_0, \Delta \omega}$. Apart from these two poles, the product $\rho H_{\omega_0, \Delta \omega}$ is analytic everywhere on the upper half plane. (b) Average electric and (c) magnetic LDOS near a Ag halfspace centered at different frequencies around its peak (3.65eV), compared to their respective bounds. Taking bandwidth to zero gives the single-frequency limit found in earlier works \cite{miller_polimeridis_reid_hsu_delacy_joannopoulos_soljacic_johnson_2016, miller_ilic_christensen_reid_atwater_joannopoulos_soljacic_johnson_2017}. In the opposite limit of infinitely large bandwidth that includes the entire LDOS spectrum, our bounds reproduce the electric and magnetic sum rules in \secref{sumrule}. The emitter-scatterer distance \textit{d} is set to 10nm.}  
    \label{fig:pb_limits} 
\end{figure*}

Inserting the Lorentzian of \eqref{lor} into the average LDOS (\eqref{avgdef}) and writing the LDOS in terms of the near-field scattering amplitude, $\rho = \Im s(\omega)$, the product $\rho(\omega) H(\omega)$ in the averaging integrand is given by $\Im [s(\omega) H(\omega)]$. Outside of the lower-half plane, the function $s(\omega) H(\omega)$ has two simple poles: one at the origin, which was responsible for the sum rule from \secref{sumrule}, and another at $\omega_0 + i\Delta\omega$, from the Lorentzian, as shown in \figref{pb_limits}(a). One can integrate over the contour in \figref{pb_limits}(a) and use Cauchy's residue theorem to equate the all-frequency integral of \eqref{avgdef} to the evaluation of two complex-frequency quantities:
\begin{align}
\langle \rho \rangle = \Im s(\omega_0+i\Delta \omega) + 2H_{\omega_0, \Delta \omega}(0) \alphaL, \label {eq:lorsum}
\end{align}
where $s(\omega_0 + i\Delta\omega)$ is the near-field scattering amplitude evaluated at the single complex frequency $\omega_0 + i\Delta\omega$, and $\alphaL $ is the electrostatic constant appearing in the sum rule defined in  \eqref{Vint}. In the limit of zero bandwidth, $H_{\omega_0, \Delta \omega}(0)$ equals zero, and \eqref{lorsum} comprises only the first term, which represents single-frequency LDOS: $\rho = \Im s(\omega)$; conversely, as the bandwidth goes to infinity, $\Im s(\omega_0+i\Delta \omega)$ decays rapidly (as we will show below) and the second term converges to the sum rule (\eqref{Vint}). Between these extremes, the two terms comprising $\langle \rho \rangle$ combine to represent a bandwidth-averaged response. 

At the complex frequency $\omega_0+i\Delta\omega$, the positive imaginary part of the wavenumber $k$ means that the incident field emanating from the dipolar source is exponentially decaying, as can be understood from the outward-going wave $e^{ikr}/r \rightarrow e^{i\omega_0 r /c} e^{-\Delta \omega r / c}$. The decaying source is mathematically equivalent to a scattering problem in which the frequency is real-valued but material loss is increased in both the scatterer and the background~\cite{hashemi_qiu_mccauley_joannopoulos_johnson_2012}. Through either viewpoint, one can see that large broad-bandwidth response is inherently more difficult to achieve than large single-frequency response, due to an inherent bandwidth-induced dissipation.

By expressing the weighted integral of LDOS in terms of residues evaluated at single \textit{complex} frequencies, \eqref{lorsum} is now conceptually similar to single-frequency response functions at real frequencies for which we have developed an energy-conservation/passivity-based approach to upper bounds~\cite{miller_polimeridis_reid_hsu_delacy_joannopoulos_soljacic_johnson_2016, miller_ilic_christensen_reid_atwater_joannopoulos_soljacic_johnson_2017}. The key idea as applied here is that \eqref{lorsum} is the imaginary part of a function that is \emph{linear} in the scattering amplitude $s(\omega_0 + i\Delta\omega)$, while representing the total (bandwidth-averaged) power lost by the dipolar source, to either far-field radiation or near-field absorption. By contrast, absorption itself is dissipation in the medium, computed with the imaginary part of the susceptibility and the field intensity $|\mathbf{E}|^2$, a function that is \emph{quadratic} in the fields. Since absorption must be smaller than total LDOS (absorption + radiation), this implies that the quadratic functional must be smaller than the linear functional, a \emph{convex} optimization constraint that necessarily bounds how large the fields and induced currents can be. We will defer to \citeasnoun{miller_polimeridis_reid_hsu_delacy_joannopoulos_soljacic_johnson_2016} for a more detailed discussion of such single-frequency optimization, and emphasize below the new developments in the case of a complex frequency.

To bound the first term, $s(\omega_0 + i\Delta\omega)$, in \eqref{lorsum} (the second term is the known electrostatic constant), it is helpful to rewrite the near-field scattering amplitude not in terms of the field at the source location, but instead in terms of the fields \emph{within} the scatterer, at the complex frequency $\omega = \omega_0 + i\Delta\omega$. We will see that the amplitude as well as a passivity constraint can be written most succinctly in terms of the material susceptibility $\chi(\omega)$, which is the difference between the scatterer permittivity/permeability and that of the background:
\begin{align}
    \chi(\omega) = \nu(\omega) - \nu_0(\omega) = 
    \begin{pmatrix}
        (\varepsilon - 1) \mathcal{I} & \\
        & (\mu - 1) \mathcal{I}
    \end{pmatrix}.
\end{align}
(In the \SM\ we treat the most general scenario in which the susceptibility is a 6$\times$6 tensor that can be magnetic, anisotropic, nonreciprocal, and spatially inhomogeneous.) If we consider the LDOS for a single dipole orientation $j$, and momentarily drop the $j$ notation for simplicity, the scattering amplitude is, per \eqref{s(w)}, $s(\omega) = (1/\pi \omega) \xi^T(\xv_0) \psi_s(\xv_0)$. The scattered field at the dipole location, $\psi_s(\xv_0)$, is given by the convolution of the free-space Green's function $\tens{\Gamma}$ with the polarization currents $\phi = \chi \psi$ in the scatterer; then, reciprocity~\cite{rothwell_cloud_2009} can connect the free-space Green's function to the incident field from the dipole itself (a procedure we followed at real frequencies in \citeasnoun{miller_polimeridis_reid_hsu_delacy_joannopoulos_soljacic_johnson_2016}). After defining a modified incident field, $\widetilde{\psi}_{\rm inc} = \begin{pmatrix} \Ev_{\rm inc} & -\Hv_{\rm inc} \end{pmatrix}^T$, the near-field scattering amplitude can be written as $s(\omega) = \left[(1/\pi\omega) \int_V \widetilde{\psi}_{\rm inc}^T \chi \psi \right]$.

Finding a convex constraint that encodes energy conservation---requiring absorbed power (properly normalized) to be smaller than total power expended---requires some care at complex frequencies. One cannot analytically continue the absorbed and scattered powers into the UHP, as they are not analytic everywhere there (originating from their quadratic field dependence). To recover the notion of an energy-conservation constraint, we start with \emph{passivity}, which states that everywhere in the UHP, the product $\Im (\omega \chi)$ must be positive-definite~\cite{welters_avniel_johnson_2014}:
\begin{align}
    \Im \left[\omega \chi(\omega)\right] > 0 \qquad {\rm for } \quad \Im \omega > 0,
    \label{eq:passive}
\end{align}
where $\Im (\omega\chi) = [\omega\chi - (\omega\chi)^\dagger]/2i$, and \eqref{passive} extends into the complex plane the notion that passivity implies positive(-definite) imaginary susceptibilities. From passivity, we define two positive functionals: an integral of the (positive) quantity $\Im[\psi^\dagger (\omega \chi) \psi]$ within the scatterer volume, and an integral of the (positive) quantity $\Im[\psi_s^\dagger (\omega \chi) \psi_s]$ outside the scatterer volume. Through repeated application of the divergence theorem and the complex-frequency Maxwell equations, we can define two new functionals, $\varphi_A(\omega)$ and $\varphi_E(\omega)$. At real frequencies the functionals equal absorption and extinction, but we define them here in the UHP:
\begin{align}
    \varphi_A(\omega) &= \frac{1}{2} \Im \int_V \psi^\dagger \left(\omega \nu \right) \psi \nonumber \\
    \varphi_E(\omega) &= \frac{1}{2} \Im \int_V \psi^\dagger_{\rm inc} \left[\left(\omega\nu - \left(\omega\nu_0\right)^\dagger \right) \psi - \omega \nu_0 \psi_{\rm inc} \right]. \label {eq:absext}
\end{align} 
In the \SM\, we show that these functionals indeed satisfy an absorption/extinction-like constraint everywhere in the UHP: $\varphi_A(\omega) < \varphi_E(\omega)$ for $\Im \omega > 0$. This constraint is precisely the type of convex constraint needed for a bound, providing a mechanism to derive one at complex frequencies. Given the expression above for $s(\omega)$, and the constraint $\varphi_A < \varphi_E$, we formulate the upper bound as the solution of a (convex) optimization problem at $\omega = \omega_0 + i\Delta\omega$:
\begin{equation}
    \begin{aligned}
        & \underset{\psi(\xv,\omega)}{\text{maximize}} & & \Im s(\omega) = \Im \left[ \frac{1}{\pi\omega} \int_V  \widetilde{\psi}^T_{\rm inc}(\omega) \chi(\omega) \psi(\omega) \right] \\
        & \text{subject to}       & &  \varphi_A(\omega) \leq  \varphi_E(\omega). \label {eq:optimization}
    \end{aligned}
\end{equation}
\Eqref{optimization} has a unique, globally optimal solution. The optimal field distribution $\psi(\omega)$ and scattering amplitude $s(\omega)$ can be found through variational calculus, following a similar procedure to that developed in \citeasnoun{miller_polimeridis_reid_hsu_delacy_joannopoulos_soljacic_johnson_2016} and detailed in the \SM. A crucial term that emerges is a material-dependent ``material figure of merit,'' $f(\omega)$. For bulk (non-2D), nonmagnetic materials with scalar electric susceptibilities $\chi(\omega)$, the material figure of merit (FOM) is given by
\begin{align}
    f(\omega=\omega_0 + i\Delta\omega) =  \frac{\left|\omega\chi\right|^2 + \left|\omega\chi\right| \Delta \omega}{\left|\omega\right| \Im\left(\omega\varepsilon\right)}.
    \label{eq:f(w)}
\end{align}
The optimal field is proportional to this material function, as well as to the conjugate of the incident field. Now reintroducing the average over dipole orientation $j$, the optimal field yields a frequency-averaged LDOS bound (\SM): 
\begin{align}
    \langle \rho \rangle  \leq & \, \frac{f(\omega)}{\pi\left|\omega \right|} \sum_j \int_V \psi_{\textrm{inc},j}^\dagger(\omega) \psi_{\textrm{inc},j}(\omega) \,{\rm d}V \nonumber \\
    &+ 2H_{\omega_0, \Delta \omega}(0) \alphaL, \label {eq:lorbound}
\end{align}
where the complex frequency $\omega = \omega_0 + i\Delta\omega$ encodes the center frequency and bandwidth of interest.

\Eqref{lorbound} shows that near-field LDOS, averaged in a half-width-at-half-max bandwidth $\Delta \omega$ around a center frequency $\omega_0$, is fundamentally limited by the field of a dipole in free space, and by the frequency-dependent material composition of the scatterer(s). The volume integral of the incident field can be further simplified by enclosing the scatterer within some bounding shape of high symmetry over which the integral can be calculated analytically. A typical example is that of an emitter above a structured (or randomly textured~\cite{gersten_1980, wood_1981}) surface, in which case the scatterer can be enclosed in a halfspace that is a separation distance $d$ from the emitter. Near-field interactions ($|\omega|d/c \ll 1$) are dominated by the rapidly decaying evanescent fields emanating from the sources, which implies that the overall shape of the scatterer, aside from its dimensionality, has little effect on the volume integral in {\eqref{lorbound}}. Enclosing \emph{any} structure by a halfspace and keeping only the dominant near-field term in the integral in {\eqref{lorbound}}, we obtain a simple, shape-independent, analytical expression. (All remaining terms, which are non-divergent and typically small, are included in the \SM.) The bound scaling is very different when the incident field is generated by a \emph{magnetic} dipole rather than an electric one, and thus we can separately derive for total LDOS $\rho$, electric LDOS $\rho^E$, and magnetic LDOS $\rho^H$, the general bandwidth-averaged bounds (\SM):
\begin{equation}
\begin{aligned}
    \frac{\langle \rho \rangle}{\rho_0 (|\omega|)}, \frac{\langle \rho^E \rangle}{\rho_0 (|\omega|)} &\leq \frac{1}{8|k|^3d^3} \left(f(\omega) e^{-2 d \Delta \omega /c} + 2\frac{\Delta \omega}{|\omega|} \right), \\
    \frac{\langle \rho^H \rangle}{\rho_0 (|\omega|)} &\leq \frac{1}{4|k|d} f(\omega) e^{-2 d \Delta \omega /c}, \label{eq:nearbound}
\end{aligned} 
\end{equation}
where we have defined a complex-valued wavenumber $k = \omega/c$, the function $f(\omega)$ is the material FOM from \eqref{f(w)}, and for the second term of the first line we have inserted the halfspace bounds for $\alphaL$ from \eqref{hsbound}. 

\eqreftwo{lorbound}{nearbound} are foundational results of our paper. No geometrical engineering of resonances or coupling can overcome their limits. For any structure and bandwidth, the bound of \eqref{nearbound} depends only on the frequency range of interest, the material properties at those frequencies, and the emitter--scatterer separation.

\Figref{pb_limits} compares the LDOS near a silver halfspace to the bounds of \eqref{nearbound}. Center frequencies ranging from $\omega_0 = \SI{3.65}{eV}/\hbar$ to $\omega_0 = \SI{4.2}{eV}/\hbar$ (with corresponding wavelengths from $\SI{340}{nm}$ to $\SI{295}{nm}$) are considered, near the surface-plasmon frequency of silver. One can show analytically that in the zero-bandwidth, near-field ($kd \ll 1$) limit, the electric LDOS above a nonmagnetic, surface-plasmon-resonant interface should approach the bound of \eqref{nearbound}, while the magnetic LDOS above the same interface should approach its respective bound within a factor of two. (This can be shown starting from asymptotic expressions in \citeasnoun{joulain_carminati_mulet_greffet_2003}.) Such close approaches in the zero-bandwidth limit are visible in both \figref{pb_limits}(b,c). In the large-bandwidth limit, the bounds converge to the sum rules of \secref{sumrule}, ensuring that they are ``tight'' (i.e. there is no smaller upper bound) in that regime as well. To simplify the ultrahigh-bandwidth computations ($\Delta \omega \gtrsim \SI{100}{eV}/\hbar$), we use a 5-pole Drude-Lorentz multi-oscillator model for silver that closely approximates tabulated susceptibility data~\cite{palik_2003} (comparison included in \SM).  There is an interesting peak in the bound at moderate bandwidths that arises due to the large, lossy permittivity of silver at about $\SI{5}{eV}$; standard quasistatic theory~\cite{Miller2014} would predict that a halfspace is non-optimal for such a bandwidth, but that perhaps another structure is optimal. A key question prompted by these bounds is whether nonplanar, designed structures---or perhaps even randomly corrugated structures, can approach the bounds at frequencies away from the surface-plasmon resonance. In \figref{mono}(c), it was observed that double cones approach within almost of factor of 2 of their upper bounds (using \eqref{lorbound} for their specific geometry) over a large range of bandwidths, further suggesting that such structural design should be possible.

The convergence of the electric/magnetic LDOS bounds to their respective sum rules, in the infinite-bandwidth limit, can be verified directly from \eqref{nearbound}. The first term in the $\rho^E$ bound goes to zero as $\Delta \omega \rightarrow \infty$ due to the $e^{-2d\Delta\omega/c}$ factor, in which case one can rewrite the bound as $\left\langle \rho^E \right\rangle \leq 1 / (8 \pi^2 \Delta \omega d^3) = (2 / \pi \Delta \omega) (1/16\pi d^3) = (2 / \pi \Delta \omega) \int_0^\infty \rho^E {\rm d}\omega$, where the last term is precisely the average electric LDOS in the large-bandwidth limit [since $H(\Delta\omega\rightarrow\infty) = 1/\pi\Delta\omega$]. Similarly, the $\rho^H$ bound only contains the $e^{-2d\Delta\omega/c}$ factor for nonmagnetic materials, and hence the bound tends to $\left\langle \rho^H \right\rangle \leq 0$, in agreement with the sum rule. By construction, the bounds agree with their respective sum rules for large bandwidths.

The power--bandwidth limit of \eqref{lorbound} applies equally well to 2D materials characterized by a spatial conductivity $\sigma(\omega)$ , with the substitution $\omega\chi(\omega) \rightarrow i\delta_S(\xv) \sigma(\omega)$, where $\delta_S(\xv)$ is a delta function on the surface of the (not necessarily planar~\cite{christensen_jauho_wubs_mortensen_2015}) 2D material. In doing so, the delta function transforms the volume integral in \eqref{lorbound} to a surface integral over the incident field. We can enclose the 2D scatterer in a high-symmetry enclosure: for a 2D plane enclosure, and keeping only the highest-order terms in the near-field limit ($|\omega|d/c \ll 1$), we find that the LDOS above the 2D material is bounded above by (\SM):
\begin{equation}
\begin{aligned}
    \frac{\langle \rho \rangle}{\rho_0 (|\omega|)}, \frac{\langle \rho^E \rangle}{\rho_0 (|\omega|)} &\leq \frac{3}{8|k|^4 d^4} f(\omega) e^{-2 d \Delta \omega /c} + \frac{1}{4|k|^3 d^3}\frac{\Delta \omega}{|\omega|} , \\
    \frac{\langle \rho^H \rangle}{\rho_0 (|\omega|)} &\leq \frac{1}{4|k|^2d^2} f(\omega) e^{-2 d \Delta \omega /c}, \label{eq:nearbound2D}
\end{aligned} 
\end{equation}
where for 2D materials the material FOM is
\begin{align}
    f(\omega) = \frac{| \sigma(\omega) |^2}{\Re{\sigma(\omega)}} \label{eq:f(w)2D}
\end{align}
(In SI units, there would be an additional factor of the free-space impedance $Z_0$ multiplying $|\sigma|^2 / \Re \sigma$.) There are two distinct features that emerge for 2D materials: the material FOM is $|\sigma(\omega)|^2 / \Re \sigma(\omega)$, and the electric- and magnetic-LDOS bounds have terms that scale as $1/d^4$ and $1/d^2$, instead of $1/d^3$ and $1/d$ in the bulk-material bounds. The different distance scaling is a natural consequence of integrating the norm of the Green's function over an area instead of a volume. Yet there is an interesting contrast embedded within the bounds for $\avg{\rho}$ and $\avg{\rho^E}$: their first term, dominant over narrow bandwidths, scales as $1/d^4$, whereas the second term, dominant over wider bandwidths and corresponding to the sum rule, scales as $1/d^3$. The faster scaling with $1/d$ of the first term suggests a scenario in which the average LDOS over a narrow bandwidth may be larger than the sum-rule would seem to allow. One possibility is that the bound is ``loose'' and that the $1/d^4$ scaling is artificial, but in multiple previous studies~\cite{koppens_chang_abajo_2011, miller_ilic_christensen_reid_atwater_joannopoulos_soljacic_johnson_2017} of single-frequency behavior, $1/d^4$ scaling has been observed in the LDOS near 2D materials. Another possibility is that such large response is only possible over a narrow bandwidth, though the connection of broadband response to single-complex-frequency response would seem to suggest that large response likely is not restricted to single frequencies. Finally, perhaps the most likely possibility is that such a bound is achievable, and simply requires negative (scattered) LDOS at frequencies outside the range of interest. There is no requirement that scattered LDOS be positive at all frequencies, since a scatterer can suppress all modes and reduce the total LDOS to nearly zero. The idea of exploiting such suppression to achieve anomalously large response over some desired bandwidth is intriguing.

Our derivations provide general insight into optimal structures that would reach the bounds. First, as noted above, it is critical to have as much material as possible in the near-field region of the source---that material enables the polarization currents that ultimately drive the large response. Sharp tips, though potentially exhibiting strong resonances, are not ideal. Second, the convexity-based optimization provides not only the maximal bandwidth-averaged response, but also the optimal fields $\psi$ that would generate such response. In all cases, those optimal fields, throughout the volume of the scatterer, are proportional to the incident fields. Hence one would want to generate, perhaps via computational design~\cite{jensen2011,Lalau-Keraly2013,liang_johnson_2013,piggott2015}, resonances with the same phase and amplitude profile as the fields emanating from the dipolar sources. Finally, through a volume-integral-equation (VIE) framework~\cite{miller_polimeridis_reid_hsu_delacy_joannopoulos_soljacic_johnson_2016}, one can reinterpret our bounds as upper limits that occur when the incident field couples only to a single VIE mode at the optimal resonance location. Thus if one can completely avoid exciting other modes, then the upper bound is in fact guaranteed to be achieved.

\section{Cross density of states} \label{sec:cdos}
The previous section developed the theoretical framework for power--bandwidth limits in the context of LDOS. We translate that framework to other near-field optical response functions, starting with ``cross density of states'' (CDOS)~\cite{caze_pierrat_carminati_2013}, which measures field correlations between two points of a structured environment. In addition to fundamental interest as a correlation function, CDOS is also the critical term in the frequency integrand for resonant near-field dipolar energy transfer, as in F{\"o}rster energy transfer~\cite{dung_knoll_welsch_2002, martin-cano_2010, gonzaga-galeana_zurita-sanchez_2013}, as well as for quantum entanglement and super-radiative coupling between qubits~\cite{kastel_fleischhauer_2005, kastel_laser_2005, dzsotjan_sorensen_fleischhauer_2010, martin-cano_2011, gonzalez-tudela_martin-cano_2011}. Whereas LDOS is given by the Green's function for identical source and measurement points, CDOS is given by the Green's function between different source and measurement points, $\xv_0$ and $\xv$, respectively~\cite{caze_pierrat_carminati_2013}:
\begin{align}
    \rho_{ij}(\xv,\xv_0,\omega) &= \Im \left[ \frac{1}{\pi\omega} \Gamma_{\textrm{s},{ij}}(\xv,\xv_0,\omega) \right], \label{eq:rhoc}
\end{align}
where $i$ and $j$ are the measurement and source polarizations, respectively, and we again use the ``s'' subscript to denote the scattered-field contribution (subtracting off the known free-space contribution). By analogy to \eqref{Vint}, there is a sum-rule constant for CDOS, which we denote $\alphaC$ (defined as $\frac{1}{2} \Re \left[\Gamma_{\textrm{s},{ij}}(\xv,\xv_0)\big\rvert_{\omega=0}\right]$), and by the similarity to LDOS one can follow a similar procedure to derive power--bandwidth limits. The key new feature is that instead of the single separation distance between the emitter and scatterer controlling the bound, now there are two relevant separation distances: the distance between the emitter and the scatterer, denoted $d_1$, and the distance between the scatterer and the measurement point, $d_2$. 

To bound \eqref{rhoc} averaged over any bandwidth, we first rewrite the Green's function in terms of the polarization currents of the scatterer(s). This leads to an overlap integral between the polarization currents induced by the field incident from the source position with a (parity-reversed) secondary field incident from the measurement position. Ideally, the polarization response is maximally aligned to both fields ({\SM}), in which case the maximal response is proportional to the square root of the energy of each ``incident'' field, $\psi_{\textrm{inc},1}$ and $\psi_{\textrm{inc},2}$. Applying the band-averaged bound approach modified as described above, we arrive at the following bound on average CDOS for bulk, non-magnetic materials (\SM):  
\begin{align}
    \langle \rho_{ij} \rangle  \leq & \, \frac{f(\omega)}{\pi\left|\omega \right|} \Biggl \{ \left( \int_V \psi_{\textrm{inc},1,i}^\dagger(\omega) \psi_{\textrm{inc},1,i}(\omega) \,{\rm d}V \right) \nonumber \\
    &  \left( \int_V \psi_{\textrm{inc},2,j}^\dagger(\omega) \psi_{\textrm{inc},2,j}(\omega) \,{\rm d}V \right)\Biggr \}^{1/2}   \nonumber \\
    & + 2H_{\omega_0, \Delta \omega}(0) \alphaC, \label {eq:lorcbound}
\end{align}
where the complex frequency $\omega = \omega_0 + i\Delta\omega$ encodes the center frequency and bandwidth of interest. In the near-field regime, ($|\omega|d/c \ll 1$), \eqref{lorcbound} further simplifies (\SM, neglecting $\alphaC$ for small to moderate bandwidths):
\begin{equation}
\begin{aligned}
    \frac{\langle \rho_{ij} \rangle}{\rho_0 (|\omega|)}, \frac{\langle \rho_{ij}^E \rangle}{\rho_0 (|\omega|)} &\leq \frac{1}{12|k|^3 \sqrt{d_1^3 d_2^3}} f(\omega)  e^{-\left(d_1+d_2\right) \Delta \omega /c} , \\
    \frac{\langle \rho_{ij}^H \rangle}{\rho_0 (|\omega|)} &\leq \frac{1}{6|k|\sqrt{d_1 d_2}} f(\omega) e^{-\left(d_1+d_2\right) \Delta \omega /c}, \label{eq:nearcbound}
\end{aligned} 
\end{equation}
where we separate the electric- and magnetic-source contributions to the CDOS. Now the electric bounds depend on the source--scatterer and measurement--scatterer separation distances to the three-halves power, instead of the cubic dependence for LDOS when the source and measurement points are identical. Note that the material dependence of the bound is encoded in the same material figure of merit, $f(\omega)$ as defined in \eqref{f(w)}, suggesting the universal role it may play in determining the maximal broadband response of any material.

\section{Radiative heat transfer} \label{sec:RHT}
Near-field radiative heat transfer (NFRHT) can be substantially larger than far-field radiative heat transfer, via photon-tunneling evanescent-wave energy transfer, and has generated much interest for applications such as thermophotovoltaics~\cite{Whale2002,Laroche2006,Basu2009,Bermel2010,rodriguez2011}. In RHT, there are two bodies at temperatures $T_1$ and $T_2$, with minimal separation distance $d$. The net radiative heat flux between the two bodies~\cite{joulain_mulet_marquier_carminati_greffet_2005} is given by $H_{1 \rightarrow 2} (\omega) = \Phi(\omega) \left[\planck(\omega,T_1) - \planck(\omega,T_2)\right]$, where $\planck(\omega,T)$ denotes the mean energy of the harmonic oscillator (without the zero-point energy $\hbar \omega/2$) and $\Phi$ is a temperature-independent flux rate from incoherent sources in body 1 radiating to body 2. 

Since $\planck$ is positive for all frequencies, we can bound the difference $\planck(T_1) - \planck(T_2)$ by its maximum value, which is simply $\planck(T_1)$ (taking $T_1 > T_2$), i.e. ($\Phi$ is non-negative at all frequencies for passive media):
\begin{align}
   H_{1 \rightarrow 2} = & \int_0^\infty  \Phi(\omega)\left[\planck(\omega,T_1) - \planck(\omega,T_2)\right] \,{\rm d}\omega \\
   & \leq \int_0^\infty  \Phi(\omega)\planck(\omega,T) \,{\rm d}\omega \label{eq:RHTsum}
\end{align}
where $T = T_1$ and the equality holds if body 2 is at absolute zero.

In order to apply contour-integration techniques (described in \secref{power}) to bound \eqref{RHTsum}, one would need to extend the mean energy $\planck$ to negative frequencies, and $\planck$ has to decay fast enough such that the integral in \eqref{RHTsum} is finite for all real frequencies. While it is bounded and decays for large positive frequencies, it diverges when extended to negative frequencies. To avoid such asymmetry, one could add the vacuum energy $\hbar\omega/2$ and work with $\Theta_v=\frac{1}{2}\hbar \omega \coth(\hbar \omega / 2 k_B T)$~\cite{joulain_mulet_marquier_carminati_greffet_2005} ($k_B$ denotes the Boltzmann constant), which is symmetric about the origin. However, $\Theta_v$ diverges linearly for large frequencies and would enforce dramatic restrictions on the flux rate $\Phi$ to fall rapidly at high frequencies. Even if convergence were not an issue, $\Theta_v$ contains infinitely many singularities along the imaginary axis. This would reduce a contour integral evaluation of \eqref{RHTsum} (albeit with $\Theta_v$ instead of $\Theta$) to a sum of infinitely many residues (at the ``Matsubara frequencies''~\cite{bruus_flensberg_2004}), which is cumbersome to handle. We can avoid all of these issues (asymmetry, non-convergence, and the Matsubara sum) in a single stroke by virtue of the following fact: for any temperature $T$, the mean-energy spectrum $\Theta(\omega,T)$ is simultaneously bounded above and closely approximated (at frequencies with non-negligible contributions) by a Lorentzian function centered at zero with bandwidth $\sqrt{2} k_B T / \hbar$, $H_{0, \sqrt{2} k_B T / \hbar}(\omega)$, properly scaled such that it coincides with $\planck$ at zero frequency:
\begin{align}
   \planck(\omega,T) \leq \sqrt{2}\pi \hbar \left( \frac{k_{B}T}{\hbar} \right )^2 H_{0, \sqrt{2} k_{B}T / \hbar}(\omega),
\label{eq:RHTineq}
\end{align}
where for this line $H$ again denotes the Lorentzian window function defined in \eqref{lor}. There is close agreement between  $H_{0, \sqrt{2} k_{B}T / \hbar}(\omega)$ and $\planck(\omega,T)$, with the total energy as measured by the integral $\int_0^\infty {\rm d}\omega$ larger for the Lorentzian by only $3\sqrt{2}/\pi \approx 1.35$ for all temperatures (cf. \SM). (Note that unlike the spectrum of a blackbody, whose peak wavelength is nonzero and scales inversely with temperature, the mean-energy spectrum $\planck(\omega)$ peaks at zero frequency.)

Since the integral in \eqref{RHTineq} is the Lorentzian-averaged flux rate, it might be tempting to close the contour in the UHP to relate the integral to a sum of residues, in the spirit of \secref{power}. However, unlike LDOS, the flux rate $\Phi$ is \emph{not} directly given by the real/imaginary part of a scattering amplitude. But we can transform the problem by generalized reciprocity~\cite{kong_1975}, recasting the flux rate from a surface integral of fields generated by volume sources to a volume integral of fields generated by sources along a surface, revealing a surprising similarity to LDOS and ultimately leading to NFRHT bounds. 

The heat flux between the two bodies is given by the power flow through a separating surface $S$:
\begin{align}
    H_{1 \rightarrow 2}(\omega) &= \frac{1}{2} \Re \int_S \left( \Ev \times \cc{\Hv} \right) \cdot \mathbf{\hat{n}} \nonumber \\
    &= \frac{1}{4} \int_S \psi^{\dagger} \Lambda \psi \label{eq:tent}
\end{align}
where $\Lambda = \begin{pmatrix}& -\mathbf{\hat{n}} \times \\  \mathbf{\hat{n}} \times &  \end{pmatrix}$ is a real symmetric matrix. The fields can be expressed~\cite{chew_1995} as convolutions of the system Green's function $\tens{\Gamma}(\xv,\xv')$ with the thermal sources $\phi(\xv')$: $\psi(\xv) = \int_V \tens{\Gamma}(\mathbf{\xv},\mathbf{\xv'})\phi({\mathbf{\xv'}}) \,{\rm d}\mathbf{\xv'}$. The incoherence of the thermal sources can be incorporated via the fluctuation-dissipation theorem in a standard substitution~\cite{joulain_mulet_marquier_carminati_greffet_2005}. The key step is that we then use generalized reciprocity~\cite{kong_1975} (cf. {\SM}) to interchange the source positions in body 1 with the measurement position on the surface $S$ separating the bodies. This transforms the problem to sources over a surface in free space (or a homogeneous background) radiating into one of the bodies, and the net flux rate is an energy-like quadratic form evaluated in the body, for a specific linear combination of electric and magnetic dipoles (cf. \SM). In the ideal scenario, the dipoles are only emitting into body 1; in any case, the radiation into body 1 is bounded above by the total radiation. In the near field, the total radiation is essentially exactly equal to the scattered-field LDOS, such that we can directly bound (via Cauchy--Schwarz arguments) the flux rate at any frequency by the product of the electric and magnetic LDOS (\SM): $\Phi(\omega) \leq 4 \int_S \sqrt{ \rho^E(\omega) \rho^H(\omega) }$. To bound the bandwidth-averaged flux rate, $\langle \Phi \rangle$, the product of the electric and magnetic LDOS prevents direct identification of a complex-analytic quantity, but we can again use Cauchy--Schwarz for a bound in terms of the individually bandwidth-averaged $\rho^E$ and $\rho^H$ (ideally, they would exhibit the same lineshape, in which case the Cauchy--Schwarz substitution would be an equality), such that we can write (\SM):
\begin{align}
  \langle \Phi \rangle \leq 4c \int_S \sqrt{\langle \rho^E \rangle \langle \rho^H \rangle}.\label{eq:rhtbound} 
\end{align}
In the case of near-field RHT, the center frequency and bandwidth are fixed by the temperature as discussed above. Using \eqref{rhtbound}, we can bound NFRHT in \eqref{RHTineq}:
\begin{align}
   H_{1 \rightarrow 2} \leq 2\sqrt{2}\pi \hbar c \left( \frac{k_{B}T}{\hbar} \right )^2 \int_S \sqrt{\langle \rho^E \rangle \langle \rho^H \rangle}_{0, \sqrt{2} k_{B}T / \hbar}.\label{eq:RHTsumbound} 
\end{align}
\Eqref{RHTsumbound} is the culmination of our transformations: the flux/oscillator-energy product in the integrand of \eqref{RHTsum}, which is not easily extensible to negative frequencies nor analytic over the upper-half plane, is bounded above by (and, for optimal designs, equal to) an integral over a bounding surface between the bodies of the geometric mean of the bandwidth-averaged electric and magnetic LDOS. Then, for bulk, nonmagnetic materials, we can directly insert the LDOS bounds, \eqref{nearbound} (with the second term slightly modified to allow for a two-halfspaces enclosure), into \eqref{RHTsumbound} to obtain a near-field bound on net radiative power transfer (with suitable generalizations for magnetic and/or 2D materials). The material figure of merit of \eqref{f(w)} again plays a key role; in the case of near-field RHT, the temperature determines the bandwidth. Moreover, because the center frequency is zero, the material figure of merit is evaluated at the purely imaginary frequency $i \sqrt{2} k_B T / \hbar$; since susceptibilities are real and positive for imaginary frequencies in the UHP~\cite{landau_lifshitz_1984}, the material FOM (now written as a function of temperature) can be greatly simplified:
\begin{align}
   f(T) = \chi(i\sqrt{2} k_{B}T / \hbar). \label{eq:fom_imag}
\end{align}
If we denote $r_1$ and $r_2$ as the distances from a surface point to bodies 1 and 2, respectively, and $f_1(T)$ and $f_2(T)$ as the corresponding material figures of merit, then the bound for an arbitrary separating surface $S$ is:
\begin{align}
   H_{1 \rightarrow 2} \leq & \frac{\rho_0(\sqrt{2} k_{B}T / \hbar) \hbar c^3}{2}   \int_S  \Biggl \{ \Biggl ( \frac{f_1(T) e^{-\frac{2\sqrt{2} k_{B}Tr_1}{\hbar c}} + 4.21...}{r_1^3} \\
   & + \frac{f_2(T) e^{-\frac{2\sqrt{2} k_{B}Tr_2}{\hbar c}} + 4.21...}{r_2^3}\Biggl ) \nonumber \\ 
  & \left( \frac{f_1(T) e^{-\frac{2\sqrt{2} k_{B}Tr_1}{\hbar c}}}{r_1} +  \frac{f_2(T) e^{-\frac{2\sqrt{2} k_{B}Tr_2}{\hbar c}}}{r_2}  \right) \Biggr \}^{1/2}, \label{eq:RHTsumbound2} 
\end{align}
where the constant terms proportional to 4.21 arise from the LDOS constant $\alphaL$ for two halfspaces (\SM). For a planar bounding surface halfway between the two bodies at $d/2$ for a minimal separation $d$, the integral in \eqref{RHTsumbound2} can be done analytically, simplifying the bound to maximum heat transfer per unit surface area $A$:
\begin{align}
     \frac{H_{1 \rightarrow 2}}{A} \leq \frac{2}{\pi^2 \hbar}  \left( \frac{k_{B}T}{d} \right )^2  e^{-\frac{\sqrt{2}k_{B}Td}{\hbar c}}  [ f_1(T) + f_2(T) ] \label{eq:RHTsumbound3} 
\end{align}
where we have dropped constant terms (which are small relative to $f_1$ and $f_2$ for all practical materials and temperatures of interest). \Eqref{RHTsumbound3} represents the first general bound to near-field radiative heat transfer. 

We can more easily interpret \eqref{RHTsumbound3} by recasting the expression as a product of dimensionless enhancement factors with the far-field blackbody limit, $H_{\rm BB} = \sigma T^4$, where $\sigma$ is the Stefan--Boltzmann constant. Using the thermal de Broglie wavelength, $\lambda_T = \pi^{2/3} \hbar c / k_B T$, algebraic manipulations yield an equivalent alternative,
\begin{align}
    \frac{H_{1 \rightarrow 2}}{A} \leq \sigma T^4 \left \lbrace \beta e^{-\frac{\sqrt{2}k_{B}Td}{\hbar c}} \left(\frac{\lambda_T}{d}\right)^2 \left[ f_1(T) + f_2(T) \right] \right\rbrace
    \label{eq:RHTsumbound4}
\end{align}
where $\beta = 120 / \pi^{16/3} \approx 0.268$ is of order 1. \Eqref{RHTsumbound4} succinctly identifies two maximum possible enhancements beyond the blackbody limit. First, a distance-dependent enhancement $(\lambda_T / d)^2$ emerges, which accounts for the increased amplitudes of evanescent waves at shorter separations. This enhancement factor is intermediate between that appearing in the bounds for electric and magnetic LDOS ($1/d^3$ and $1/d$ respectively), as RHT is equivalent to a combination of electric and magnetic dipolar radiation in free space. The thermal wavelength $\lambda_T$ (which is $\sim 10\mu m$ at room temperature) sets the threshold for the near-field regime, highly sensible since the blackbody radiation limit holds only when the length scales involved are greater than $\lambda_T$. The second enhancement factor is a material-dependent factor $f_1(T) + f_2(T)$, which accounts for material-based resonant enhancements. Per the material FOM of \eqref{fom_imag}, the larger the susceptibility is at the complex frequency set by the temperature of the emitter, the larger the possible response is. The bounds of \eqreftwo{RHTsumbound3}{RHTsumbound4} cannot be overcome by any metamaterial, metasurface, or other design approaches.

\section{Optimal materials} \label{sec:materials}
Embedded throughout the LDOS, CDOS, and near-field RHT bounds is a material metric $f(\omega)$, defined in \eqreftwo{f(w)}{f(w)2D}, that indicates the intrinsic capability of any material to exhibit large optical response over a frequency bandwidth $\Delta\omega$ around a center frequency $\omega_0$. This material metric enables comparison of \emph{any} material---dielectric/metal, 2D/3D, lossless/lossy---many of whose capabilities cannot be understood through single-frequency bounds or sum rules.
\begin{figure*} [h]
    \includegraphics[width=1\linewidth]{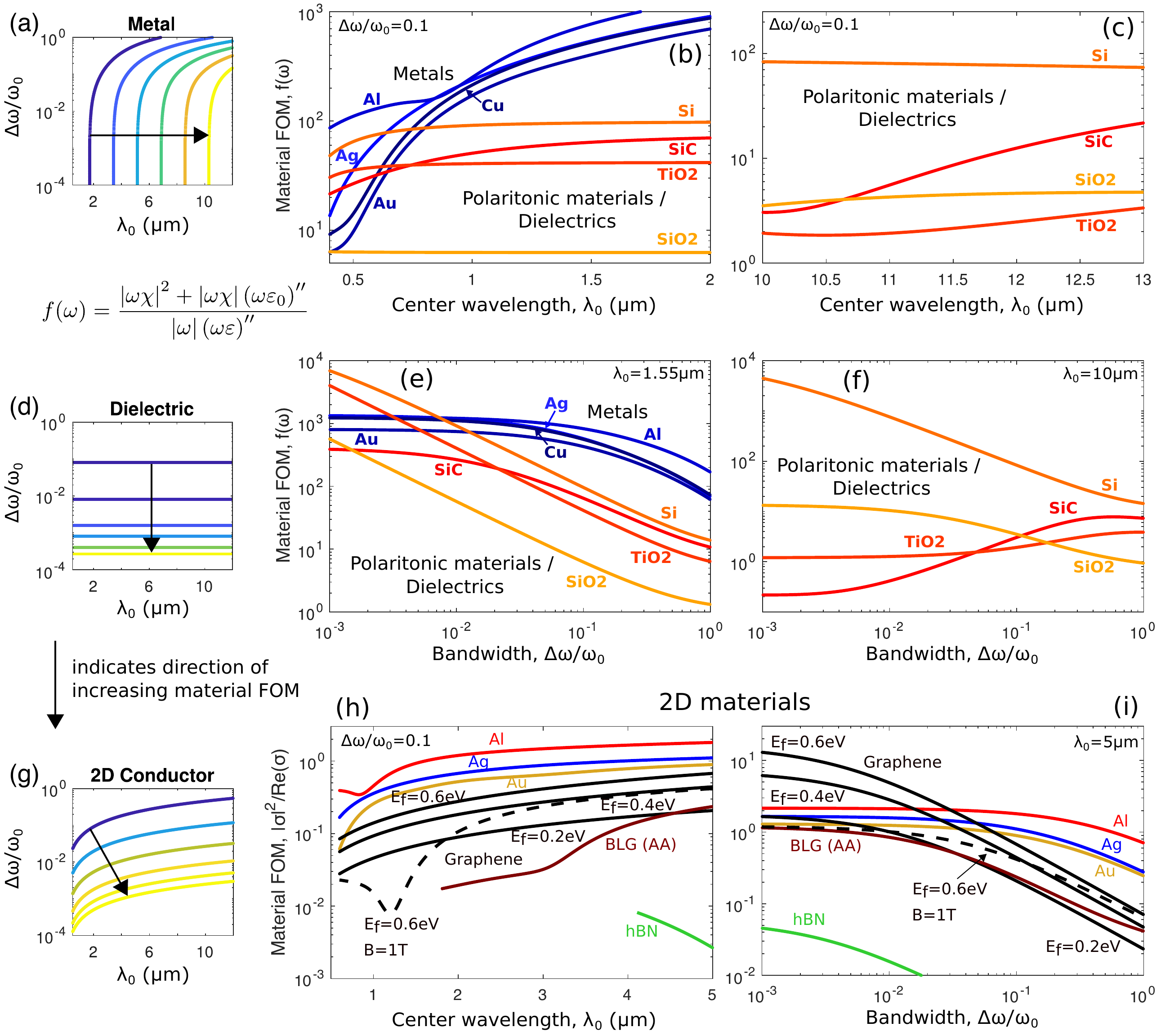} 
    \caption{(a,d,g) Isocurves of material FOM for a Drude metal (with material loss rate $\gamma=0.1\omega_p$), a lossless dielectric (of susceptibility $\chi=9$), and a Drude 2D material (with $\gamma=0.01\omega_p$). The arrows indicate increasing material FOM in each case. (b) Comparison of material FOM for various bulk metals and polaritonic materials / dielectrics, keeping the bandwidth-to-center-frequency ratio $\Delta \omega / \omega_0$ fixed to 0.1. For the modestly large 10\% relative bandwidth, the large susceptibilities of metals compensates for their material loss, generally resulting in greater maximum response. Part (c) compares surface-phonon-polariton-supporting materials at mid-IR wavelengths. (e),(f) Comparison of material FOM for varying bandwidths relative to the center wavelengths of 1.55 and 10 $\mu$m. At very narrow bandwidths, dielectrics offer greater possible response than metals. (h,i) Comparison of material FOM for 2D materials for different choices of center wavelength and $\Delta \omega / \omega_0$. (2D Al, Ag, and Au properties derived from their bulk counterparts.)}  
    \label{fig:fom} 
\end{figure*} 

Sum rules, such as \eqref{canonical}, typically have little-to-no dependence on material parameters, suggesting that different materials only alter resonant bandwidths, without impacting total optical response. Yet this is misleading on two fronts: (1) it only applies over infinite bandwidth; over any finite bandwidth, material properties play an important role in maximal response, and (2) sum rules require susceptibilities that satisfy Kramers--Kronig relations, diminishing to zero at high frequencies. The decay-to-zero requirement, though physically reasonable, means that even ``dielectric'' media (semiconductors, insulators, etc.) have a plasma-like response at large enough frequencies. Such response contributes to sum rules, often in a large way due to the negative susceptibility. This obscures the behavior of, for example, a transparent dielectric at optical frequencies, by accounting for transitions that occur at UV and X-ray frequencies. Thus sum rules miss finite-bandwidth effects and dramatically overestimate dielectric-material interactions. At the other end of the continuum, single-frequency bounds~\cite{miller_johnson_rodriguez_2015, miller_polimeridis_reid_hsu_delacy_joannopoulos_soljacic_johnson_2016, miller_ilic_christensen_reid_atwater_joannopoulos_soljacic_johnson_2017, yang_miller_christensen_joannopoulos_soljacic_2017,Yang2018} apply at any given frequency, but use material loss as the intrinsic system limitation, and thereby diverge for materials with vanishingly small imaginary susceptibilities (such as dielectrics). The material FOM embedded in the power--bandwidth limits does not have any of these limitations: it can account for finite bandwidths, it does not require susceptibilities that asymptotically approach zero at large frequencies, and it provides a finite bound for lossless materials for any nonzero bandwidth. Hence $f(\omega)$ is a simple expression that enables comparison among the multitude of possible optical materials.
\begin{figure*} [bth]
    \includegraphics[width=1\linewidth]{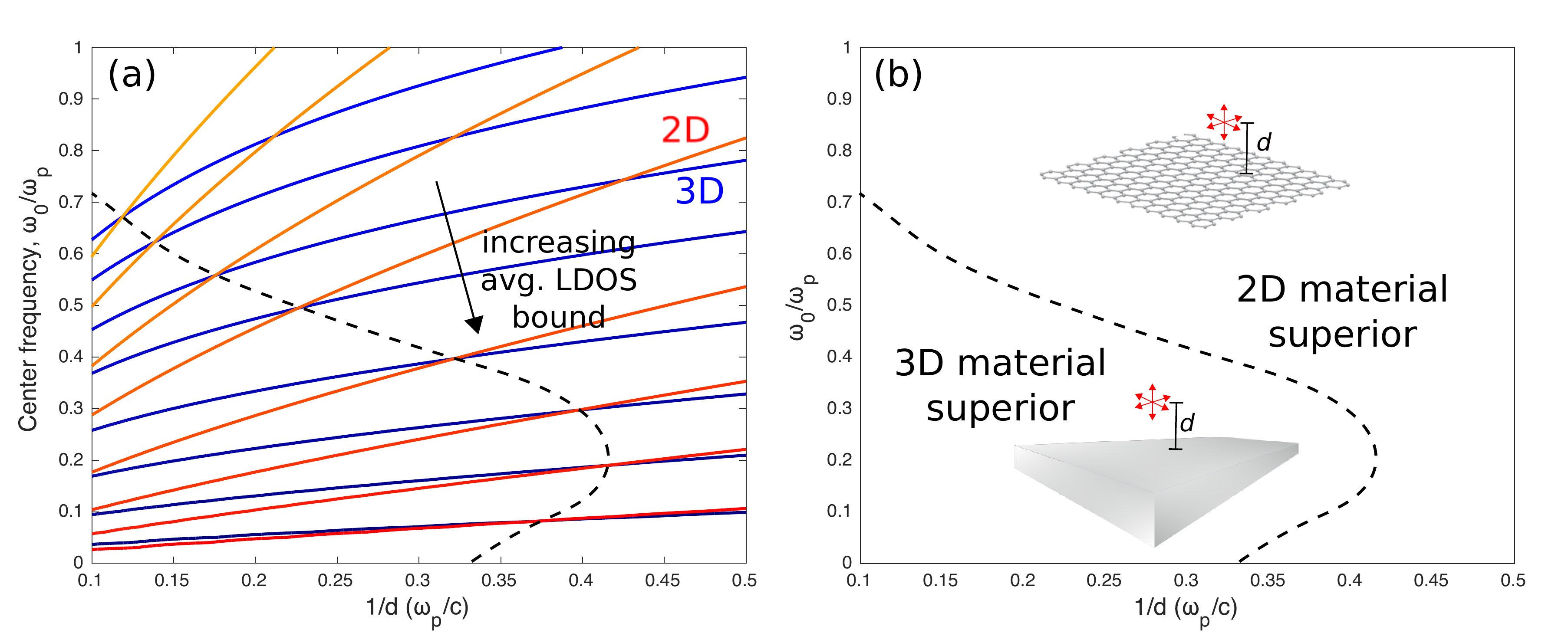} 
    \caption{Comparison of average LDOS bounds for  Drude-like 3D and 2D materials with identical decay constants $\omega_p$ and material loss rates $\gamma=0.01\omega_p$. (a) Isocurves of average LDOS bound, which increase as both center frequency and distance decrease. (b) For a small enough emitter-scatterer distance $d$, 2D materials are clearly superior, whereas for modest center frequencies and larger distances 3D materials exhibit a larger maximum response. Such behavior arises from the fact that the bounds for 2D materials scale as $1/d^4$, in contrast to $1/d^3$ scaling for bulk materials. The dotted curve delineates the regions within which 3D or 2D materials are superior. The bandwidth-to-center-frequency ratio $\Delta \omega / \omega_0$ is set to 0.1.}
    \label{fig:2d3dfom} 
\end{figure*} 

To gain intuition about the material FOM, we consider the small-bandwidth limit in which $\Delta\omega \ll \omega_0$. We delineate two types of (bulk, 3D) materials: lossy materials, with a nonzero $\Im \chi(\omega)$ in the small-bandwidth limit, and lossless materials, with $\Im \chi(\omega) \approx 0$ (and $\Im \chi(\omega) \ll \Delta \omega / \omega_0$ even in the small-bandwidth limit), as would characterize many transparent materials at optical frequencies. In the small-bandwidth limit, the material FOM is approximately given by
\begin{equation}
    f(\omega) \approx
    \begin{dcases}
        \frac{|\chi(\omega)|^2}{\Im{\chi(\omega)}} & \text{lossy (e.g. metals)} \\              
        \frac{\omega_0}{\Delta\omega}\frac{\left[\chi(\omega)\right]^2}{\chi(\omega)+1} & \text{lossless (dielectrics)} \\
        \frac{|\sigma(\omega)|^2}{\Re \sigma(\omega)}  & \text{2D materials},
    \end{dcases} \label{eq:foms} 
\end{equation}
where we have retained the full material FOM for 2D materials since it is already simple. For high-index lossless materials, the expression would simplify even further
\begin{equation}
    f(\omega) \approx \frac{\omega_0}{\Delta\omega} \chi(\omega) \qquad \text{lossless, high-index}. \\
    \label{eq:foms2} 
\end{equation}
For small to moderate bandwidths, a natural dichotomy emerges: lossy materials are inherently restricted by material loss in $\Im \chi(\omega)$, whereas lossless materials are inherently limited by the relative bandwidth $\Delta \omega / \omega_0$. Intuitively, in simple single-mode interactions, one could interpret the figures of merit as dictating that lossy materials have maximum responses proportional to $|\chi(\omega)|$, over a bandwidth proportional to $|\chi(\omega)| / \Im \chi(\omega)$, whereas lossless materials have maximum responses proportional to $\chi(\omega)$, over bandwidths proportional to $\omega_0 / \Delta \omega$. The intuitive interpretation about lossy-material bandwidths is supported by previous results in quasistatic plasmonic frameworks~\cite{wang_shen_2006, raman_shin_fan_2013}. Of course, we make no assumption of single-mode or quasistatic behavior, and our scattering framework is valid for any number of resonances as well as more complex phenomena such as Fano interactions~\cite{fano_1961} and exceptional points~\cite{kato_1995, heiss_2004}. And perhaps more importantly, it enables consideration of lossless and lossy media on equal footing. As discussed in the introduction, the maximum response of lossless media has been impossible to accurately capture with either the sum-rule or the single-frequency-bound approaches known today. In the complex-frequency approach, bandwidth naturally adds a form of ``loss'' to the system, yielding finite bounds that vary smoothly with bandwidth.

\Figref{fom} compares the material FOM for a large variety of materials at optical frequencies. To evaluate the material susceptibilities and conductivities at complex frequencies, we use analytic models (e.g. Lorentz--Drude oscillators) that can be continued into the complex plane, and ensure that they are accurate over the range of bandwidths considered. On the left side of the figure, we model the material FOM for canonical material types: (a) a Drude metal, $\chi(\omega) = -\omega_p^2 / (\omega^2 + i\gamma\omega)$, for plasma frequency $\omega_p$ and loss rate $\gamma$, (d) a lossless, constant-susceptibility ($\chi(\omega) = 9$) material, and (g) a Drude 2D material, with conductivity $\sigma(\omega) = i\omega_p / (\omega + i\gamma)$. One can see that these three material types show very different characteristic dependencies of their FOM on frequency and bandwidth. The Drude-metal FOM is nearly independent of small-to-moderate bandwidths, as expected from \eqref{foms}---for metals, intrinsic loss is the limiting factor. The FOM of a Drude metal increases with the center wavelength (of the frequency band of interest), $\lambda_0$, since the increasing wavelength increases the magnitude of the susceptibility. By contrast, a constant-permittivity ``dielectric'' has nearly opposite dependencies. The figure of merit is independent of center wavelength, and highly dependent on the bandwidth. Because the bandwidth is the source of loss, there is a tradeoff between average response and bandwidth. Finally, 2D Drude conductors are somewhere in between. Loss originates from both the material parameter $\gamma$ as well as the bandwidth, with increasing FOM towards the lower-right-hand corner of \figref{fom}(g): small bandwidth and large wavelength (for a large conductivity). These simplified metal/dielectric/2D conductor profiles capture well the key dependencies of the FOM for real materials: the plots in \figref{fom}(b,c,h) follow the same trends as those in \figref{fom}(a,d,g): metal~\cite{palik_2003} FOM increases with wavelength, whereas dielectrics (Si \cite{chandler-horowitz_amirtharaj_2005, green_2008} and SiC \cite{francoeur_menguc_vaillon_2010, larruquert_2011}) and polaritonic materials (SiO2 \cite{malitson_1965, popova_tolstykh_vorobev_1972, kitamura_pilon_jonasz_2007} and TiO2 \cite{devore_1951, siefke_2016}) that support surface phonon-polaritons at mid-IR frequencies~\cite{maier_2007} do not depend appreciably on wavelength. Conversely, the plots in \figref{fom}(e,f,i) show the effects of increasing bandwidth, with metal material FOMs nearly unchanged but those of the dielectrics and polaritonic materials decreasing nearly linearly. The material FOM of 2D conductors increases with both wavelength and smaller bandwidths. We consider the 2D conductivities of graphene for various Fermi levels~\cite{jablan_buljan_soljacic_2009}, magnetic biasing~\cite{hanson_2008}, and AA-type bilayer stacking (BLG)~\cite{wang_xiao_mortensen_2016}, hBN~\cite{brar_jang_2014}, and metals Ag, Al, and Au with conductivities set by a combination~\cite{abajo_manjavacas_2015} of bulk properties and interlayer atomic spacing.  

An intriguing prediction that emerges from the LDOS and CDOS power--bandwidth limits is that the 2D-material bounds increase more rapidly for smaller separations ($\sim 1/d^4$) than for bulk materials ($\sim 1/d^3$), suggesting that 2D materials should overtake bulk materials as optimal, with the precise transition depending on the bound prefactors and, crucially, the relative 2D/bulk material figures of merit. In \figref{2d3dfom}, we consider Drude models for both a 2D conductivity ($\sigma = i\omega_p/(\omega + i\gamma)$) and a bulk-material susceptibility ($\chi = -\omega_p^2 / (\omega^2 + i \gamma \omega)$), and plot isocontours for the material FOMs of each in (a). In \figref{2d3dfom}(b), we trace out the region of frequency and bandwidth for which the bulk, 3D material has a larger maximal response, and the region for which the 2D material offers larger maximal response. This line will be different for every 2D/bulk-material pair, and is determined by \eqreftwo{nearbound}{nearbound2D} and their CDOS analogs.

\begin{figure} [tbh]
    \includegraphics[width=1\linewidth]{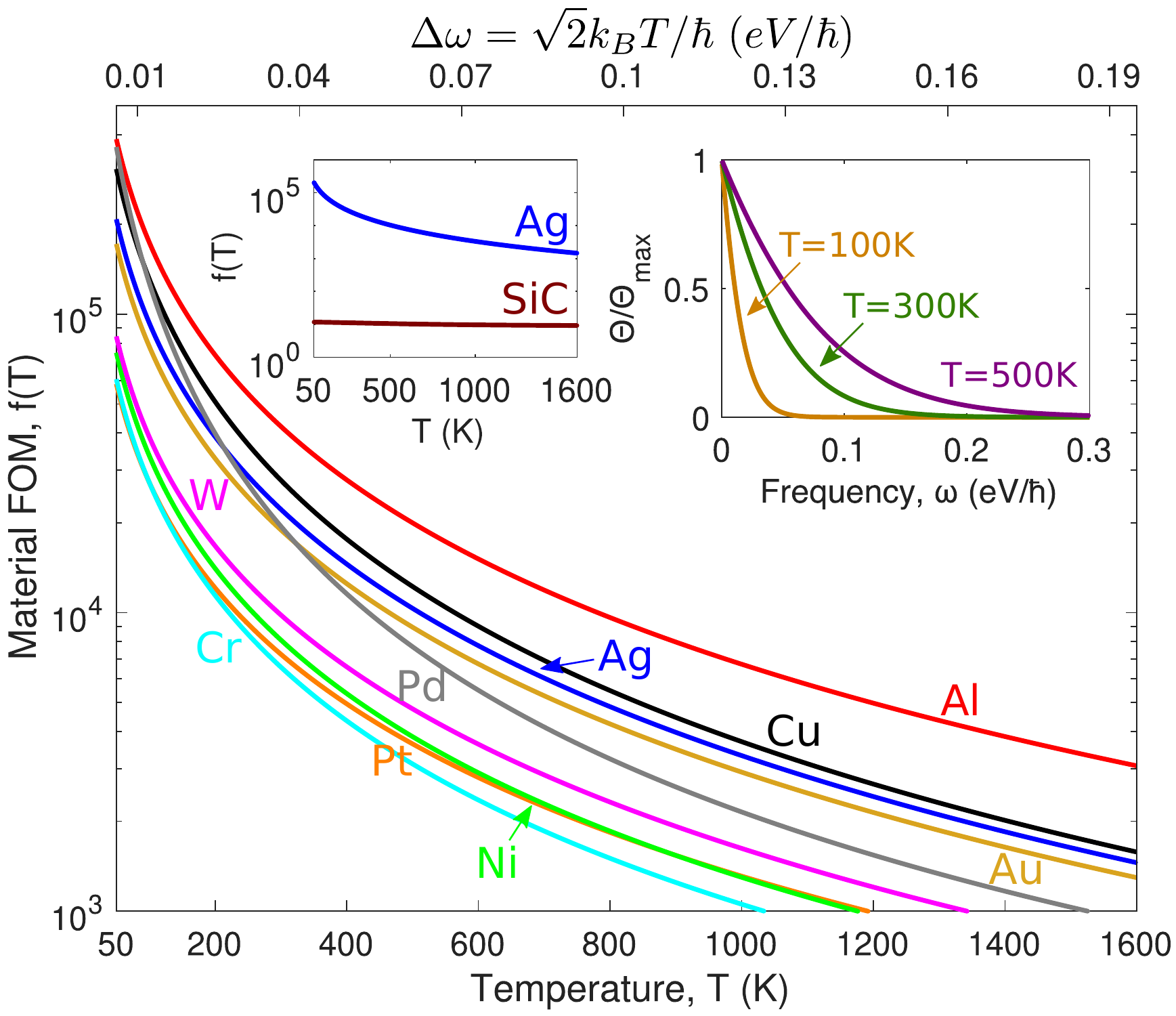} 
    \caption{Comparison of material FOM $f(T)$ in the context of NFRHT for a variety of conventional metals. Left inset shows orders-of-magnitude difference between silver and SiC, a polar dielectric that supports surface-phonon polaritons at infrared frequencies yet has a small material FOM due to its non-Drude-like permittivity.  The right inset shows the mean energy spectrum $\planck$ (normalized to its maximum value) at different temperatures. The decaying FOM reflects increasing loss from bandwidth, which broadens with temperature. For practical range of temperatures as shown here, metals are superior under this metric compared to dielectrics. Generally, $f(T)$ favors materials with large plasma frequency $\omega_p$ and small material loss rate $\gamma$.}  
    \label{fig:rhtfom} 
\end{figure} 

The above discussion about the material FOM as a function of bandwidth is particularly relevant for LDOS and CDOS, where different source configurations lead to different center frequencies and bandwidths. For near-field RHT, as shown in \secref{RHT}, the center frequency is fixed (at zero), the bandwidth is uniquely determined by the temperature, and the material FOM takes a particularly simple form, $f(T) = \chi(i\sqrt{2}k_B T / \hbar)$. At practical temperatures of interest, the material FOM is determined by the susceptibility at an infrared frequency, except along the imaginary frequency axis instead of the real line. Dielectrics have near-constant susceptibilities over this range; surprisingly, polar dielectrics, which have resonant transitions in the infrared, corresponding to surface-phonon-polariton modes, have no resonant peaks along the imaginary axis, and exhibit dielectric-like permittivities and material FOMs. For plasmonic materials, we can use a Drude model to describe the material FOM, since $k_{B}T / \hbar$ is typically an infrared frequency. For a Drude model with plasma frequency $\omega_p$ and loss rate $\gamma$, the material FOM is
\begin{align}
   f(T) = \frac{\hbar^2 \omega_p^2}{\sqrt{2}k_{B}T  \, (\hbar \gamma + \sqrt{2}k_{B}T)}  \label{eq:fom_dr} 
\end{align}
Materials with large plasma frequency and small $\gamma$ are ideal for maximum heat transfer.

\Figref{rhtfom} shows the material FOM for common plasmonic materials, all exhibiting a \emph{decrease} with temperature. Although this may appear counterintuitive, it does not imply that maximal NFRHT decreases with temperature, as \eqref{RHTsumbound4} has a separate $T^4$ multiplicative factor. Instead, such behavior reflects the fact that increasing temperature has the same effect as increasing material loss due to the larger bandwidth over which NFRHT occurs. We can understand this through two alternative vantage points: the larger bandwidth increases the imaginary part of the complex frequency at which the equivalent scattering problem occurs, and moving higher into the UHP increases loss; alternatively, \secref{sumrule} showed than LDOS quantities are subject to sum rules, and thus increasing bandwidth necessarily reduces the average response. For non-dispersive dielectrics, the material FOM does not depend on temperature and takes on small values relative to metals (which usually have large susceptibilities at small frequencies), as shown in the inset of \figref{rhtfom}. 

We can derive a particularly simple form of the NFRHT bound, \eqref{RHTsumbound4}, for plasmonic materials with $f$ given by \eqref{fom_dr}. For many such materials, at practical temperatures of interest, the material loss rate $\gamma$ for many materials is much larger than $\gamma \gg k_{B}T / \hbar$. In the near field, the separation distance $d$ is much smaller than $\lambda_T$, such that the exponential factor in \eqref{RHTsumbound4} is approximately 1. Assuming that both bodies consist of the same material, we can plug \eqref{fom_dr} into \eqref{RHTsumbound4} to arrive at the following bound:
\begin{align}
    \frac{H_{1 \rightarrow 2}}{A} \leq \frac{2\sqrt{2}}{\pi^2}   \frac{\omega_p^2 k_{B}T}{\gamma d^2} = \sigma T^4 \left \lbrace \sqrt{2} \beta \left( \frac{\lambda_T}{d} \right)^2 \frac{\omega_p}{\gamma}\frac{\omega_p}{k_{B}T / \hbar} \right \rbrace.
    \label{eq:rht_dl} 
\end{align}
Thus we see that there are three enhancements relative to a blackbody: the near-field distance enhancement, an enhancement from the ratio of the plasma frequency to the loss rate, and an enhancement from the ratio of the plasma frequency to an effective thermal frequency.

The material FOM extends in a natural way to anisotropic, magnetic, and even spatially inhomogeneous media, as shown in the \SM. Nonlocality, wherein the polarization field at a position $\xv$ depend on the electromagnetic field at another $\xv'$, can also be incorporated for certain hydrodynamic models~\cite{ciraci_hill_2012, mortensen_raza_2014, raza_bozhevolnyi_wubs_mortensen_2015}. An intriguing open question is whether density functional theory models can be bounded in a similar fashion. Such bounds could motivate and clarify the search for new ``quantum materials.''

\section{Extensions and Summary} \label{sec:extensions}
We have established a framework for identifying upper bounds to near-field optical response over any frequency bandwidth of interest, with an emergent material FOM that enables quantitative comparisons of any material. We derived bounds for three optical-response functions: the local density of states (LDOS), a measure of the spontaneous-emission enhancement for any electric or magnetic dipole (or atomic dipolar transition), cross density of states (CDOS), a field-correlation function, and radiative heat transfer, a measure of energy transfer from thermal fluctuations. The property of these response functions that is critical to our framework is the fact that they can be related to the imaginary part of a function that is analytic in the upper half of the complex-frequency plane. Here we explore how our complex-analytic framework can be extended to other optical response functions.

There are a few near-field quantities that map closely to LDOS. First, atomic Lamb shifts due to inhomogeneous electromagnetic environments~\cite{fussell_mcphedran_sterke_2005, chang_rivera_joannopoulos_soljacic_kaminer_2017} are given by frequency integrals of 
\begin{align}
    \Im \Gamma_{ij}(\xv,\xv,\omega),
\end{align}
multiplied by frequency-dependent prefactors that include the atomic frequencies and position matrix elements. Hence the sum rules and power--bandwidth limits derived here can be directly extended to the emitter--environment coupling rate in the Lamb shift. Second, Raman scattering~\cite{Boyd2003} is a process in which a pump wave interacts with a molecular transition, and subsequent emission that is potentially enhanced by the electromagnetic environment. It appears possible to bound this interaction above by the product of the LDOS at the separate pump and emission frequencies, in which case the framework herein can be applied for sum rules and bounds. We will discuss the derivation and bounds to this process in a separate publication~\cite{Michon2018}. 


A more complex case is that of free-electron radiation (e.g. Smith-Purcell, Cherenkov, etc.), in which a free electron at high speed (of order $c$) interacts with a structured medium to generate electromagnetic radiation. The incident electromagnetic field of a free electron is proportional to a modified Bessel function. One difficulty that arises in considering sum rules and power--bandwidth limits in this case is that the modified Bessel function has a logarithmic frequency dependence at the origin, rendering it difficult to apply standard contour-integral techniques as we have done here. In two dimensions, a constant-velocity free electron emits (evanescent) plane waves, and sum rules and bounds appear to emerge in a straightforward way. The three-dimensional case may be more difficult, however.

Another complication emerges when the dipolar sources are embedded within the scatterers of interest. This is typical of Casimir force, which is a momentum-transport quantity arising from vacuum-induced fluctuations. It appears possible to derive sum rules for such a quantity by exploiting the same generalized reciprocity~\cite{kong_1975} that we used for near-field radiative heat transfer. We will consider sum rules, power--bandwidth limits, and interesting physical consequences for Casimir physics in an upcoming publication~\cite{Shim2018}.

Finally, we consider extensions of this framework to cases when the incident field is not generated by a localized dipolar or free-electron source, but instead by a plane wave. At first glance, it would appear that the conventional optical theorem~\cite{newton_1976, bohren_clothiaux_huffman_1983, lytle_carney_schotland_wolf_2005, jackson_2013} provides a simple analytic quantity to serve as the basis for the contour-integral and energy-conservation approaches developed here. Yet the bounds derived by such an approach yield a term that grows exponentially with bandwidth (the opposite of the exponential decay seen in \eqref{lorbound}). Such a dependence is not physical---known sum rules~\cite{gordon_1963} would contradict it---and is instead an indication that the energy constraints developed herein, based on the positive-definite quantities in \eqref{absext}, may not be optimal for plane-wave sources. Modified constraints may be required to develop power--bandwidth limits for plane-wave scattering.

The bounds derived here, and those suggested above, suggest a tremendous design opportunity in near-field nanophotonics. For various canonical structures, there are frequency ranges at which they come close to reaching the bounds, but there are also wide frequency ranges at which there is a sizeable gap. New designs, and new approaches such as large-scale computational ``inverse design''~\cite{jensen2011,Lalau-Keraly2013,liang_johnson_2013,piggott2015}, offer the prospect for overcoming the gap, and revealing the physical principles underlying optimal operation. 

\section{Acknowledgments}
We thank Yi Yang, Chia-Wei Hsu, and Thomas Christensen for helpful discussions. We thank Haejun Chung for codes to optimize material permittivity functions. H.S. and O.D.M. were supported by the Air Force Office of Scientific Research under award number FA9550-17-1-0093. L.F. was supported by a Shanyuan Overseas scholarship from the Hong Kong Shanyuan foundation at Nanjing University. S.G.J was supported by the Army Research Office under contract numbers W911NF-18-2-0048 and W911NF-13-D-0001.

\providecommand{\noopsort}[1]{}\providecommand{\singleletter}[1]{#1}%

\end{document}


\preprint{APS/123-QED}

\title{Supplementary Materials: Fundamental limits to near-field optical response, over any bandwidth}

\author{Hyungki Shim}
\affiliation{Department of Applied Physics and Energy Sciences Institute, Yale University, New Haven, Connecticut 06511, USA}
\affiliation{Department of Physics, Yale University, New Haven, Connecticut 06511, USA}
\author{Lingling Fan}
\affiliation{Department of Applied Physics and Energy Sciences Institute, Yale University, New Haven, Connecticut 06511, USA}
\affiliation{School of Physics and National Laboratory of Solid State Microstructures, Nanjing University, Nanjing 210093, China}
\author{Steven G. Johnson}
\affiliation{Department of Mathematics, Massachusetts Institute of Technology, Cambridge, Massachusetts 02139, USA}
\author{Owen D. Miller}
\affiliation{Department of Applied Physics and Energy Sciences Institute, Yale University, New Haven, Connecticut 06511, USA}

\date{\today}

\maketitle

\tableofcontents

\section{Sum rules and power--bandwidth limits for non-vacuum background}
If the emitter and scatterer are embedded in a background material with a scalar, non-dispersive permittivity $\epsb$ and permeability $\mub$, LDOS is modified as follows:
\begin{align}
\rho(\omega) =\Im \sum_j \left[ \frac{1}{\pi \omega} \left(\epsb \cc{\pv}_j \cdot \Ev_{\textrm{s},j}(\xv_0)  + \mub \cc{\mv}_j \cdot \Hv_{\textrm{s},j}(\xv_0) \right)\right] = \Im \Tr \left[ \frac{1}{\pi\omega} \nu_b\tens{\Gamma}_{\rm s}(\xv_0,\xv_0,\omega) \right], \label{eq:bgrho}
\end{align}
where $\nu_b$ is a diagonal $6\times 6$ matrix:
\begin{align}
 \nu_b = \begin{pmatrix}
        \epsb \mathcal{I} & \\
        & \mub \mathcal{I}
    \end{pmatrix}.
\end{align}
Then, to update the sum rules and power--bandwidth limits throughout the paper, $\nu_b$ (or $\epsb$ and $\mub$ individually) is directly inserted into Eqs. (3, 5, 9, 17, 21) and the expression for $\alphaL$ after Eq.~(4) of the main text.

\section{High-frequency scaling behavior of LDOS} 
This section analyzes the high-frequency asymptotic behavior of the frequency-resolved LDOS, to show that its all-frequency integral converges to a finite value. In order to do so, we need to rewrite LDOS in terms of the polarization currents induced in the scatterer $V$ by a point dipole source in its vicinity. By encoding the material properties of the scatterer in the induced polarization $\Pv$ instead of the scattered field $\Esv$, we can better understand the high-frequency scaling behavior of LDOS and hence the validity of its sum rule as will be demonstrated here. Starting from Eq.~(2) of the main text, LDOS can be expressed in terms of the free-space Green's function $\mathbf G_{EE}$, omitting the subscript in this section for clarity (we only consider electric LDOS in the presence of a nonmagnetic material for simplicity, but the high-frequency asymptotics are unchanged even if we include magnetic LDOS and/or consider magnetic materials):
\begin{align}
    \rho^E(\omega) &= \frac{1}{\pi \omega}\Im \sum_j \pv_j(\mathbf x_0) \cdot \Esv{_{,j}}(\mathbf x_0) \nonumber \\
                    &= \frac{1}{\pi\omega} \Im  \sum_j \int_V \pv_j^T(\mathbf x_0) \mathbf G(\mathbf x_0,\mathbf x) \Pv(\mathbf x) \nonumber \\
                    &= \frac{1}{\pi\omega} \Im  \sum_j \int_V \Pv^T(\mathbf x) \mathbf G^T(\mathbf x_0,\mathbf x) \pv_j(\mathbf x_0) \nonumber \\
                    &= \frac{1}{\pi\omega} \Im  \sum_j \int_V \Pv^T(\mathbf x) \mathbf G(\mathbf x,\mathbf x_0) \pv_j(\mathbf x_0) \nonumber \\
                    &= \frac{1}{\pi\omega} \Im \int_V \Eiv^T(\mathbf x) \Pv(\mathbf x)
    \label{eq:ldos}
\end{align}
where $\mathbf G(\mathbf x,\mathbf x_0) \pv_j(\mathbf x_0)$ is the electric field incident on $\mathbf x$ from an electric dipole oriented along $\pv_j$ at $\mathbf x_0$. In the derivation, we have also taken the transpose of the integrand (which does not affect $\rho$, a scalar) and then used the relation $\mathbf G^T(\mathbf x_0,\mathbf x) = \mathbf G(\mathbf x,\mathbf x_0)$, which follows from reciprocity~\cite{rothwell_cloud_2009}. Note that \eqref{ldos} represents the electric component of the expression in Eq.~(20) in the main text, and the steps leading to \eqref{ldos} show explicitly how the expression in Eq.~(20) is derived as well.

In the high-frequency limit, the polarization field must decay towards zero (the bound charges cannot respond to such high frequencies), and on physical grounds~\cite{landau_lifshitz_1984} the decay must occur in proportion to $1/\omega^2$. Conventionally, the decay constant is chosen to be a ``plasma frequency" $\omega_p$ that is physically meaningful for metals but applies to dielectrics as well. Because the scatterer becomes transparent at high frequencies, the Born approximation applies and the polarization field will be directly proportional to the incident field:
\begin{align}
    \Pv(\omega \rightarrow \infty) = -\epso \frac{\omega_p^2}{\omega^2} \Eiv.
    \label{eq:PInf}
\end{align}
Inserting \eqref{PInf} into \eqref{ldos} results in a term of the form $\Im \int_V \Eiv^2$, which can be solved using the free-space Green's function $\mathbf G$ by noticing that the conventional definition for LDOS averages over all dipole orientations:
\begin{align}
    \Im \int_V | \Eiv | ^2 \rightarrow \Im \sum_{ij} \int_V |G_{ij}|^2.
    \label{eq:gfsquare}
\end{align}
The incident field is given by Green's function multiplied by the dipole polarization (with unit magnitude). The Green's function is~\cite{chew_1995}:
\begin{align}
    G_{ij} = \frac{k^2 e^{ikr}}{4\pi r} \left[\left(1 + \frac{i}{a} - \frac{1}{a^2}\right)\delta_{ij} + \left(-1 - \frac{3i}{a} + \frac{3}{a^2}\right) \frac{x_i x_j}{r^2} \right]
    \label{eq:gf}
\end{align}
where $a = kr$, $k=\omega/c$. As a side remark, $s(\omega)$ (the complex extension of LDOS) decays exponentially in the upper half complex-$\omega$ plane due to the term $e^{ikr}$, leaving us to only worry about real frequencies to confirm the validity of LDOS sum rule. In the limit as $\omega \rightarrow \infty$, we can keep only the highest order term in $\omega$ for $G_{ij}^2$, resulting in the following asymptotic expansion (dropping uninteresting numerical prefactors --- we only care about the scaling behavior of $\omega$):
\begin{align}
    \rho^E(\omega \rightarrow \infty) \sim \omega \int_V \sin(2\omega r/c).
    \label{eq:asymp}
\end{align}
An interesting remark about the $\sin(2\omega r/c)$ term is that it arises from taking the square of individual components of $\mathbf G$ instead of its Frobenius norm~\cite{trefethen_bau_1997}, which would get rid of the oscillatory term $e^{ikr}$ and hence $\sin(2\omega r/c)$ in \eqref{asymp}. This is closely related to the fact that we take our dipole source to be real ($\pv$ = $\cc{\pv}$), as it allows us to rewrite LDOS as a volume-integral expression in terms of $\Eiv$ in \eqref{ldos}. For various computations, it appears that $\int_V \sin(2\omega r/c)$ tends to scale as $\sin(\omega)/\omega^2$, rendering $\rho^E(\omega \rightarrow \infty) \sim \sin(\omega)/\omega$.

\begin{figure} [t!]
    \includegraphics[width=0.5\linewidth]{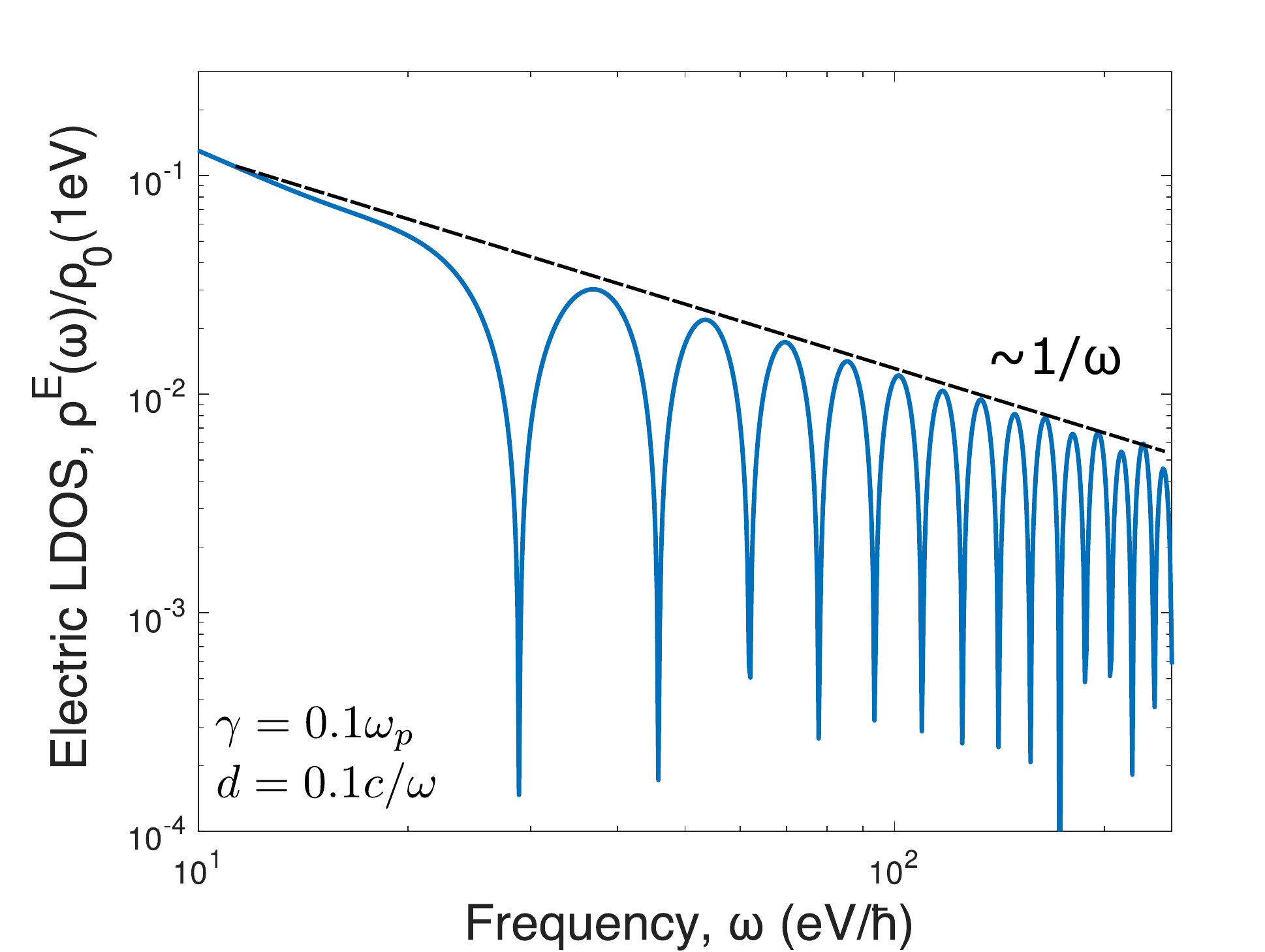} 
    \caption{High-frequency behavior of electric LDOS for a Drude metal with loss rate $\gamma=0.1\omega_p$ and distance $d=0.1c/\omega$ from halfspace. For frequencies much larger than the plasma frequency $\omega_p$, it falls off as ${1/\omega}$ (in an oscillatory manner).}  
    \label{fig:highfldos} 
\end{figure}

While the high-frequency scaling behavior of LDOS depends on the scatterer geometry, we can still investigate under what conditions $\rho(\omega \rightarrow \infty)$ falls off ``sufficiently rapidly" for $\int_{0}^\infty \rho(\omega) \,{\rm d}\omega$ to be finite. Recalling that $\int_{0}^\infty \sin(x)/x^p \,{\rm d}x$ is convergent for $p>0$, our requirement for a valid sum rule is as follows (assuming an oscillatory term):
\begin{align}
    \rho(\omega \rightarrow \infty) \sim O({1/\omega^{\varepsilon}}), \quad \varepsilon>0.
    \label{eq:asymp2}
\end{align}
What \eqref{asymp2} says is that $\int_V \sin(2\omega r/c)$ should fall off as $\sin(\omega)/\omega$ or faster for LDOS to decay at large enough frequencies. As a prototypical example, \figref{highfldos} shows that electric LDOS falls off as $1/\omega$ for a halfspace. In fact, our sum rule for LDOS appears to hold for all geometries (subject to the limiting procedure for singular geometries discussed in the main text).

\section{Sum rule derivation}
In this section, we lay out in detail how the sum rule in Eq~(4) can be derived starting from Eq~(2) and Eq~(3) of the main text. Eq~(3) defines $s(\omega)$ as a complex-valued function that is analytic in the upper half of the complex-$\omega$ plane. Our goal is to construct a closed contour $C$ that includes the real line and is analytic inside $C$, so that we can apply Cauchy's integral theorem to conclude that the integral over $C$ vanishes. However, $s(\omega)$ has a simple pole at $\omega=0$, which forces us to avoid the origin and instead integrate over the infinitesimally small upper semicircle (``bump'' at the origin as shown in Fig.~2(a) of the main text). Since the scattered fields decay exponentially at positive imaginary frequencies, the integral over the infinite semicircle vanishes. This leaves us  with the integral over the real line (except at the singularity at $\omega=0$) and that over the ``bump.'' The latter can be straightforwardly evaluated in polar coordinates, resulting in the following (principal value of the) integral of $s(\omega)$ over all real frequencies (assuming real-valued dipole amplitudes as in the main text):
\begin{align}
    \int_{-\infty}^\infty s(\omega) \,{\rm d}\omega &= \lim_{\omega \to 0} \pi i  \sum_j \frac{1}{\pi} \xi_j^T(\xv_0) \psi_{\textrm{s},j}(\xv_0)  \nonumber \\
  &= i \sum_j [\xi_j^T(\xv_0) \psi_{\textrm{s},j}(\xv_0)]\rvert_{\omega=0}. \label{eq:sumrule}
\end{align}
The zero-frequency limit follows from the fact that our ``bump" can be chosen arbitrarily small in radius. Recalling from Eq. (2) that $\rho(\omega)=\Im s(\omega)$ on the real axis and that $\rho(\omega)=\rho(-\omega)$, we obtain the sum rule for LDOS after taking the imaginary part on both sides of \eqref{sumrule}:
\begin{align}
    \int_{0}^\infty \rho(\omega) \,{\rm d}\omega = \frac{1}{2} \Re \sum_j [\xi_j^T(\xv_0) \psi_{\textrm{s},j}(\xv_0)]\rvert_{\omega=0}. \label{eq:sumrule2}
\end{align}
\Eqref{sumrule2} is our sum rule for LDOS, identical to Eq.~(4) in the main text since $\sum_j \xi_j^T(\xv_0) \psi_{\textrm{s},j}(\xv_0) = \Tr \tens{\Gamma}_{\rm s}(\xv_0,\xv_0,\omega)$ for summation over $j=\{x,y,z\}$.
 
In the main text, we have assumed a scalar permittivity $\varepsilon$ as it reduces $\alphaL = \frac{1}{2}\Re \sum_j [\xi_j^T(\xv_0) \psi_{\textrm{s},j}(\xv_0)]\rvert_{\omega=0}$ to simple analytical form for certain geometries. However, we did not make such an assumption in deriving \eqref{sumrule2}, and so the definition of $\alphaL$ is valid for tensor permittivities as well.

\section{Positivity of $\alphaL$}
While not immediately obvious, the sum-rule constant $\alphaL$ is a positive quantity for arbitrary geometries and materials. For simplicity, we only consider electric fields---the result presented here carries over to magnetic LDOS after appropriate replacements ($\epsten \rightarrow \tens{\mu}$, $\Ev \rightarrow \Hv$). To prove the positivity of $\alphaL$, it is useful to express it in terms of the polarization currents induced, and hence the field, in the scatterer $V$ (following the procedure in \eqref{ldos}):
\begin{align}
    \alphaL = \frac{1}{2}\Re \sum_j [\pv_j^T(\xv_0) \Ev_{\textrm{s},j}(\xv_0)]\rvert_{\omega=0} =  \frac{1}{2}\Re \int_V \Eiv^T(\mathbf x) \Pv(\mathbf x)\rvert_{\omega=0}.  \label{eq:alphavol}
\end{align}
In what follows, the fields are all evaluated in the electrostatic regime.

Expressing $\Ev$ as a sum of the incident and scattered components $\Eiv$ and $\Esv$ respectively, where $V^{+}$ denotes the entire volume outside the scatterer (where the susceptibility is zero):
\begin{align}
    \int_V \Eiv^T\Pv = \int_V \Eiv^T\chiten\Ev &= \int_V \Eiv^T\chiten\Eiv + \int_V \Eiv^T\chiten\Esv \\
    &= \int_V \Eiv^T\chiten\Eiv + 2 \int_V \Eiv^T\chiten\Esv + \int_{V+V^{+}}\Esv^T\epsten\Esv.   \label{eq:pos1}
\end{align}
In going from the first to the second line, we have used the following relation, which can be derived by expressing the integral over $V^{+}$, $\int_{V^{+}} \Esv^T\epsten\Esv$, in terms of the fields inside the scatterer $V$ through repeated use of the divergence theorem:
\begin{align}
    \int_V \Eiv^T\chiten\Esv=-\int_{V+V^{+}}\Esv^T\epsten\Esv. \label{eq:pos2}
\end{align}

Rewriting \eqref{pos1} in a more suggestive form,
\begin{align}
 \int_V \Eiv^T\chiten\Ev &= \int_{V}\left( \Esv + \chiten \epsten^{-1}\Eiv \right)^T \epsten \left( \Esv + \chiten \epsten^{-1}\Eiv \right) + \int_{V}  \Eiv^T \left(\chiten - \chiten^2 \epsten^{-1}\right) \Eiv + \int_{V^{+}}
\Esv^T\epsten\Esv \\
&= \int_{V}\left( \Esv + \chiten \epsten^{-1}\Eiv \right)^T \epsten \left( \Esv + \chiten \epsten^{-1}\Eiv \right) + \int_{V} \Eiv^T \epso\chiten\epsten^{-1} \Eiv + \int_{V^{+}}\Esv^T\epsten\Esv \\
&> 0 \label{eq:pos3}
\end{align}
since $\chiten$ and $\epsten$ are assumed to be hermitian and positive-definite (they also commute, since $\chiten \epsten = \left(\epsten - \epso \mathcal{I} \right) \epsten = \epsten \left(\epsten - \epso \mathcal{I}\right) = \epsten \chiten$ and similarly for their inverses and squares). In going from the first to the second line, we used the fact that $\chiten - \chiten^2 \epsten^{-1} =   \chiten \left[\mathcal{I} - \left(\epsten - \epso \mathcal{I} \right) \epsten^{-1}\right]=\epso\chiten\epsten^{-1}$. \Eqref{pos3} thus shows that $\alphaL>0$.

An intuitive way of seeing that $\alphaL$ is positive is to consider an infinitesimally small volume of a scatterer. Since we know that $\alphaL$ is by definition zero in free space, the presence of an arbitrarily small scatterer should also have $\alphaL$ arbitrarily close to zero and positive (and hence positive for \emph{any} nonzero volume). Otherwise, the presence of a scatterer with negative $\alphaL$ would imply from our monotonicity theorem that further shrinking its volume (until it becomes negligible in spatial extent) will cause $\alphaL$ to become more and more negative---in contradiction with the fact that $\alphaL=0$ in the absence of a scatterer.

\section{Derivation of $ \alphaL$ for canonical geometries}
If the scattered fields are known at zero frequency, we can use the expression from \eqref{sumrule2} to directly evaluate $\alphaL$:
\begin{align}
    \alphaL = \frac{1}{2}\Re \sum_j [\xi_j^T(\xv_0) \psi_{\textrm{s},j}(\xv_0)]\rvert_{\omega=0} \label{eq:Vscat}
\end{align}
where the dipole source is at $\mathbf x_0$ and $\psi_{\textrm{s},j}$, here evaluated at zero frequency, denotes the scattered field due to $\xi_j$, a unit dipole polarized along $\hat{\bf{j}}$. In what follows, we use the image charge method to exactly determine the electro/magnetostatic scattered field for canonical geometries. Except for a halfspace, these geometries do not admit closed-form expressions for $ \alphaL$ for finite $\varepsilon$, and so for non-halfspace geometries we assume conductive materials (that are perfect conductors at zero frequency). For simplicity, we focus on the electric LDOS in this section. Duality simplifies the corresponding computations for magnetic LDOS.

\subsection{Halfspace}
Consider a plane interface separating the two semi-infinite half spaces---vacuum with permittivity $\epso=1$ at $z>0$ and a dielectric with (electrostatic) permittivity $\varepsilon$ at $z<0$. In the electrostatic regime, a dipole ${\pv}_j$ a distance $d$ above the interface will create an image dipole with some orientation at a distance $d$ below the interface. Since the LDOS due to randomly oriented dipoles equals that for dipoles each along one of the $x,y,z$-axis,
\begin{align}
    \frac{1}{2}\sum_j {\pv}_j \cdot \Esv{_{,j}} (\mathbf x_0) &= \frac{1}{8\pi (2d)^3}\sum_j  {\pv}_j \cdot [3 ({\pv}'_j \cdot \hat{\mathbf r})\hat{\mathbf r}-{\pv}'_j]  \nonumber \\
  &= \frac{1}{16 \pi d^3}\frac{\varepsilon-1}{\varepsilon+1} \label{eq:hpimage}
\end{align}
where ${\pv}'_j$ is the image dipole due to ${\pv}_j$ and $\hat{\mathbf r}$ the unit vector pointing from ${\pv}'_j$ to  ${\pv}_j$ ($\hat{\mathbf r}=\hat{\mathbf z}$ here). The effect of the dielectric medium with permittivity $\varepsilon$ at $z<0$ is to introduce the prefactor $\frac{\varepsilon-1}{\varepsilon+1}$ (which simplifies to 1 for perfect conductors, as expected) in the image charges~\cite{jackson_2013} in order to satisfy the requirement of continuous tangential fields at $z=0$. From ${\pv}'_x = -\frac{\varepsilon-1}{\varepsilon+1}{\pv}_x$, ${\pv}'_y = -\frac{\varepsilon-1}{\varepsilon+1}{\pv}_y$, ${\pv}'_z = \frac{\varepsilon-1}{\varepsilon+1}{\pv}_z$, we obtain the final result in \eqref{hpimage}. Adding the corresponding term for magnetic LDOS, we arrive at $ \alphaL$ for a halfspace in Eq.~(5) of the main text.

\subsection{Sphere}
A dipole ${\pv}_j$ is placed at a distance $h$ from the center of a perfectly conducting sphere of radius $R$ such that it is outside of the sphere ($h>R$). For simplicity, assume that the dipole lies along the $z$-axis with the origin coinciding with the center of the sphere. For vanishing tangential fields on the spherical surface, the image dipole ${\pv}'_j$ will be at a distance $\frac{R^2}{h}$ from the center of sphere (also along $z$-axis) such that ${\pv}'_x = -\left(\frac{R}{h}\right)^3{\pv}_x$, ${\pv}'_y = -\left(\frac{R}{h}\right)^3{\pv}_y$, ${\pv}'_z =\left(\frac{R}{h}\right)^3{\pv}_z$. Denoting the distance between the source and image dipoles as $r=h\left[1-\left(\frac{R}{h}\right)^2 \right]$,
\begin{align}
    \frac{1}{2}\sum_j {\pv}_j \cdot \Esv{_{,j}} (\mathbf x_0) &= \frac{1}{8\pi r^3}\sum_j  {\pv}_j \cdot [3 ({\pv}'_j \cdot \hat{\mathbf z})\hat{\mathbf z}-{\pv}'_j]  \nonumber \\
  &= \frac{1}{2\pi h^3 \left[1-\left(\frac{R}{h}\right)^2 \right]^3}\left(\frac{R}{h}\right)^3. \label{eq:spimage}
\end{align}
Since $h=R+d$ where $d$ is the distance from the dipole to the nearest surface of the sphere, we arrive at the following result:  
\begin{align}
\alphaL=\frac{R^3}{2\pi(R+d)^6 \left( 1- \left(\frac{R}{R+d} \right)^2 \right)^3}. \label{eq:spsumrule}
\end{align}
While we assumed the dipole to be on the $z$-axis, \eqref{spsumrule} holds for any dipole a distance $d$ away from the surface (as long as we sum over random orientations) from spherical symmetry.

\subsection{Two halfspaces}
The derivation of $\alphaL$ for two perfectly conducting halfspaces closely mimics the halfspace case, but now there are two boundary conditions to safisfy---the tangential fields must vanish at both interfaces. Taking the two halfspaces to be at $z<0$ and $z>d$ (and vacuum with permittivity $\epso$ in between the two halfspaces), consider a dipole ${\pv}_j$ placed in between the two halfspaces at $z=\frac{d}{2}$. This will create image dipoles on each halfspace a distance $d$ from the dipole source. However, the image dipole in one halfspace will cause a non-vanishing tangential field at the interface of the other halfspace, so that these image charges leads to another set of image charges (now a distance $2d$ from the dipole source on both sides). Iterating this procedure, we obtain the following expressions for the parallel ($z$-axis) and perpendicular ($x,y$-axis) components:
\begin{align}
 \frac{1}{2}\sum_j {\pv}_j \cdot \Esv{_{,j}} (\mathbf x_0) =\frac{1}{4\pi d^3}\begin{cases}
              2\left(1+ \frac{1}{2^3} + \frac{1}{3^3} + ... \right) \label{eq:twohalf} & j= z \\              
             1 - \frac{1}{2^3} + \frac{1}{3^3} - ... & j=x,y \\
           \end{cases} 
\end{align}
where the alternating/constant sign for the perpendicular fields follow from the fact that the image dipole parallel/orthogonal to the interfaces is opposite/same in sign to the dipole that created it (which itself is an image dipole until one arrives at the original dipole source). As expected on physical grounds, the infinite summations corresponding to infinite number of image charges converge to finite values given by $\zeta(3)= \sum\limits_{n=1}^{\infty}\frac{1}{n^3}=1.20...$ and $\eta(3)= \sum\limits_{n=1}^{\infty}\frac{(-1)^{n-1}}{n^3}=0.90...$ (where $\zeta$ and $\eta$ are Riemann zeta and Dirichlet eta functions respectively~\cite{abramowitz_stegun_1974}). Using these relations in \eqref{twohalf} gives us $\alphaL$ for two halfspaces separated by $d$:
\begin{align}
\alphaL=\frac{2.10...}{2\pi d^3} \label{eq:dhssumrule}
\end{align}
where the prefactor $2.10...$ is equivalent to $\zeta(3)+\eta(3)$.

\section{Generalization of monotonicity theorem for anisotropic materials}
In deriving the monotonicity theorem in Sec. IB of the main text, we have taken the permittivity and permeability to be isotropic. In this section, we generalize Eq.~(8) of the main text for the case of anisotropic media, following the prescription given in~\cite{kottke_farjadpour_johnson_2008}. For simplicity, we consider electric fields only---extension to the magnetic case is straightforward after appropriate replacements ($\epsten \rightarrow \tens{\mu}$, $\Ev \rightarrow \Hv$, $\Dv \rightarrow \Bv$). Working in a local coordinate system for each point on the geometrical boundary dividing regions of permittivity $\epsten_1$ (scatterer) from $\epsten_0$ (background), we define $F=(D_1,E_2,E_3)$ such that $F_1 = D_{\perp}$ and $\mathbf{F}_{2,3} = \Ev_{\parallel}$. We can then write $\Delta \alphaL$ under outward deformation of the scatterer:
\begin{align}
\int \Delta h_n \mathbf{F}\cdot\left[\tens{\tau}\left(\epsten_{1}\right)-\tens{\tau}\left(\epsten_{0}\right)\right]\cdot\mathbf{F} \rm{d}A, \label{eq:tenmono}
\end{align}
where $\tens{\tau}$ is a $3\times 3$ matrix given by:
\begin{align}
 \tens{\tau}(\epsten)=\left(\begin{array}{ccc}
-\frac{1}{\varepsilon_{11}} & \frac{\varepsilon_{12}}{\varepsilon_{11}} & \frac{\varepsilon_{13}}{\varepsilon_{11}}\\
\frac{\varepsilon_{21}}{\varepsilon_{11}} & \quad\varepsilon_{22}-\frac{\varepsilon_{21}\varepsilon_{12}}{\varepsilon_{11}}\quad & \varepsilon_{23}-\frac{\varepsilon_{21}\varepsilon_{13}}{\varepsilon_{11}}\\
\frac{\varepsilon_{31}}{\varepsilon_{11}} & \varepsilon_{32}-\frac{\varepsilon_{31}\varepsilon_{12}}{\varepsilon_{11}} & \varepsilon_{33}-\frac{\varepsilon_{31}\varepsilon_{13}}{\varepsilon_{11}}\end{array}\right).\label{eq:ptens-definition}
\end{align}
For scalar, isotropic permittivities $\epsten_1$ and $\epsten_0$, \eqref{tenmono} reduces to Eq.~(8) of the main text. Thus, a monotonicity theorem holds in the tensor-material case \emph{if} the matrix $\tens{\tau}(\epsten_1) - \tens{\tau}(\epsten_0)$ is positive-definite.
 
\section{Monotonicity theorem for conductive materials}
The monotonicity theorem derived in the main text applies for any finite $\varepsilon$, and the fact that it works for arbitrarily large $\varepsilon$ suggests that such a theorem may apply in the perfect-conductor (at zero frequency case). In this section, we provide an alternative derivation of the monotonicity theorem for the perfect-conductor case. From Sec. IB of the main text, we wish to show that $\Delta \alphaL= [\frac{1}{2}\xi(\mathbf x_0) \cdot \Delta \psi_{\textrm{s}}(\mathbf x_0)]\rvert_{\omega=0} >0$. Working at zero frequency throughout this section, imagine a small change in the geometry of the scatterer such that the LDOS dipole source induces a polarization current $\Pv_{ind}$ in a volume $\delta V = \delta x_n \rm{d}A$ small compared to $V$, where $\delta x_n$ indicates the size of deformation normal to the scatterer boundary. Under this condition, $\Pv_{ind} = \sigma \hat{\mathbf n} = \epso \Ev$ where $\hat{\mathbf n}$ denotes the normal to the conducting surface and $\sigma$ the surface charge density. This follows from the fact that for electric conductors, the electric field is normal to the conducting surface such that $\Ev = \frac{\sigma}{\epso}\hat{\mathbf n}$. Likewise, $\Mv_{ind} = \mu_0 \Hv$ for magnetic conductors. Based on these relations, we arrive at the following expression (absorbing $\epso$ and $\mu_0$ in the definition of $\tens{G}$, which represents the Green's function in the presence of scatterer $V$):
\begin{align}
\label{eq:deltascat}
\Delta \psi_{\textrm{s}}(\mathbf x_0) &=\frac{1}{2} \int \Delta h_n \tens{\Gamma}(\mathbf x_0, \mathbf x)\phi_{ind}(\mathbf x) \rm{d}A \nonumber \\
&=\frac{1}{2} \int \Delta h_n \tens{\Gamma}(\mathbf x_0, \mathbf x) \psi(\mathbf x) \rm{d}A.
\end{align}
Using \eqref{deltascat} and taking transpose on both sides (to exploit reciprocity~\cite{rothwell_cloud_2009}), we can derive $\delta F$ as follows:
\begin{align}
   \Delta \alphaL &=\frac{1}{2} \int \Delta h_n \xi^T(\mathbf x_0)\tens{\Gamma}(\mathbf x_0, \mathbf x) \psi(\mathbf x) \rm{d}A \nonumber \\
  &=\frac{1}{2} \int \Delta h_n \psi^T(\mathbf x) \tens{\Gamma}(\mathbf x, \mathbf x_0) \xi(\mathbf x_0) \rm{d}A \nonumber \\ 
  &=\frac{1}{2} \int \Delta h_n | \psi(\mathbf x) |^2 \rm{d}A > 0. \label{eq:mono}
\end{align}
\Eqref{mono} proves our monotonicity theorem for perfect conductors. Notice that the derivation presented here closely mirrors that of \eqref{ldos} in that we take the transpose of the integrand, use reciprocity (at zero frequency, the ``off-diagonal" $3\times3$ matrices $\tens{G}_{EH}$ and $\tens{G}_{HE}$ vanish and we do not have to worry about sign issues associated with them in taking the transpose of $\tens{\Gamma}$), and the relation $ \psi(\mathbf x) = \tens{\Gamma}(\mathbf x,\mathbf x_0) \xi(\mathbf x_0)$ (but now $\tens{\Gamma}$ is not the free-space Green's function, but one in the presence of scatterer V \textit{before} shape deformation).

\section{Convex constraints in the UHP}
In this section, motivate our definition of $\varphi_{A}$ and $\varphi_{E}$ in Eq.~(19) of the main text, the complex extensions of the non-radiative and total power respectively. To this end, we start from the complex-frequency Maxwell's equations:
\begin{align}
    \underbrace{\begin{pmatrix}
        & \nabla \times \\
        -\nabla \times &
    \end{pmatrix}}_\Theta
    \underbrace{\begin{pmatrix}
        \Ev \\
        \Hv
    \end{pmatrix}}_\psi
    + i \omega \underbrace{\begin{pmatrix}
        \varepsilon & \\
                    & \mu
    \end{pmatrix}}_\nu
    \begin{pmatrix}
        \Ev \\
        \Hv
    \end{pmatrix}
    =
    \underbrace{\begin{pmatrix}
        \mathbf J_e \\
        \mathbf J_m 
    \end{pmatrix}}_J
\end{align}
where $\Theta$ is Hermitian and complex-symmetric. If we separate the fields into incident and scattered components, $\psiInc$ and $\psiScat$ respectively, with the incident field as the solution of $\Theta \psiInc + i\omega\nu = J$ ($\nu_0$ is the background medium), then the scattered field is the solution to
\begin{align}
    \Theta \psiScat + i\omega \eps \psiScat = - i\omega \chi \psiInc
\end{align}
where $\chi = \nu - \nu_0$.

At complex frequencies, passivity requires
\begin{align}
    \Im \left(\omega \chi\right) > 0,
\end{align}
and, consequently, $\Im \left(\omega \eps \right) > 0$ as well. We can use this simple fact to show that two quantities (identical to the absorbed and scattered powers at real frequencies) are positive-definite. We partition the space into two regions, the scatterer occupying the volume $V$, and the external region $V_{\rm out}$. We assume the outer region is finite, which can be accomplished using perfectly matched layers~\cite{berenger_1994, sacks_kingsland_1995}. Then by the passivity condition, two quantities that must be greater than zero are:
\begin{align}
    \varphi_{A} &= \frac{1}{2} \Im \int_V \psi^\dagger \left(\omega \chi\right) \psi \nonumber \\
    \varphiScat &= \frac{1}{2} \Im \int_{V_{\rm out}} \psiScat^\dagger \left(\omega \nu\right) \psiScat \label{eq:phinr}
\end{align}
where $\varphi_{A}$ and $\varphiScat$ are complex extensions of absorbed (non-radiative) and scattered power in that they reduce to the corresponding real-valued powers at real frequencies. 

Through two applications of the divergence theorem we can rewrite $\varphiScat$ in terms of quantities over the \emph{interior} volume $V$,
\begin{align}    
    \varphiScat &= \frac{1}{2} \Im \int_{V_{\rm out}} \psiScat^\dagger \left[ -i \Theta \psiScat\right] \nonumber \\
      &= -\frac{1}{2} \Re \int_S \hat{n}_{\rm in} \cdot \mathbf E_{\rm s} \times \cc{\mathbf H_{\rm s}} \nonumber \\
      &= \frac{1}{2} \Re \int_S \hat{n} \cdot \mathbf E_{\rm s} \times \cc{\mathbf H_{\rm s}} \nonumber \\
      &= -\frac{1}{2} \Re \int_V \psiScat^\dagger \Theta \psiScat \nonumber \\
      &= -\frac{1}{2} \Im \int_V \psiScat^\dagger \left(\omega \nu \psiScat + \omega \chi \psi_{\rm inc}\right)
\end{align}
where the overline denotes complex conjugation and $\hat{n}$ a unit normal pointing outwards from the volume V. Note that in the first line we used the fact that $\chi=0$ in the exterior volume. This is the key to the derivation: the scattered power is the power radiated to the exterior by currents in the interior volume, and this separation of sources and power flow ensures positivity. 

If we define $\varphi_{E}=\varphi_{A}+\varphiScat$ (motivated by extinction), then we have:
\begin{align}
    \varphi_{E} &= \frac{1}{2} \Im \int_V \psi^\dagger \omega \chi \psi - \psiScat^\dagger \left(\omega \nu \psiScat + \omega \chi \psi_{\rm inc}\right) \nonumber  \\
      &= \frac{1}{2} \Im \int_V \psi^\dagger \omega \chi \psi - \psi^\dagger \omega \nu \psi - \psiInc^\dagger \omega \nu \psiInc + \psi^\dagger \omega \nu \psiInc + \psiInc^\dagger \omega \nu \psi - \psi^\dagger \omega \chi \psiInc + \psiInc^\dagger \omega \chi \psiInc \nonumber  \\
      &= \frac{1}{2} \Im \int_V \psi^\dagger \omega \nu_0 \psiInc + \psiInc^\dagger \omega \nu \psi - \psi^\dagger \omega \nu_0 \psi  - \psiInc^\dagger \omega \nu_0 \psiInc \nonumber  \\
      &= \frac{1}{2} \Im \int_V \psi^\dagger \left(\omega \nu_0 - \left(\omega \nu_0\right)^\dagger \right) \psiInc + \psiInc^\dagger \omega \chi \psi  - \psi^\dagger \omega \nu_0 - \psiInc^\dagger \omega \nu_0 \psiInc
\end{align}
Finally, with some notational simplification (using double prime to denote the imaginary part of a matrix), we have
\begin{align}
      \varphi_{E} = \frac{1}{2} \Im \int_V \psiInc^\dagger \left(\omega \chi\right) \psi + \Re \int_V \psi^\dagger \left(\omega \nu_0\right)'' \psiInc - \frac{1}{2} \int_V \psi^\dagger \left(\omega \nu_0\right)'' \psi + \psiInc^\dagger \left(\omega \nu_0\right)'' \psiInc 
      \label{eq:phitot}
\end{align}
We can see that when the frequency is real and the background is lossless ($\nu_0''=0$), the last three terms are zero and we are left with the first term, which is identical to the extinction at real frequencies. Again, this motivates the interpretation of $ \varphi_{E}$ as representing a complex extension of total power.

Combining \eqref{phinr} and \eqref{phitot} ,We can write the constraint $\varphi_{A} \leq \varphi_{E}$ as
\begin{align}
    \frac{1}{2} \int_V \psi^\dagger \left(\omega \nu\right)'' \psi \leq \frac{1}{2} \Im \int_V \psiInc^\dagger \left(\omega \chi\right) \psi + \Re \int_V \psi^\dagger \left(\omega \nu_0\right)'' \psiInc - \frac{1}{2}\int_V \psiInc^\dagger \left(\omega \nu_0\right)'' \psiInc
\end{align}
or, equivalently,
\begin{align}
   \varphi_{A}=\frac{1}{2} \int_V \psi^\dagger \left(\omega \nu\right)'' \psi \leq \frac{1}{2} \Im \int_V \psiInc^\dagger \left[\omega \nu - \left(\omega \nu_0\right)^\dagger\right] \psi - \frac{1}{2}\int_V \psiInc^\dagger \left(\omega \nu_0\right)'' \psiInc = \varphi_{E}
    \label{eq:cons}
\end{align}
In the main text, $\varphi_{A}$ and $\varphi_{E}$ are defined as the left- and right-hand sides of \eqref{cons}. While this differs slightly from the definition given in \eqref{phinr} (where $\varphi_{E}=\varphi_{A}+\varphiScat$) , such difference does not matter --- complex extension of power need not be uniquely defined as long as the different extensions agree at real frequencies. In fact, the two ways of defining $\varphi_{A}$ and $\varphi_{E}$ give the same constraint in \eqref{cons}.

\section{Formulation of power--bandwidth bounds}  \label{sec:forpb}
Returning to the optimization problem of $\Im s(\omega_0 + i\Delta \omega)$ appearing in average LDOS (see Eq.~(16) of the main text), we wish to formulate it in terms of a quantity linear in (total) field, which is then amenable to convex optimization. To this end, we follow the procedure in \eqref{ldos} and define a modified incident field $\widetilde{\psi}_{\rm inc} = \begin{pmatrix} \Ev_{\rm inc} & -\Hv_{\rm inc} \end{pmatrix}^T$ (the minus sign in front of $\Hv_{\rm inc}$ appears from using reciprocity to the magnetic Green's function). After using the following relation between polarization currents $\phi$ and fields $\psi$, $\phi=\chi \psi$, we obtain the following expression (where $\omega = \omega_0 + i\Delta \omega$): 
\begin{align}
\Im s(\omega) = \Im \left[ \frac{1}{\pi\omega} \int_V  \widetilde{\psi}^T_{\rm inc}(\omega) \chi(\omega) \psi(\omega) \right].
\label{eq:tilde}
\end{align}
In light of \eqref{tilde}, maximizing $\Im s(\omega_0 + i\Delta \omega)$ subject to energy-conservation constraints in the UHP reduces to the following optimization problem (evaluated at a \emph{complex} frequency):
\begin{equation}
    \begin{aligned}
        & \underset{\psi}{\text{maximize}} & & \Im \left[ f^T \left(\omega \chi\right) \psi \right] \\
    & \text{subject to}                & & \psi^\dagger \left(\omega \nu\right)'' \psi \leq \Im \left[ \psiInc^\dagger \left(\omega\nu - \left(\omega\nu_0\right)^\dagger \right) \psi \right]  - \psiInc^\dagger \left(\omega \nu_0\right)'' \psiInc
    \end{aligned}
\end{equation}
where the volume integrals are implicit in the vector-vector multiplications and $ f = \frac{1}{\pi\omega^2}\widetilde{\psi}_{\rm inc}$. For electric LDOS, $\widetilde{\psi}_{\rm inc}$ only contains electric fields, which greatly simplifies the problem. But, we keep the term as it is for sake of generality, simplifying only after the optimal solution is derived. Before solving the optimization problem, it is helpful to rewrite it in terms of generic variables:
\begin{equation}
    \begin{aligned}
    & \underset{\psi}{\text{maximize}} & & \Im \left[ a^T \psi \right] \\
    & \text{subject to}                & & \psi^\dagger b \psi \leq \Re \left[c^\dagger \psi\right] - d
    \end{aligned}
\label{eq:genopt}    
\end{equation}
where 
\begin{align*}
    a &= \left(\omega \chi\right)^T f \\
    b &= \left(\omega \nu\right)'' \\
    c &= i \left[ \left(\omega \nu\right)^\dagger - \omega \nu_0 \right] \psiInc \\
    d &= \psiInc^\dagger \left(\omega \nu_0\right)'' \psiInc
\end{align*}
and $b=b^\dagger$ (guaranteeing that $\psi^\dagger b \psi$ is real). A straightforward Lagrange multiplier approach yields the optimal current $\psi$ in terms of a Lagrange multiplier $\lambda$:
\begin{align}
    \psi = \frac{b^{-1}}{2} \left(c - \frac{i}{\lambda} \cc{a}\right).
\end{align}
Enforcing the constraint in Eq.~(6), per the KKT conditions~\cite{mikosch_j._jorge_2006}, requires the Lagrange multiplier to satisfy
\begin{align}
    \lambda^2 = \frac{a^T b^{-1} \cc{a}}{c^\dagger b^{-1} c - 4d}
\label{eq:lambda}   
\end{align}
from which we can obtain the figure of merit:
\begin{align}
    \Im\left[a^T \psi\right] &= \Im \left[ \frac{1}{2} a^T b^{-1} c - \frac{i}{2\lambda} a^T b^{-1} \cc{a} \right] \\
&= \frac{1}{2} \Im \left[ a^T b^{-1} c \pm i\sqrt{\left(a^T b^{-1} a^*\right) \left(c^\dagger b^{-1} c - 4d\right)} \right]    
\end{align}
where we have inserted the expression for $\lambda$ from \eqref{lambda} in going from the first to the second line. In the common case where $\nu''$ is a homogeneous, scalar quantity, we can simplify the above expression:
\begin{align}
    \Im\left[a^T \psi\right] = \frac{1}{2b} \Im \left[ a^T c \pm i\sqrt{a^\dagger a \left(c^\dagger c - 4bd\right)} \right].
\label{eq:root}  
\end{align}
Since we want to maximize $\Im \left[a^T \psi\right]$, we want to choose the positive sign in \eqref{root}. Also, after some algebraic manipulation,
\begin{align}
    c^\dagger c - 4bd = \left| \omega \chi \right|^2 \psiInc^\dagger \psiInc
\end{align}
when the material values are scalar constants (in general, $|\omega|^2 \psiInc^\dagger \chi^\dagger \chi \psiInc$). Rewriting in terms of physical parameters, 
\begin{align}
    \Im s(\omega) \leq \frac{1}{2} \left( \left\| \left(\omega \chi\right) \left(\left(\omega \nu\right)''\right)^{-1}\left(\omega\chi\right)^\dagger \right\| + \left\| \left(\omega \chi\right)^\dagger \left(\left(\omega \nu\right)''\right)^{-1}\left(\omega\chi\right) \right\| + 2 \left\| \left(\omega \chi\right) \left(\left(\omega \nu\right)''\right)^{-1} \left(\omega \nu_0\right)''\right\| \right) \sqrt{f^\dagger f \psiInc^\dagger \psiInc}.
    \label{eq:app_general_bound}
\end{align}
Notice that \eqref{app_general_bound} applies to general $6\times6$ susceptibility tensor that can be magnetic, anisotropic, nonreciprocal, and even spatially inhomogeneous. For the case of scalar, homogeneous materials that we are considering, \eqref{root} can be further simplified as:
\begin{align}
  \Im s(\omega) \leq \left( \frac{\left|\omega\chi\right|^2 + \left|\omega\chi\right| \left(\omega\nu_0\right)''}{\left(\omega\nu\right)''} \right) \sqrt{f^\dagger f \psiInc^\dagger \psiInc}
    \label{eq:bound_gen}
\end{align}
Upon plugging in $\psiInc = \left(\begin{smallmatrix}\Eiv \\ \Hiv \end{smallmatrix}\right)$, \eqref{bound_gen} gives us the contribution for the upper bound of average LDOS arising from the Lorentzian pole (see Eq.~(15) of main text). Since \eqref{bound_gen} bounds $\Im s(\omega_0 + i\Delta \omega)$ in [Eq.~(16) of main text], we arrive at an upper bound for frequency-averaged LDOS:
\begin{align}
    \langle \rho \rangle  \leq &\frac{f(\omega)e^{-2 d \Delta \omega /c}}{\pi\left|\omega \right|} \sum_j \int_V \psi_{\textrm{inc},j}^\dagger(\omega) \psi_{\textrm{inc},j}(\omega) \,{\rm d}V + 2H_{\omega_0, \Delta \omega}(0) \alphaL. \label {eq:genbound}
\end{align}
\Eqref{genbound} is the result used in Eq.~(23) of main text. For simplicity, we now consider nonmagnetic materials in the presence of an electric dipole ($ \Hiv=\mathbf 0$). Calculating the incident electric field in the spirit of \eqref{gfsquare} from the free-space Green's function in \eqref{gf}, we arrive at the following bound that keeps all the terms in the expansion (where we bound $\alphaL$ by a halfspace enclosing the scatterer $V$ with a minimum emitter-scatterer separation $d$):
\begin{align}
\langle \rho \rangle  \leq \rho_0(|\omega|)\frac{f(\omega) \left| k \right|^3}{8\pi} \int_{V} \sum_{n=2}^{6} \frac{C_n e^{-ar}}{|k|^{n} r^n} + \frac{\Delta \omega /\pi}{\omega_0^2 + (\Delta\omega)^2} \frac{1}{8\pi d^3} \label{eq:lorbound}
\end{align}
where $f(\omega) = \frac{1}{2\left|\omega\right|} \left( \left\| \left(\omega \chi\right) \left(\left(\omega \nu\right)''\right)^{-1}\left(\omega\chi\right)^\dagger \right\| + \left\| \left(\omega \chi\right)^\dagger \left(\left(\omega \nu\right)''\right)^{-1}\left(\omega\chi\right) \right\| + 2 \left\| \left(\omega \chi\right) \left(\left(\omega \nu\right)''\right)^{-1} \left(\omega \nu_0\right)''\right\| \right)$, which simplifies to $f(\omega) = \left( \frac{\left|\omega\chi\right|^2 + \left|\omega\chi\right| \left(\omega\epso \right)''}{\left|\omega\right| \left(\omega\varepsilon\right)''} \right)$ for scalar, homogeneous materials, is a material-dependent dimensionless metric, $a = 2 \Im \omega / c$, and $r$ denotes the distance from the dipolar source. The coefficients $C_n$'s are given by:
\begin{align*}
    C_2 &= 2 \\
    C_3 &=4\frac{\Delta\omega}{|\omega|} \\
    C_4 &=6-4\left[ \left(\frac{\omega_0}{|\omega|} \right)^2 - \left(\frac{\Delta\omega}{|\omega|} \right)^2 \right] \\
    C_5&=12\frac{\Delta\omega}{|\omega|} \\
    C_6&=6.
\end{align*}
We defer the calculations of the integrals appearing in \eqref{lorbound} to the next section, where we will see how we can use various approximations to simplify their expressions for canonical geometries. We can also extend \eqref{lorbound} to the case of magnetic materials as well as magnetic LDOS by modifying the free-space Green's function to accommodate for magnetic fields and/or dipoles. 

\section{Power--bandwidth bounds}
In our power--bandwidth LDOS bound (\eqref{lorbound}), we end up with an integral that has 5 terms:
\begin{align}
    C_6 \int \frac{e^{-ar}}{r^6}, \ C_5 \int \frac{e^{-ar}}{r^5}, \ C_4 \int \frac{e^{-ar}}{r^4}, \ C_3 \int \frac{e^{-ar}}{r^3}, \ C_2 \int \frac{e^{-ar}}{r^2} \label{eq:bnd_terms}
\end{align}
where $a = 2 \Im \omega / c$ is proportional to the bandwidth over which averaging occurs. The $1/r^n$ contribution to the integrand in the first three terms decays sufficiently quickly that the exponential term is not required for convergence. The term $e^{-ar}$ can simply be bounded above by $e^{-ad}$ and brought out of the integral; for $(\Delta \omega) d / c \ll 1$, such a simplification is also nearly tight. Thus, we can use bounds for the first term:
\begin{align}
    \int \frac{e^{-ar}}{r^6} \leq e^{-ad} \int \frac{1}{r^6} = \pi e^{-ad}
    \begin{cases}
        \frac{1}{6d^3} & \textrm{halfspace} \\
        \frac{1}{2d^4} & \textrm{2D plane} \\
        \frac{1}{3d^3} & \textrm{3D spherical shell} \\
        \frac{1}{d^4} & \textrm{2D spherical surface},
    \end{cases}
\end{align}
and the second term:
\begin{align}
    \int \frac{e^{-ar}}{r^5} \leq e^{-ad} \int \frac{1}{r^5} = \pi e^{-ad}
    \begin{cases}
        \frac{1}{3d^2} & \textrm{halfspace} \\
        \frac{2}{3d^3} & \textrm{2D plane} \\
        \frac{2}{d^2} & \textrm{3D spherical shell} \\
        \frac{4}{d^3} & \textrm{2D spherical surface},
    \end{cases}
\end{align}
and the third term:
\begin{align}
    \int \frac{e^{-ar}}{r^4} \leq e^{-ad} \int \frac{1}{r^4} = \pi e^{-ad}
    \begin{cases}
        \frac{1}{d} & \textrm{halfspace} \\
        \frac{1}{d^2} & \textrm{2D plane} \\
        \frac{4}{d} & \textrm{3D spherical shell} \\
        \frac{4}{d^2} & \textrm{2D spherical surface}.
    \end{cases}
\end{align}
Thus we are left with the final two terms of \eqref{bnd_terms} (the $1/r^n$ integrals are calculated separately in \secref{gf}). For those terms, we cannot extract the exponential from the integrand, as the remaining integral is divergent for the 3D cases. Thus, we compute the integrand \emph{with} the $e^{-ad}$ term in the integrand.

With that term, we find the bounds:
\begin{align}
    \int \frac{e^{-ar}}{r^3} \leq 2\pi
    \begin{cases}
        (ad+1) \operatorname{E_1}(ad) & \textrm{halfspace} \\
        \frac{e^{-ad}}{d} & \textrm{2D plane} \\
        2 \operatorname{E_1}(ad) & \textrm{3D spherical shell} \\
        2 \frac{e^{-ad}}{d} & \textrm{2D spherical surface}.
    \end{cases}
\end{align}
and
\begin{align}
    \int \frac{e^{-ar}}{r^2} \leq 2\pi
    \begin{cases}
        \frac{e^{-ad}}{a} & \textrm{halfspace} \\
        \operatorname{E_1}(ad) & \textrm{2D plane} \\
        \frac{2e^{-ad}}{a} & \textrm{3D spherical shell} \\
        2e^{-ad} & \textrm{2D spherical surface}.
    \end{cases}
\end{align}
where $\operatorname{E_1}(x) = -\operatorname{Ei}(-x)$ and $\operatorname{Ei}(-x)$ is the exponential integral.

For the halfspace, for example, we can put all of this together to bound \eqref{lorbound} above by
\begin{align}
    \frac{\langle \rho \rangle}{\rho_0(|\omega|)} \leq \frac{f(\omega)e^{-ad}}{8} \left[ \frac{C_6}{6| k |^3 d^3} + \frac{C_5}{3| k |^2 d^2} + \frac{C_4}{| k | d} + 2C_3 (ad+1)\operatorname{E_1}(ad) e^{ad} + \frac{2C_2 | k |}{a} \right].
\end{align}
For a spherical shell, we get the bound
\begin{align}
     \frac{\langle \rho \rangle}{\rho_0(|\omega|)} \leq  \frac{f(\omega)e^{-ad}}{8} \left[ \frac{C_6}{3| k |^3 d^3} + \frac{2C_5}{| k |^2 d^2} + \frac{4C_4}{| k | d} + 4C_3\operatorname{E_1}(ad) e^{ad} + \frac{4C_2 | k |}{a} \right].
\end{align}
For a 2D plane:
\begin{align}
     \frac{\langle \rho \rangle}{\rho_0(|\omega|)} \leq \frac{f(\omega)e^{-ad}}{8} \left[ \frac{C_6}{2| k |^4 d^4} + \frac{2C_5}{3| k |^3 d^3} + \frac{C_4}{| k |^2 d^2} + \frac{2C_3}{| k | d} + 2 C_2\operatorname{E_1}(ad) e^{ad} \right].
\end{align}
Finally, for a 2D spherical surface:
\begin{align}
     \frac{\langle \rho \rangle}{\rho_0(|\omega|)} \leq  \frac{f(\omega)e^{-ad}}{8} \left[ \frac{C_6}{| k |^4 d^4} + \frac{4C_5}{| k |^3 d^3} + \frac{4C_4}{| k |^2 d^2} + \frac{4C_3}{| k | d} + 4 C_2 \right].
\end{align}
For simplicity in the above equations, we have dropped the last term in \eqref{lorbound} coming from $\alphaL$, which is negligible for bandwidths $\Delta \omega$ that are smaller than typical frequencies $|\omega|$ considered.

\section{Green's-function-relevant integrals: halfspaces and spheres}  \label{sec:gf}
We consider integrals of various-order $1/r^n$ and $e^{-ar}/r^n$ terms, that arise repeatedly in our various bound formulations.

\subsection{Planar integrals}
Here we consider integrals from a point to either a 2D planar surface or a 3D halfspace. For the halfspace, the integral of some function $f(x)$ can be written
\begin{align}
    \int f(x) \,{\rm d}V \rightarrow 2\pi \int_d^\infty \,{\rm d}z \int_0^\infty \,{\rm d}\rho \left[ \rho f(\rho,z) \right]. \label{eq:plaint} 
\end{align}
\Eqref{plaint} can be applied to a 2D planar surface simply by removing the integral over $z$ on the right-hand side, and thus the 2D-plane integrals below emerge as intermediate steps of the halfspace solution.

\tocless\subsubsection{$\int \frac{1}{r^6} \,{\rm d}V$}
The integral of $1/r^6$ is
\begin{align}
    \int \frac{1}{r^6} \,{\rm d}V = 2\pi \int \,{\rm d}z \int \,{\rm d}\rho \frac{\rho}{\left(\rho^2 + z^2\right)^3} = \frac{\pi}{2} \int \,{\rm d}z \frac{1}{z^4} = \frac{\pi}{6} \left(-\frac{1}{z^3}\right)\bigg\rvert_d^\infty = 
    \begin{cases}
        \frac{\pi}{6d^3} & \textrm{halfspace} \\
        \frac{\pi}{2d^4} & \textrm{2D plane}
    \end{cases}
\end{align}

\tocless\subsubsection{$\int \frac{1}{r^5} \,{\rm d}V$}
The integral of $1/r^5$ is
\begin{align}
    \int \frac{1}{r^5} \,{\rm d}V = 2\pi \int \,{\rm d}z \int \,{\rm d}\rho \frac{\rho}{\left(\rho^2 + z^2\right)^(5/2)} = \frac{2\pi}{3} \int \,{\rm d}z \frac{1}{z^3} =
    \begin{cases}
        \frac{\pi}{3d^2} & \textrm{halfspace} \\
        \frac{2\pi}{3d^3} & \textrm{2D plane}
    \end{cases}
\end{align}

\tocless\subsubsection{$\int \frac{1}{r^4} \,{\rm d}V$}
The integral of $1/r^4$ is
\begin{align}
    \int \frac{1}{r^4} \,{\rm d}V = 2\pi \int \,{\rm d}z \int \,{\rm d}\rho \frac{\rho}{\left(\rho^2 + z^2\right)^2} = \pi \int \,{\rm d}z \frac{1}{z^2} =
    \begin{cases}
        \frac{\pi}{d} & \textrm{halfspace} \\
        \frac{\pi}{d^2} & \textrm{2D plane}
    \end{cases}
\end{align}

\tocless\subsubsection{$\int \frac{e^{-ar}}{r^3} \,{\rm d}V$}
The integral of $e^{-ar}/r^3$ is
\begin{align}
    \int \frac{e^{-ar}}{r^3} \,{\rm d}V &= 2\pi \int \,{\rm d}z \int \,{\rm d}\rho \frac{\rho e^{-a\sqrt{\rho^2+z^2}}}{\left(\rho^2 + z^2\right)^{3/2}} \nonumber \\
    &= 2\pi \int_d^\infty \,{\rm d}z \left[a \operatorname{Ei}(-az) + \frac{e^{-az}}{z} \right] \nonumber \\
    &= 
    \begin{cases}
        2\pi \left[\left(ad + 1\right)\operatorname{E_1}(ad) - e^{-ad} \right] & \textrm{halfspace} \\
        2\pi \left[\frac{e^{-ad}}{d} - a\operatorname{E_1}(ad) \right] & \textrm{2D plane},
    \end{cases}
\end{align}
where $\operatorname{Ei}$ denotes the exponential integral and $\operatorname{E_1}(x) = -\operatorname{Ei}(-x)$ for positive real $x$. These expressions can be further simplified via the bounds
\begin{align}
    \frac{1}{2} e^{-x} \ln \left(1 + \frac{2}{x}\right) < \operatorname{E_1}(x) < e^{-x} \ln \left(1 + \frac{1}{x}\right)
\end{align}

\tocless\subsubsection{$\int \frac{e^{-ar}}{r^2} \,{\rm d}V$}
The integral of $e^{-ar}/r^2$ is
\begin{align}
    \int \frac{e^{-ar}}{r^2} \,{\rm d}V &= 2\pi \int \,{\rm d}z \int \,{\rm d}\rho \frac{\rho e^{-a\sqrt{\rho^2+z^2}}}{\rho^2 + z^2} \nonumber \\
    &= -2\pi \int_d^\infty \,{\rm d}z \operatorname{Ei}(-az) \nonumber \\
    &= 
    \begin{cases}
        2\pi \left[\frac{e^{-ad}}{a} - d\operatorname{E_1}(ad)\right] & \textrm{halfspace} \\
        2\pi \operatorname{E_1}(ad) & \textrm{2D plane},
    \end{cases}
\end{align}
with the same notation and possible simplification as in the previous section.

\tocless\subsubsection{$\int \frac{e^{-ar}}{r^4} \,{\rm d}V$}
Although we don't use it in the power--bandwidth limits, it is useful to compare the integral of $e^{-ar}/r^4$ to $e^{-ad}$ times the integral for $1/r^4$. The integral of $e^{-ar}/r^4$ is
\begin{align}
    \int \frac{e^{-ar}}{r^4} \,{\rm d}V &= 2\pi \int \,{\rm d}z \int \,{\rm d}\rho \frac{\rho e^{-a\sqrt{\rho^2+z^2}}}{(\rho^2 + z^2)^2} \nonumber \\
    &= \pi \int_d^\infty \,{\rm d}z \left[ -a^2 \operatorname{Ei}(-az) + \frac{e^{-az}(az-1)}{z^2} \right] \nonumber \\
    &= 
    \begin{cases}
        \pi \left[\frac{e^{-ad}}{d}\left(1 + ad\right) - a^2 d \operatorname{E_1}(ad)\right] & \textrm{halfspace} \\
        \pi a^2 \operatorname{E_1}(ad) + \frac{e^{-ad}(ad-1)}{z^2} & \textrm{2D plane}.
    \end{cases}
\end{align}
Note that for $ad$ small ($ad \ll 1$), the halfspace bound simplifies to $\pi e^{-ad}/d$, which is exactly what we get when we pull $e^{-ad}$ out of the integrand and multiply it by the integral of $1/r^4$. Hence for the typical case of $ad$ small, pulling the exponential out of the integrand hardly loosens the bound.

\subsection{Spherical integrals}
Here we consider integrals from a point to either a 2D spherical surface or a 3D spherical shell of infinite thickess. Then, the integral of some function $f(x)$ can be written
\begin{align}
    \int f(x) \,{\rm d}V \rightarrow 4\pi \int_d^\infty \,{\rm d}r \left[ r^2 f(r) \right].
     \label{eq:sphint} 
\end{align}
\Eqref{sphint} can be applied to a 2D spherical surface by removing the integral over $r$ on the right-hand side, and thus the 2D-spherical-surface integrals below emerge as intermediate steps of the shell solution. For a smaller solid angle $\Omega$, one can always make the replacement
\begin{align}
    4\pi \rightarrow \Omega
\end{align}

\tocless\subsubsection{$\int \frac{1}{r^6} \,{\rm d}V$}
The integral of $1/r^6$ is
\begin{align}
    \int \frac{1}{r^6} \,{\rm d}V = 4\pi \int \,{\rm d}r \frac{1}{r^4} =  
    \begin{cases}
        \frac{4\pi}{3d^3} & \textrm{3D spherical shell} \\
        \frac{4\pi}{d^4} & \textrm{2D spherical surface}
    \end{cases}
\end{align}

\tocless\subsubsection{$\int \frac{1}{r^5} \,{\rm d}V$}
The integral of $1/r^5$ is
\begin{align}
    \int \frac{1}{r^5} \,{\rm d}V = 4\pi \int \,{\rm d}r \frac{1}{r^3} =  
    \begin{cases}
        \frac{2\pi}{d^2} & \textrm{3D spherical shell} \\
        \frac{4\pi}{d^3} & \textrm{2D spherical surface}
    \end{cases}
\end{align}

\tocless\subsubsection{$\int \frac{1}{r^4} \,{\rm d}V$}
The integral of $1/r^4$ is
\begin{align}
    \int \frac{1}{r^4} \,{\rm d}V = 4\pi \int \,{\rm d}r \frac{1}{r^2} =  
    \begin{cases}
        \frac{4\pi}{d} & \textrm{3D spherical shell} \\
        \frac{4\pi}{d^2} & \textrm{2D spherical surface}
    \end{cases}
\end{align}

\tocless\subsubsection{$\int \frac{e^{-ar}}{r^3} \,{\rm d}V$}
The integral of $e^{-ar}/r^3$ is
\begin{align}
    \int \frac{e^{-ar}}{r^3} \,{\rm d}V &= 4\pi \int_d^\infty \,{\rm d}r \frac{e^{-ar}}{r} \nonumber \\
    &= 
    \begin{cases}
        4\pi \operatorname{E_1}(ad) & \textrm{3D spherical shell} \\
        4\pi \frac{e^{-ad}}{d} & \textrm{2D spherical surface},
    \end{cases}
\end{align}
with the same notation and simplifications as the corresponding halfspace sections.

\tocless\subsubsection{$\int \frac{e^{-ar}}{r^2} \,{\rm d}V$}
The integral of $e^{-ar}/r^2$ is
\begin{align}
    \int \frac{e^{-ar}}{r^2} \,{\rm d}V &= 4\pi \int_d^\infty e^{-ar} \,{\rm d}r \nonumber \\
    &= 
    \begin{cases}
        \frac{4\pi}{a} e^{-ad} & \textrm{3D spherical shell} \\
        4\pi e^{-ad} & \textrm{2D spherical surface},
    \end{cases}
\end{align}
with the same notation and possible simplification as the corresponding halfspace sections.

\section{LDOS from Drude-Lorentz model vs. Palik data}
In Fig.~4 of the main text, we have used the Drude-Lorentz (DL) multi-oscillator model to describe the average LDOS and bounds for Ag, since it behaves nicely for complex frequencies in the UHP. In principle, however, we can analytically continue Palik data~\cite{palik_2003} for permittivity of Ag to the UHP (but using DL model is simpler as it explicitly provides an analytic expression for permittivity).

\begin{figure*} [t!]
    \includegraphics[width=1\linewidth]{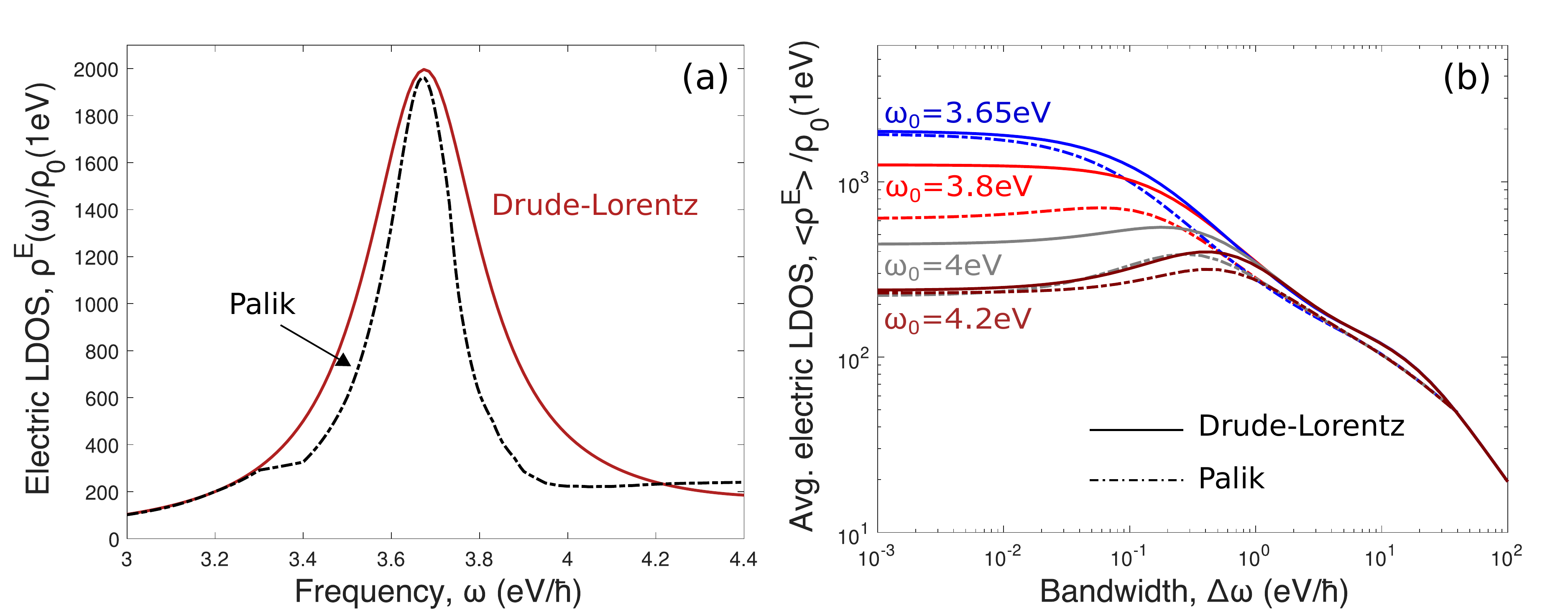} 
    \caption{(a) Electric LDOS from Palik data for permittivity of Ag compared to that based on Drude-Lorentz multi-oscillator model. (b) Average electric LDOS at various center frequencies, which show close agreement at center frequencies of 3.65eV and 4.2eV. At frequencies between these two, the difference in average LDOS reflects the fact that the DL model does not match Palik data very well at these frequencies. The emitter-scatterer distance \textit{d} is set to 10nm, and the scatterer here is Ag halfspace.}  
    \label{fig:dlvsexp} 
\end{figure*}

\Figref{dlvsexp}(a) shows how the DL model fit compares to LDOS for bulk Ag (halfspace). In \figref{dlvsexp}(b), we compute the average LDOS for both the DL model and Palik data. Notice that for the latter, since we do not have data for complex frequencies, we have numerically integrated LDOS (weighted by the Lorentzian in Eq.~(15) of main text) over real frequencies. Although the two results are extremely close to each other at center frequencies of 3.65eV and 4.2eV, with greater deviation at 3.8eV and 4.0eV due to the error incurred in approximating the Palik data by a DL model. 

\section{Lorentzian approximation to the mean Planck energy $\Theta(\omega)$}
\begin{figure*} [t!]
    \includegraphics[width=1\linewidth]{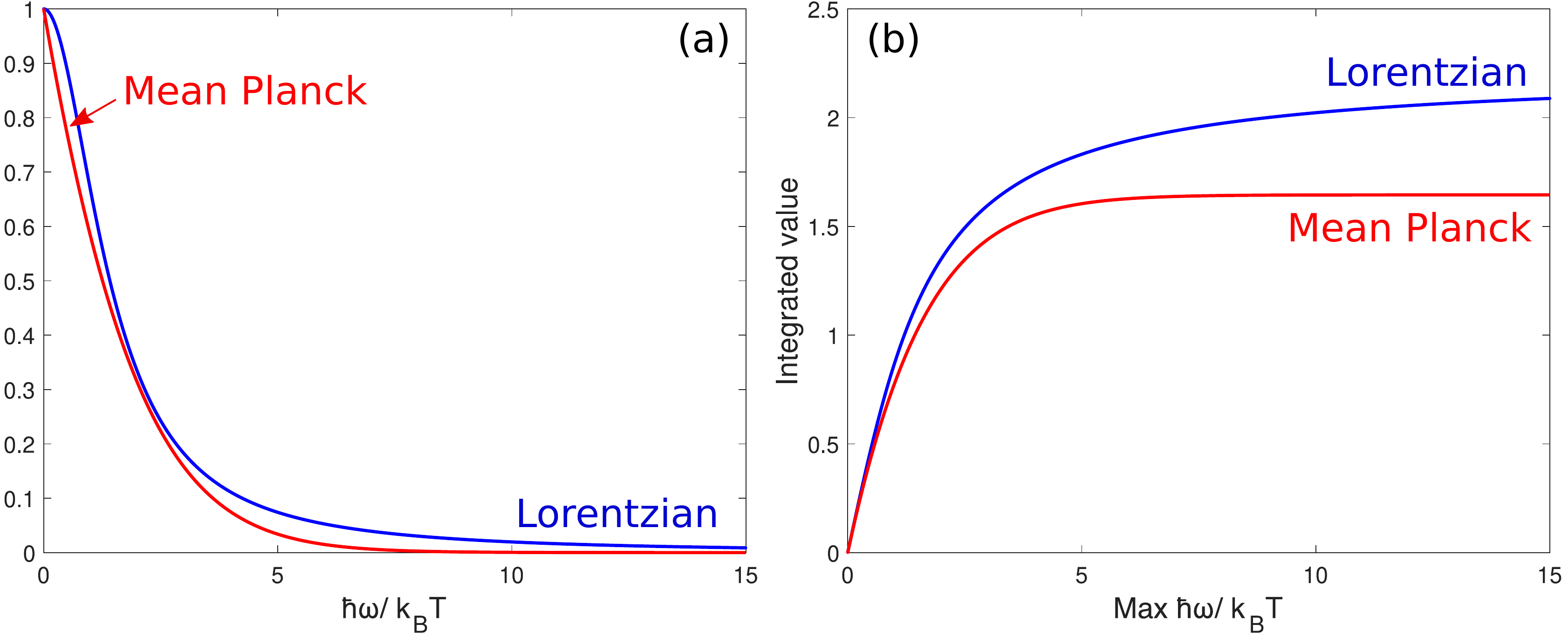} 
    \caption{(a) Plot of the mean Planck energy and its Lorentzian approximation, both normalized to unity at the origin. The x-axis is a dimensionless parameter proportional to frequency, given a temperature set by the problem. (b) Comparison of the integral of these two curves up to some maximum range. Although not clear from the figure, their integral over all positive frequencies converges and are very close---suggesting that the Lorentzian is indeed a close bound to the mean energy.}  
    \label{fig:plvslor} 
\end{figure*}

In bounding near-field RHT, we have chosen to replace the mean Planck energy $\Theta(\omega)$ by a Lorentzian, which is more amenable to contour-integration techniques (involved in bounding RHT). As long as the Lorentzian strictly is larger than $\Theta$ for all frequencies, the Lorentzian approximation is indeed a bound for the actual RHT. However, a more ideal bound is one that minimizes their difference, and in  \figref{plvslor} we compare how closely our Lorentzian approximates the mean energy. 

As \figref{plvslor}(a) illustrates, there is close agreement between the two curves (normalized such that they both have unit magnitude at the origin). In  \figref{plvslor}(b), the integrated values are not fully converged, but one can show analytically that the area for the (peak-normalized-to-unity) Lorentzian and the mean energy is $\sqrt{2} \pi/2 \approx 2.22 $ and $\pi^2 / 6 \approx 1.64$, respectively. Thus, the Lorentzian has about 1.35 times more area than for the mean energy $\Theta$ as quoted in the main text. Although their high-frequency asymptotic behavior differs qualitatively (polynomial vs. exponential decay), the Lorentzian approximation provides a tight bound to $\Theta$ and thus comes in handy for RHT bounds.

\section{Power--bandwidth bounds for near-field RHT}
Given object 1 at temperature $T_1$ and object 2 at $T_2$, we can write the radiative heat flux between the two bodies as $H_{1 \rightarrow 2}(\omega) = \Phi(\omega) \left[\planck(\omega,T_1) - \planck(\omega,T_2)\right]$, where $\planck(\omega,T)$ denotes the Planck spectrum and $\Phi$ is a temperature independent flux rate from incoherent sources in body 1 radiating to body 2. In order to derive upper limits to the heat flux, we need to bound the structure-dependent flux rate $\Phi(\omega)$ (since $\planck(\omega,T)$ is a known, broad spectrum). 

Starting from the heat flux between two bodies expressed in terms of the electromagnetic Poynting vector,
\begin{align}
    H_{1 \rightarrow 2}(\omega) &= \frac{1}{2} \Re \int_S \left( \Ev \times \cc{\Hv} \right) \cdot \mathbf{\hat{n}} \nonumber \\
    &= \frac{1}{4} \int_S \psi^{\dagger} \Lambda \psi \label{eq:tent}
\end{align} 
we can relate the fields to the fluctuation currents in volume 1, $V_1$, in terms of the Green's functions (in index notation):
\begin{align}
H_{1 \rightarrow 2}(\omega) &=  \frac{1}{4} \int_S  \cc{\psi_i}(\xs) \Lambda_{ij}(\xs)\psi_j (\xs) \nonumber \\
&= \frac{1}{4} \int_S \int_{V_1(\xv)} \int_{V_1(\xv')} \cc{\Gamma_{ik}}(\xs, \xv) \cc{\phi_{k}}(\xv) \Lambda_{ij}(\xs) \Gamma_{jl}(\xs, \xv') \phi_{l}(\xv').
\end{align}
The fluctuation--dissipation theorem~\cite{joulain_mulet_marquier_carminati_greffet_2005} gives the ensemble-average of the currents (taking $\epso=\mu_0=1$ for simplicity): 
\begin{align}
    \left\langle \cc{\phi_i}(\xv') \phi_j(\xv'') \right \rangle_{FD} = \frac{4}{\pi\omega} \planck(\omega,T) \Im \chi_{ij}(\omega) \delta(\xv' - \xv'').
\end{align}
where $\textsubscript{FD}$ indicates that we are averaging over the fluctuation currents (not to be confused with averaging over frequency bandwidth). Since the Planck spectrum is known, we will simply work with the energy flux $\Phi(\omega)$ in what follows. 

Inserting the average fluctuations into the power, taking the transpose, and using generalized reciprocity~\cite{kong_1975} ($\tens{\Gamma}(\xs,\xv) = \tens{\Gamma}^{'T}(\xv,\xs)$ where $'$ denotes flipping the sign of $3\times3$ off-diagonal matrices $\tens{G}^{EH}$ and $\tens{G}^{HE}$) plus Cholesky decomposition on the Hermitian, positive-definite matrix $\Im \chi$ (expressed as $LL^{\dagger}$) leads to 
\begin{align}
    \Phi(\omega) &= \frac{1}{\pi\omega}\int_S \int_{V_1} \cc{\Gamma_{ik}^T}(\xs, \xv) \Lambda_{ji}(\xs)  \Im \chi_{lk}(\omega) {\Gamma_{jl}^T}(\xs, \xv) \nonumber \\
      &= \frac{1}{\pi\omega}\int_S \int_{V_1} \cc{\Gamma_{ki}^{'}}(\xv, \xs) \Lambda_{ji}(\xs) L_{l\alpha}L_{\alpha k}^{\dagger}  \Gamma_{lj}^{'}(\xv, \xs) \nonumber \\
      &=  \frac{1}{\pi\omega} \sum_{\{i,j\}}\int_S \int_{V_1} c_{ij}(\xs)\cc{ L_{\alpha k}^{T} \Gamma_{ki}^{'}(\xv, \xs)} L_{\alpha l}^{T}  \Gamma_{lj}^{'}(\xv, \xs)
      \label{eq:RHT_cross_terms}
\end{align}
where $\{i,j\}$ indicates the allowed dipole orientations by $\Lambda_{ji}$, which is nonzero only for certain combinations of $i,j$ and whose structure is reflected in the coefficients $c_{ij}$. Note that if the fields were from the same dipole (remember $ \Gamma_{kj}^{'}(\xv, \xs)$ represents the field due to a $\phi_j$ up to a sign), then the integral over $V_1$ in \eqref{RHT_cross_terms} would correspond to power absorbed in $V_1$ due to that dipole (note also the $\Im \chi_1$ prefactor). Through the Cauchy--Schwarz inequality, we can bound \eqref{RHT_cross_terms} exactly in terms of such dissipation:
\begin{align}
    \Phi(\omega) &\leq \frac{1}{\pi\omega} \sum_{\{i,j\}}\int_S |c_{ij}| \sqrt{ \int_{V_1} \left |L_{\alpha k}^{T} \Gamma_{ki}^{'}(\xv, \xs) \right |^2 \int_{V_1} \left |L_{\alpha l}^{T}  \Gamma_{lj}^{'}(\xv, \xs) \right |^2 } \nonumber \\
    &\leq \frac{1}{\pi\omega} \sum_{\{i,j\}}\int_S |c_{ij}| \sqrt{ \left( \int_{V_1} \psi_{\xi_i(\xs)}(\xv) \Im \chi  \, \psi_{\xi_i(\xs)}(\xv) \right) \left( \int_{V_1} \psi_{\xi_j(\xs)}(\xv) \Im \chi \, \psi_{\xi_j(\xs)}(\xv) \right) } 
\end{align}
where $ \psi_{\xi_i(\xs)}(\xv)$ denotes the field at $\xv$ due to a dipolar source $\xi_i$ at $\xs$. We can relate this to power absorption due to each individual dipole by inserting an $\omega/2$ prefactor:
\begin{align}
    \Phi(\omega) \leq \frac{2}{\pi\omega^2} \sum_{\{i,j\}}\int_S |c_{ij}| \sqrt{\left( \frac{\omega}{2} \int_{V_1} \Im \chi \left |\psi_{\xi_i(\xs)}(\xv) \right |^2 \right) \left( \frac{\omega}{2} \int_{V_1} \Im \chi \left |\psi_{\xi_j(\xs)}(\xv) \right |^2 \right)} 
\end{align}
where the terms in parentheses are now exactly the powers absorbed in $V_1$ by each dipole. Thus we can write
\begin{align}
    \Phi(\omega) \leq \frac{2}{\pi\omega^2} \sum_{\{i,j\}}\int_S |c_{ij}| \sqrt{P_{V_1,\xi_i(\xs)}(\omega) P_{V_1,\xi_j(\xs)}(\omega) }.
\end{align}
The power radiated by a given dipole into $V_1$ is less than or equal to the total power radiated by that dipole. If we denote the polarization-resolved LDOS as $\rho_i = (6/\pi\omega) \Im \phi_i^\dagger \psi$, with the factor of 6 introduced because we are no longer averaging over dipole orientation, then
\begin{align}
    P_{\rm rad} = \frac{\pi \omega^2}{6} \rho_i.
\end{align}
Here we will make our first approximation: that in the near-field cases of interest, the LDOS $\rho_i$ is much larger than its free-space value, i.e. $\rho_i \gg \rho_0$, in which case we can approximate $\rho_i$ as the \emph{scattered-field} LDOS. Henceforth $\rho_i$ will denote the scattered-field LDOS. The inequality becomes
\begin{align}
    \Phi(\omega) \leq \frac{1}{3} \sum_{\{i,j\}}\int_S |c_{ij}|  \sqrt{\rho_{\xi_i(\xs)}(\omega) \rho_{\xi_j(\xs)}(\omega) }.
\end{align}
Now, we want to integrate over frequency and multiply by a bandwidth weighting function $H_{\omega_0,\Delta\omega}(\omega)$. We will hold off on the weight function for a moment, and again use Cauchy--Schwarz to simplify the integrals:
\begin{align}
     \int_{-\infty}^\infty \Phi(\omega) {\rm d}\omega \leq \frac{1}{3} \sum_{\{i,j\}} \int_S  |c_{ij}| \sqrt{ \left( \int_{-\infty}^\infty {\rm d}\omega \rho_{\xi_i(\xs)}(\omega) \right) \left( \int_{-\infty}^\infty {\rm d}\omega \rho_{\xi_j(\xs)}(\omega) \right) }            
\end{align}
Noting that the additional term in the integrand $H_{\omega_0,\Delta\omega}(\omega)$ would pass through to the integrals under the square roots, we can finally bound:
\begin{align}
    \langle \Phi \rangle \equiv \int_{-\infty}^\infty \Phi(\omega)H_{\omega_0,\Delta\omega}(\omega) {\rm d}\omega \leq \frac{1}{3} \sum_{\{i,j\}} \int_S |c_{ij}|\sqrt{\langle \rho_{\xi_i(\xs)} \rangle \langle \rho_{\xi_j(\xs)} \rangle}.\label{eq:rhtbound0} 
\end{align}
While \eqref{rhtbound0} depends on the shape of the bounding surface, we can further simplify by noticing that for each point $\xs$ on the bounding surface, we are free to choose our coordinate system such that $\mathbf{\hat{n}}$ is along the $z$-axis. Then, since $ \Lambda = \begin{pmatrix}& -\mathbf{\hat{z}} \times \\  \mathbf{\hat{z}} \times &  \end{pmatrix}$ has only four non-zero elements all with unit magnitude, we have that  $ \sum_{\{i,j\}} |c_{ij}| = 4$). Also, by virtue of the off-diagonal structure of $\Lambda$, an electric dipole ($i=1,2,3$) will always be accompanied by a magnetic dipole ($j=4,5,6$) and vice versa. Using the fact that $\langle \rho_{\pv_j(\xs)} \rangle \leq 3 \langle \rho^E \rangle$ and similarly for $ \langle \rho_{\mv_j(\xs)} \rangle$ (also including $c$ for correct dimensions---equivalent to keeping $\epso, \mu_0$),
\begin{align}
  \langle \Phi \rangle \leq 4c \int_S \sqrt{\langle \rho^E \rangle \langle \rho^H \rangle}.\label{eq:rhtbound} 
\end{align}
\Eqref{rhtbound} is the result used in Eq.(33) of the main text.

\section{Power--bandwidth bounds for CDOS}
Optimizing average CDOS very closely follows the derivation laid out in \secref{forpb}. In fact, the only difference (apart from replacing the sum-rule constant $\alphaL$ by $\alphaC$) is that we now have to maximize $\Im \left[ \frac{1}{\pi\omega} \int_V  \widetilde{\psi}^T_{\rm inc,2}(\omega) \chi(\omega) \psi(\omega) \right]$ where $\widetilde{\psi}_{\rm inc,2} = \begin{pmatrix} \Ev_{\rm inc,2} & -\Hv_{\rm inc,2} \end{pmatrix}^T$ is the incident field as if the dipole source were at the measurement position $\xv$ instead of the emitter position $\xv_0$ (up to a sign flip for the magnetic field). With this simple modification, We can carry over the results in \secref{forpb} (formulating the optimization problem in the form of \eqref{genopt}, just with slightly modified variables) to arrive at the following bound on average CDOS appearing in Eq.(27) of the main text (where $\omega = \omega_0 + i\Delta \omega$):
\begin{align}
    \langle \rho_{ij} \rangle  \leq & \, \frac{f(\omega)}{\pi\left|\omega \right|} \sqrt{\left( \int_V \psi_{\textrm{inc},1,i}^\dagger(\omega) \psi_{\textrm{inc},1,i}(\omega) \,{\rm d}V \right) \left(  \int_V \psi_{\textrm{inc},2,j}^\dagger(\omega) \psi_{\textrm{inc},2,j}(\omega) \,{\rm d}V \right)} + 2H_{\omega_0, \Delta \omega}(0) \alphaC \label {eq:lorcbound}
\end{align}
where ${\psi}_{\rm inc,1}$ is the usual incident field from the dipole at $\xv_0$. The calculation now proceeds similar to \eqref{gfsquare}, but now we do not sum over the dipole orientations (assuming nonmagnetic materials in the presence of an electric dipole for simplicity):
\begin{align}
    \int_V | \Ev_{\textrm{inc},j}  | ^2 \rightarrow \sum_{i} \int_V |G_{ij}|^2
    \label{eq:gfcsquare}
\end{align}
where $j$ refers to the dipole polarization at $\xv_0$ (or at $\xv$ if we consider $\Ev_{\textrm{inc},2,j}$). In general, \eqref{gfcsquare} depends on this polarization, but we can bound the integral by a polarization-independent term (where r denotes the emitter-scatterer distance and other terms appearing below are defined near the end of \secref{forpb}):
\begin{align}
    \sum_{i} \int_V |G_{ij}|^2 &= \frac{\left| k \right| ^4 e^{-ar}}{16\pi^2 r^2} \left[ \left| \left(1 + \frac{i}{a} - \frac{1}{a^2}\right) \right| ^2 + 2\Re{ \left( \left(1 + \frac{i}{a} - \frac{1}{a^2}\right) \cc{ \left(-1 - \frac{3i}{a} + \frac{3}{a^2}\right)} \right) }\frac{x_j^4}{r^4} +  \left| \left(-1 - \frac{3i}{a} + \frac{3}{a^2}\right)  \right| ^2 \frac{x_j^4}{r^4}  \right] \nonumber \\
    &\leq \frac{\left| k \right| ^4 e^{-ar}}{16\pi^2 r^2} \left[ \left| \left(1 + \frac{i}{a} - \frac{1}{a^2}\right) \right| ^2 + 2\Re{ \left( \left(1 + \frac{i}{a} - \frac{1}{a^2}\right) \cc{ \left(-1 - \frac{3i}{a} + \frac{3}{a^2}\right)} \right) } +  \left| \left(-1 - \frac{3i}{a} + \frac{3}{a^2}\right)  \right| ^2  \right] \nonumber \\
    &= \frac{\left| k \right| ^6 e^{-ar}}{4\pi^2} \sum_{n=4}^{6} \frac{B_n e^{-ar}}{|k|^{n} r^n} 
    \label{eq:gfcbound}
\end{align}
where the coefficients $B_n$'s are given by:
\begin{align*}
    B_4 &=1 \\
    B_5&=2\frac{\Delta\omega}{|\omega|} \\
    B_6&=1.
\end{align*}
Written out in full form, we can bound average CDOS for nonmagnetic materials in the presence of an electric dipole (where $r_1$ and $r_2$ respectively denote the distance from the scatter to the source and measurement positions):
\begin{align}
\langle \rho_{ij} \rangle  \leq \rho_0(|\omega|)\frac{f(\omega) \left| k \right|^3}{2\pi} \sqrt{ \left( \int_{V} \sum_{n=4}^{6} \frac{B_n e^{-ar_1}}{|k|^{n} r^n_1} \right) \left( \int_{V} \sum_{n=4}^{6} \frac{B_n e^{-ar_2}}{|k|^{n} r^n_2} \right) } + 2H_{\omega_0, \Delta \omega}(0) \alphaC. \label{eq:lorcboundint}
\end{align}
For a halfspace enclosure, we can exactly calculate all the integrals appearing in \eqref{lorcboundint} as was done in \secref{gf} to arrive at the following bound (dropping the last term in \eqref{lorcboundint} coming from $\alphaC$, which is negligible for bandwidths $\Delta \omega$ small compared to typical frequencies $|\omega|$):
\begin{align}
 \frac{ \langle \rho_{ij} \rangle}{\rho_0(|\omega|)} \leq \frac{f(\omega)e^{-a(d_1+d_2)/2}}{2} \sqrt{ \left(  \frac{B_6}{6| k |^3 d_1^3} + \frac{B_5}{3| k |^2 d_1^2} + \frac{B_4}{| k | d_1} \right) \left(  \frac{B_6}{6| k |^3 d_2^3} + \frac{B_5}{3| k |^2 d_2^2} + \frac{B_4}{| k | d_2}  \right) }  \label{eq:lorcboundfull}
\end{align}
where $d_1$ and $d_2$ denote the minimum value of $r_1$ and $r_2$ respectively. We can also extend \eqreftwo{lorcboundint}{lorcboundfull} for magnetic materials as well as magnetic LDOS by modifying the free-space Green's function to accommodate for magnetic fields and/or dipoles. It can also be extended for 2D materials by modifying the volume to a surface integral.

\providecommand{\noopsort}[1]{}\providecommand{\singleletter}[1]{#1}%